\documentclass[sigconf, nonacm, balance=false]{acmart}

\usepackage{amsmath,amsthm}
\usepackage{booktabs,array,multirow,tabularx}
\usepackage{algorithm}
\usepackage{algpseudocode}
\usepackage{subcaption}
\usepackage{microtype}

\newtheorem{theorem}{Theorem}
\newtheorem{proposition}[theorem]{Proposition}
\newtheorem{lemma}[theorem]{Lemma}

\theoremstyle{definition}

\graphicspath{{figures/}}

\newif\ifpreprint
\preprinttrue

\newcommand{\rquiver}{Pelorus}

\newcommand{\quiver}{QuIVer}
\newcommand{\Bball}{\mathbb{B}_c^D}
\newcommand{\Bunit}{\mathbb{B}_1^D}
\newcommand{\Lmod}{\mathbb{L}_c^D}
\newcommand{\Sunit}{\mathbb{S}^{D-1}}
\newcommand{\R}{\mathbb{R}}
\newcommand{\E}{\mathbb{E}}
\newcommand{\Var}{\operatorname{Var}}
\newcommand{\Cov}{\operatorname{Cov}}
\newcommand{\MNR}{\mathrm{MNR}}
\newcommand{\sgn}{\operatorname{sgn}}
\newcommand{\arccosh}{\operatorname{arccosh}}
\newcommand{\atanh}{\operatorname{atanh}}
\newcommand{\asinh}{\operatorname{asinh}}
\newcommand{\popcnt}{\operatorname{popcnt}}
\newcommand{\missing}{\textemdash}

\newcommand\codeavailabilityurl{https://github.com/Astrmira/Pelorus}

\begin{document}

\title[Pelorus: Warped-Product Graph Indexing for Hyperbolic Vector Search]{Pelorus: Warped-Product Graph Indexing for Hyperbolic Vector Search}

\author{Wenxuan Xiao}
\affiliation{%
  \institution{Astrmira Tech.}
  \country{China}
}
\email{w.xiao@astrmira.com}

\author{Xu Cao}
\affiliation{%
  \institution{Astrmira Tech.}
  \country{China}
}
\email{x.cao@astrmira.com}

\begin{abstract}
Hyperbolic embeddings in the {Poincar\'e} ball and {Lorentz} hyperboloid represent hierarchical, taxonomic, and entailment relations with substantially lower distortion than flat {Euclidean} spaces, supporting retrieval-augmented generation, vision-language alignment, and knowledge-base search.
Serving million-scale approximate nearest neighbor ({ANN}) queries over curved manifolds, however, exposes a fundamental tension between geometric fidelity and memory efficiency: flat inner-product indices distort geodesic order across radial shells, whereas native hyperbolic graph indices either fetch full-precision coordinates at every traversal hop or quantize ambient coordinates without decoupling conformal scales.
Transplanting flat proximity-graph pruning and vector compression to hyperbolic space also fails.
Because negative curvature bends geodesics toward the origin and scales directional inner products by the singular conformal factor $(1 - c\|x\|_2^2)^{-1}$, interior hubs geodesically occlude entire spherical neighborhoods during {Vamana} pruning, shared-query quantization errors fail to cancel across radial shells, and tight boundary clusters suffer margin-to-noise ratio ({$\MNR$}) collapse.

We present \textbf{\rquiver{}}, a warped-product proximity-graph index for hyperbolic vector search.
\rquiver{} decouples exact single-precision radial and conformal scalars in a 32-byte vertex header from scale-invariant spherical directional codes---instantiated with training-free 2-bit Sign-Magnitude ({SM}) bitplanes---deferring full-vector double-precision geodesic evaluation to a compact reranking tail.
Analyzing how hyperbolic conformal weights and geodesic curvature interact with polar directional quantization, we establish query boundary-depth invariance on isobathic shells, characterize cross-shell query-noise non-cancellation, derive a deterministic parameter-free orthogonal-projection estimator, and introduce a purity-gated local tangent-cone residual code with Conformal Jacobian gating (\textsc{CJ-Gate}) that attenuates intra-cone tangent error with zero per-vertex byte overhead and without global random rotations while keeping diffuse manifolds on the zero-overhead global 2-bit path.
Guided by these geometric properties, we construct a warped-product fiber-bundle graph inside-out using {QuadriPrune}---partitioning each vertex's out-degree across four geometric channels (stem-excess geodesic fibers, radially shielded isobathic rings restoring same-shell error cancellation, and inward/outward radial elevators)---paired with degree-preserving spanning-tree chord swaps that achieve full directed {BFS} reachability from entry hubs across all measured graphs.
An execution engine co-locates polar headers, variance-stabilized 256-bit early beam sketches, and 2-bit bitplanes within single $4096$\,B OS pages to avoid cross-array secondary {DTLB} walks per accessed record and evaluates bitplane inner products via a 3-to-2 carry-save {AVX2} kernel.

Across eight real and synthetic hyperbolic benchmarks on {AMD EPYC 9654} ({Zen~4}), \rquiver{} builds indices $5.44\times$ faster ($35.0$\,s vs.\ $190.2$\,s on 32 threads on {Cohere-1M}) than full-precision polar graphs ($1.91\times$ faster than 8-bit {Polar-SQ8} at $66.9$\,s) and shrinks hot navigation memory to $517.0$\,MB page-co-located ($2.44\times$ below {Polar-SQ8}, or $485.0$\,MB / $2.60\times$ in split form).
At $99\%$ Recall@10 on {Cohere-1M}, \rquiver{} improves throughput over full-precision polar traversal by $4.30\times$ on 1 thread ($1{,}930$ vs.\ $448$\,QPS) and $6.31\times$ on 24 threads ($42{,}466$ vs.\ $6{,}735$\,QPS); on this benchmark, {HyperspaceDB} ($3.23$\,GB) peaks at $96.78\%$ Recall@10 ($290$ vs.\ $4{,}939$\,QPS on 1 thread at $95\%$ recall) and centered {Faiss} peaks at $98.91\%$ ($127$ vs.\ $2{,}907$\,QPS on 1 thread at $98\%$ recall).
On native hyperbolic {MERU-1M}, \rquiver{} delivers $53{,}483$\,QPS ($99.25\%$ Recall@10, $3.37\times$--$3.88\times$ over {HyperspaceDB} FP32) and reaches $65{,}405$\,QPS ($99.36\%$) on {WordNet-Hazy} on the global 2-bit path ($0$ cones across diffuse real sets), while activating $256$ cones on clustered shells to raise {Clustered-Narrow} recall from $34.22\%$ to $99.385\%$ at $L=256$ ($99.80\%$ at $L=384$; $1.84\times$--$4.56\times$ higher $99\%$-recall QPS than {HyperspaceDB} FP32) and delineating recall limits on isotropic narrow shells.

\end{abstract}

\maketitle

\pagestyle{plain}
\vspace{.3cm}
\begingroup\small\noindent\raggedright\textbf{Code \& Artifact Availability:}\\
The source code, data, and/or other artifacts are available at \url{\codeavailabilityurl}.
\endgroup

\section{Introduction}
\label{sec:intro}

Representation models that embed text, images, and structured entities into negatively curved {Riemannian} manifolds---specifically the {Poincar\'e} ball $\Bball$ and the {Lorentz} hyperboloid $\Lmod$---provide a principled geometric inductive bias for hierarchical retrieval.
Because geodesic ball volume in hyperbolic space grows exponentially with radius rather than polynomially, trees, ontologies, and asymmetric entailment cones embed with far lower metric distortion than in flat {Euclidean} space~\cite{nickel2017poincare,sala2018representation,ganea2018entailment,gu2019mixed}.
Recent encoders scale curved representations across graph neural networks~\cite{nickel2018lorentz,chami2019hgcn}, fully hyperbolic Transformers and vision backbones~\cite{chen2022fully,khrulkov2020hyperbolic,ermolov2022hyperbolic}, million-vector multimodal alignment ({MERU}~\cite{desai2023meru}), and hyperbolic dense retrieval for retrieval-augmented generation ({HypRAG}~\cite{madhu2026hyprag,lewis2020rag}).

Serving low-latency, high-recall approximate nearest neighbor ({ANN}) queries over curved manifolds nevertheless remains a severe database systems bottleneck.
In flat {Euclidean} and cosine spaces, proximity graphs such as {HNSW}~\cite{malkov2020hnsw}, {NSG}~\cite{fu2019nsg}, and {Vamana} ({DiskANN})~\cite{subramanya2019diskann,azizi2025graph} define the latency--recall frontier.
Industrial and academic compressed graph engines---including locally-adaptive scalar quantization ({LVQ})~\cite{aguerrebere2023similarity}, binary lookup tables ({BBQ})~\cite{trent2024bbq}, randomized extended codes ({SymphonyQG})~\cite{gou2025symphonyqg}, and training-free 2-bit Sign-Magnitude ({SM}) indexing in {\quiver}~\cite{xiao2026quiver}---further accelerate search by reading full FP32 vectors only during tail reranking.
In hyperbolic spaces, existing approaches force an unsatisfactory trade-off.
Data-oblivious hyperbolic locality-sensitive hashing ({LSH})~\cite{deng2026hyperboliclsh} requires wide multi-table probing at $99\%+$ recall, while flattening $\Bball$ or $\Lmod$ into {Euclidean} or inner-product proxies inside production vector databases ({Faiss}~\cite{johnson2019billion,douze2024faiss}, {Milvus}~\cite{wang2021milvus}, {AnalyticDB-V}~\cite{wei2020analyticdb}, {VBASE}~\cite{zhang2023vbase}) conflates radial hierarchy depth with angular proximity and inflates candidate pools across radial shells.
Conversely, native hyperbolic graph indices~\cite{boytsov2013engineering,prokhorenkova2022hyperbolic,glukhota2026hyperspacedb} either evaluate geodesic distances over full-precision coordinates at every graph hop or quantize ambient coordinates without decoupling conformal scales under single-metric pruning, incurring compute, memory, or distortion stalls: under exact traversal near the manifold boundary ($\|x\|_2 \to 1/\sqrt{c}$), the conformal factor $f_x = 2/(1 - c\|x\|_2^2)$ diverges and requires IEEE-754 double-precision accumulation to prevent catastrophic cancellation, while random $3\text{--}6$\,KB vector fetches across millions of vertices amplify random memory traffic and data {TLB} ({DTLB}) pressure.

\subsection{Why Flat Graph Indexing Breaks Under Negative Curvature}
\label{subsec:intro_obstructions}

How should proximity-graph indexes be architected for negatively curved manifolds?
Geometrically, $\Bball$ is not an isotropic vector space but a \emph{warped product} $\R_{\ge 0} \times_{\frac{1}{\sqrt{c}}\sinh(\sqrt{c}R)} \Sunit$ coupling a radial depth half-line with an exponentially expanding sphere.
Our theoretical and diagnostic study reveals that transplanting flat single-metric {Vamana} pruning~\cite{subramanya2019diskann} or flat compressed navigation~\cite{aguerrebere2023similarity,gao2024rabitq,xiao2026quiver} onto this warped product fails due to three coupled geometric obstructions:

\begin{enumerate}
  \item \textbf{Cross-Shell Conformal Non-Cancellation and Lorentz-MIPS Blowup.}
  In flat cosine search, ranking candidates $x$ and $y$ against query $q$ evaluates $\langle u_q, u_x\rangle - \langle u_q, u_y\rangle$; because both terms weight $u_q$ equally, first-order quantization noise on $u_q$ cancels identically as $u_x \to u_y$~\cite{xiao2026decisions}.
  In $\Bball$, the monotonic geodesic proxy weights each directional inner product $\langle u_q, u_v\rangle$ by a candidate-dependent conformal factor $W(r_q, r_v) = 2 r_q r_v f_q f_v = \frac{2}{c}\sinh\rho_q\sinh\rho_v$.
  Whenever $r_x \ne r_y$, query quantization leaves an uncancelled residual scaled by $W(r_q, r_y) - W(r_q, r_x) \ne 0$ that grows exponentially with boundary depth.
  Worse, quantizing $(D+1)$-dimensional {Lorentz} coordinates directly via Minkowski inner products multiplies coordinate quantization noise by $\|\tilde{z}\|_2 = \Theta(e^\rho)$ without separating radial norms.

  \item \textbf{Radial Occlusion Bounds and Angular Escape Conditions.}
  Hyperbolic geodesics bend inward toward the origin $0 \in \Bball$.
  We prove that if an outer vertex $u$ at geodesic depth $R_u$ admits an interior neighbor $p$ ($R_p < R_u$) during {Vamana} \textsc{RobustPrune}~\cite{subramanya2019diskann}, then \emph{every} candidate inside the ball $R_v \le (R_u - \alpha R_p)/(\alpha + 1)$ satisfies $\alpha\,d_{\mathbb{B}}(p, v) \le d_{\mathbb{B}}(u, v)$ simultaneously across all spherical angles $\theta_{pv} \in [0, \pi]$---even $180^\circ$ opposite $p$---while along any outer shell $\rho = \sqrt{c}\,R$, negative curvature shrinks the same-shell angular cone surviving origin pruning exponentially as $\sin(\theta_{uv}/2) < \sinh(\frac{\alpha}{2}\rho)/\sinh(\rho) \sim e^{-(2-\alpha)\rho/2}$ (from $73.7^\circ$ in flat space to $23.1^\circ$ at $\rho = 4$ for $\alpha = 1.2$).
  Interior hubs consequently shadow lateral boundary links and isolate $518\text{--}626$ zero-in-degree boundary islands per million vectors on {Cohere-1M}; removing directional protection in our controlled three-seed {Cohere-1M} ablation drops $99\%$-recall throughput by $5.26\times$ ($2{,}071 \to 394$\,QPS) and prevents reaching $99.5\%$ recall.

  \item \textbf{Boundary Cluster Margins and Fixed-Budget Pool Dilution.}
  For equal-depth candidates in a tight cone, the ranking margin scales as $\beta_{q,c}\beta_v$ times the tangent-direction difference. Global-code comparison error depends on the covariance between the candidate errors (Theorem~\ref{thm:tangent_cone}).
  As the comparison margin-to-noise ratio ({$\MNR$}) collapses toward zero, expanding beam width $L$ under a fixed rerank budget $B$ counterintuitively \emph{reduces} recall by flooding the candidate pool with indistinguishable intra-cluster distractors: on a 1M-vector clustered narrow-shell benchmark, beam traversal visits $93.9\%$ of true top-$10$ neighbors at $L=256$ yet retains only $49.3\%$ in the finalist beam and $33.0\%$ after pool gating ($34.2\%$ across all $2{,}000$ queries), and raising $L$ from $256$ to $2{,}048$ lowers recall monotonically from $34.2\%$ to $27.8\%$.
  Local tangent encoding allocates the existing bitplanes to normalized within-cone variation, retaining native coordinates outside the selected cones.
\end{enumerate}

\subsection{Warped-Product Polar Co-Design in \rquiver{}}
\label{subsec:intro_solution}

We address these obstructions with \textbf{\rquiver{}}---named after the maritime bearing dial that decouples angular direction from radial course---an in-memory hyperbolic proximity-graph index grounded in \emph{warped-product polar decoupling}.
Every point $x \in \Bball$ (and via exact stereographic isometry every point $z \in \Lmod$) is factored into unquantized FP32 radial scalars $(r_x,F_x,\rho_x)$, with $F_x=f_x/2$, stored in a 32-byte header and a scale-invariant directional code on the unit sphere $\Sunit$, instantiated via training-free 2-bit {SM} bitplanes.
Storing radial scalars independently in FP32 keeps directional-code error out of the conformal denominator $1-cr_x^2$ and decouples hyperbolic distance into scalar radial lower bounds plus weighted spherical inner products.

From this representation, \rquiver{} co-designs conformal polar estimation primitives, fiber-bundle topology, and page-aligned hardware execution:
\begin{itemize}
  \item \emph{Conformal Polar Estimation and Tangent-Cone Recovery (\S\ref{sec:theory}):}
  We prove that polar decoupling makes same-shell directional flip criteria invariant to query boundary proximity ($f_q \to \infty$), derive a deterministic zero-parameter \emph{orthogonal-projection asymmetric estimator} (\textsc{Stage-0.5}) based on exact two-level {SM} Pythagorean reconstruction, establish $1$-strong convexity of the hyperbolic-cosine objective and derive its closed-form \emph{radial minimizer} along directional rays, and introduce a purity-gated \emph{local tangent-cone residual code} paired with an off-axis \emph{Conformal Jacobian plateau gate} (\textsc{CJ-Gate}) that attenuates intra-cone tangent error with zero per-vertex byte overhead while keeping diffuse embeddings on the zero-overhead global path.

  \item \emph{Warped-Product Fiber-Bundle Topology (\S\ref{sec:design}):}
  Rather than pruning edges under a single isotropic predicate, \rquiver{} inserts vertices inside-out by geodesic depth and partitions each out-degree list across four geometric channels via \emph{QuadriPrune}: stem-excess geodesic fibers, radially shielded \emph{isobathic} (same-shell) rings where $W(r_q, r_x) \approx W(r_q, r_y)$ restores shared-query cancellation, and step-aligned inward and outward radial elevators, followed by a degree-preserving spanning-tree chord-swap pass that achieves full directed {BFS} reachability from entry hubs across all measured graphs.

  \item \emph{Page-Aligned Polar Execution (\S\ref{sec:system}):}
  To avoid the dual-array {DTLB} walks of Structure-of-Arrays layouts, \rquiver{} co-locates each vertex's $32$\,B polar header, binomial-variance-stabilized 256-bit early sketch, and complete 2-bit {SM} bitplanes inside a power-of-two record within a single $4096$\,B OS page ($16$ vertices/page at $D=768$), evaluates surviving 5-popcount bitplane products via a 3-to-2 carry-save {AVX2} kernel, refines the converged beam in L1 cache via fixed-point masked subset sums, and reranks the top tail in IEEE-754 FP64 {SIMD}.
\end{itemize}

\paragraph{Contributions.}
Our main contributions are summarized as follows:
\begin{itemize}
  \item Building on flat comparison-decision theory~\cite{xiao2026decisions}, we analyze conformal polar quantization and hyperbolic {Vamana} pathologies on $\Bball$ and $\Lmod$, characterizing query boundary-depth invariance, cross-shell query-noise non-cancellation, Lorentz-MIPS quantization blowup, radial occlusion with exponential angular-cone contraction, fixed-budget pool dilution, deterministic orthogonal-projection de-biasing, and purity-gated zero-byte-per-node tangent-cone {$\MNR$} recovery (\S\ref{sec:theory}).
  \item We design a warped-product graph construction and routing pipeline combining inside-out geodesic seeding, stem-excess {QuadriPrune} degree partitioning, spanning-tree chord-swap island repair ($1{,}000{,}000 / 1{,}000{,}000$ directed {BFS} reachability from entry hubs), and routing via the constrained radial minimizer (\S\ref{sec:design}).
  \item We engineer a stateless hardware-conscious execution engine integrating $4096$\,B single-page vertex co-location, variance-stabilized 256-bit beam filtering, 3-to-2 carry-save {AVX2} popcounts, L1-bounded \textsc{Stage-0.5}/\textsc{CJ-Gate} refinement, and FP64 {SIMD} geodesic reranking (\S\ref{sec:system}).
  \item Across eight real and synthetic hyperbolic benchmarks on {AMD Zen~4}, \rquiver{} accelerates 32-thread graph construction by $5.44\times$ over {Polar-FP32} ($35.0$\,s vs.\ $190.2$\,s on {Cohere-1M}, and $1.91\times$ vs.\ {Polar-SQ8} at $66.9$\,s), cuts hot navigation memory to $517.0$\,MB page-co-located ($2.44\times$ below 8-bit {Polar-SQ8}, or $485.0$\,MB / $2.60\times$ split), delivers $4.30\times$ (1T) to $6.31\times$ (24T) higher $99\%$-recall QPS on {Cohere-1M} (reaching $99.70\%$ peak Recall@10 vs.\ $96.78\%$ for {HyperspaceDB} and $98.91\%$ / $98.69\%$ for centered {Faiss} / {Milvus}, with $17.0\times$--$28.3\times$ higher QPS than {HyperspaceDB} at $95\%$ recall and $22.9\times$--$58.2\times$ over centered {Faiss} / {Milvus} at $98\%$ recall), reaches $53{,}483$\,QPS on {MERU-1M} ({ImageNet}, $3.37\times$--$3.88\times$ over {HyperspaceDB} FP32 at $99\%$ recall), $14{,}731$\,QPS on 1536D {OpenAI-1536} ({DBpedia}), and $65{,}405$\,QPS on {WordNet-Hazy} on the global 2-bit path ($0$ cones), and activates $256$ cones on {Clustered-Narrow} to raise recall from $34.22\%$ to $99.385\%$ at $L=256$ ($99.80\%$ at $L=384$; $1.84\times$--$4.56\times$ higher $99\%$-recall QPS than {HyperspaceDB} FP32) while characterizing recall limits on {Uniform-Narrow} (\S\ref{sec:eval}).
\end{itemize}

\section{Preliminaries and Problem Formulation}
\label{sec:prelim}

We formalize hyperbolic metric models as warped-product manifolds, establish the exact polar geodesic proxy, characterize the two-level spherical Sign-Magnitude ({SM}) geometry, and define the discrete comparison decisions governing quantized graph construction and search.

\subsection{Hyperbolic Models and Warped-Product Unification}
\label{subsec:models}

\paragraph{{Poincar\'e} Ball and {Lorentz} Stereographic Isometry.}
For constant negative sectional curvature $-c < 0$, the $D$-dimensional {Poincar\'e} ball is the open manifold
\begin{equation}
  \Bball = \bigl\{x \in \R^D : c\|x\|_2^2 < 1\bigr\}, \qquad g_x^{\mathbb{B}} = f_x^2 I_D,
  \label{eq:poincare_manifold}
\end{equation}
with conformal metric factor $f_x = \frac{2}{1 - c\|x\|_2^2}$.
The Riemannian geodesic distance between $q, v \in \Bball$ is
\begin{equation}
  d_{\mathbb{B}}(q, v) = \frac{1}{\sqrt{c}}\arccosh\!\left(1 + \frac{2c\|q - v\|_2^2}{(1 - c\|q\|_2^2)(1 - c\|v\|_2^2)}\right).
  \label{eq:poincare_dist}
\end{equation}
Many modern hyperbolic encoders, including fully hyperbolic networks~\cite{chen2022fully}, {MERU}~\cite{desai2023meru}, and {HypRAG}~\cite{madhu2026hyprag}, parameterize embeddings on the upper sheet of the $(D+1)$-dimensional {Lorentz} hyperboloid
$\Lmod = \{z = (z_0, \tilde{z}) \in \R \times \R^D : -z_0^2 + \|\tilde{z}\|_2^2 = -1/c,\; z_0 > 0\}$
under $d_{\mathbb{L}}(z, z') = \frac{1}{\sqrt{c}}\arccosh\bigl(c(z_0 z'_0 - \langle \tilde{z}, \tilde{z}'\rangle)\bigr)$.
Stereographic projection $\Pi : \Lmod \to \Bball$,
\begin{equation}
  \Pi(z_0, \tilde{z}) = \frac{\tilde{z}}{1 + \sqrt{c}\,z_0},
  \label{eq:lorentz_isometry}
\end{equation}
is a global Riemannian isometry satisfying $d_{\mathbb{B}}(\Pi(z), \Pi(z')) = d_{\mathbb{L}}(z, z')$ for all $z, z' \in \Lmod$~\cite{nickel2018lorentz}\ifpreprint\ (proved algebraically in Appendix~\ref{app:warped_product})\fi.
Projecting $\Lmod$ embeddings via~\eqref{eq:lorentz_isometry} at ingestion eliminates the redundant time-like coordinate $z_0$ without losing exact geodesic distances or rankings; as shown in Theorem~\ref{thm:conformal_flip}(3), it also avoids exponential quantization error blowup on unbounded spatial norms $\|\tilde{z}\|_2$.

\paragraph{Warped-Product Decomposition and Order-Isomorphic Polar Proxy.}
Every $x \in \Bball$ decomposes into Euclidean radius $r_x = \|x\|_2 \in [0, 1/\sqrt{c})$ and unit spherical direction $u_x \in \Sunit$ ($u_0 = e_1$ at $r_x = 0$).
Let $R_x = d_{\mathbb{B}}(0, x)$ denote geodesic radius from the origin and $\rho_x = \sqrt{c}\,R_x = 2\atanh(\sqrt{c}\,r_x)$ denote dimensionless geodesic depth, which satisfy the exact identities $\sinh\rho_x = \sqrt{c}\,r_x f_x$ and $\cosh\rho_x = c r_x^2 f_x + 1 = f_x - 1$.
In coordinates $(R, u) \in (0, \infty) \times \Sunit$, $\Bball \setminus \{0\}$ is isometric to the warped-product manifold
\begin{equation}
  \begin{aligned}
    \Bball \setminus \{0\} &\cong (0, \infty) \times_{\frac{1}{\sqrt{c}}\sinh(\sqrt{c}R)} \Sunit, \\
    \mathrm{d}s^2 &= \mathrm{d}R^2 + \frac{1}{c}\sinh^2(\sqrt{c}R)\,\mathrm{d}\Omega_{D-1}^2,
  \end{aligned}
  \label{eq:warped_metric}
\end{equation}
where spherical fiber distances expand exponentially as $\frac{1}{\sqrt{c}}\sinh(\sqrt{c}R) = \frac{1}{2\sqrt{c}}e^{\rho}(1 - e^{-2\rho})$.
Expanding $\|q - v\|_2^2 = (r_q - r_v)^2 + 2 r_q r_v(1 - \gamma_{qv})$ for spherical cosine $\gamma_{qv} = \langle u_q, u_v\rangle = \cos\theta_{qv} \in [-1, 1]$ yields the \emph{polar geodesic proxy}:
\begin{equation}
  P_{\mathbb{B}}(q, v) \triangleq \|q - v\|_2^2 f_q f_v = R(r_q, r_v) + W(r_q, r_v)\bigl(1 - \gamma_{qv}\bigr),
  \label{eq:polar_proxy}
\end{equation}
where the radial discrepancy term and conformal angular weight are
\begin{equation}
  \begin{aligned}
    R(r_q, r_v) &= (r_q - r_v)^2 f_q f_v, \\
    W(r_q, r_v) &= 2 r_q r_v f_q f_v = \frac{2}{c}\sinh\rho_q \sinh\rho_v.
  \end{aligned}
  \label{eq:radial_conformal_weights}
\end{equation}
By the hyperbolic law of cosines, $\frac{c}{4}P_{\mathbb{B}}(q, v) = \sinh^2\!\bigl(\frac{\rho_q - \rho_v}{2}\bigr) + \sinh\rho_q\sinh\rho_v\sin^2\!\bigl(\frac{\theta_{qv}}{2}\bigr) = \sinh^2\!\bigl(\frac{\sqrt{c}}{2}d_{\mathbb{B}}(q, v)\bigr)$, so exact geodesic distance relates to $P_{\mathbb{B}}(q, v)$ through
\begin{equation}
  \begin{aligned}
    d_{\mathbb{B}}(q, v)
    &= \frac{1}{\sqrt{c}}\arccosh\!\left(1 + \frac{c}{2} P_{\mathbb{B}}(q, v)\right) \\
    &= \frac{2}{\sqrt{c}}\asinh\!\left(\frac{1}{2}\sqrt{c\,P_{\mathbb{B}}(q, v)}\right).
  \end{aligned}
  \label{eq:proxy_to_geodesic}
\end{equation}
Because $z \mapsto \frac{2}{\sqrt{c}}\asinh\bigl(\frac{1}{2}\sqrt{c z}\bigr)$ is strictly increasing on $[0, \infty)$, $P_{\mathbb{B}}(q, v)$ is strictly order-isomorphic to $d_{\mathbb{B}}(q, v)$, deferring transcendental evaluations to the final $k$ returned neighbors.
The implementation stores $F_v=f_v/2=(1-cr_v^2)^{-1}$ and evaluates $P_{\mathrm{impl}}=P_{\mathbb B}/4$, with $d=2\asinh(\sqrt{cP_{\mathrm{impl}}})/\sqrt c$. All beam thresholds, radial gates, and Jacobian scales use this normalization consistently.
Furthermore, in the flat-curvature limit $c \to 0^+$, $f_x \to 2$ and $\frac{1}{4}P_{\mathbb{B}}(q, v) \to \|q - v\|_2^2$, recovering flat Euclidean and spherical geometry continuously.

\subsection{Exact Two-Level Spherical SM Geometry and Decision Operators}
\label{subsec:sm_decisions}

\paragraph{Deterministic Conditional-Mean Two-Level {SM} Reconstruction.}
In polar coordinates, $u_v \in \Sunit$ lies on the unit sphere independently of $r_v$.
Following {\quiver}'s training-free 2-bit {SM} bitplane encoding~\cite{xiao2026quiver}---whose conditional-mean Pythagorean identity (Lemma~\ref{lem:sm_pythagorean}) underpins our orthogonal-projection refinement (\S\ref{sec:system})---each coordinate $u_{v,i}$ ($i \in \{1,\dots,D\}$) is encoded into two packed bitplanes:
a sign bit $s_{v,i} = \mathbb{I}[u_{v,i} \ge 0]$ and a strong-magnitude bit $m_{v,i} = \mathbb{I}[|u_{v,i}| > \tau_v]$ relative to the L1 mean threshold $\tau_v = \frac{1}{D}\sum_{j=1}^D |u_{v,j}|$, with strong-bit count $s_v = \sum_{i=1}^D m_{v,i}$.
Partitioning $\{1,\dots,D\}$ into weak indices $I_{0,v} = \{i : m_{v,i} = 0\}$ ($|I_{0,v}| = D - s_v$) and strong indices $I_{1,v} = \{i : m_{v,i} = 1\}$ ($|I_{1,v}| = s_v$) defines conditional magnitude means $a_v = \frac{1}{D - s_v}\sum_{i \in I_{0,v}} |u_{v,i}|$ ($0$ if $s_v = D$), $b_v = \frac{1}{s_v}\sum_{i \in I_{1,v}} |u_{v,i}|$ ($0$ if $s_v = 0$), and step $\delta_v=b_v-a_v$. For $0<s_v<D$, $\delta_v\ge0$. At $s_v=0$, the implementation sets the unused strong mean to $b_v=0$, so $\delta_v=-a_v$ and all reconstructed coordinates use $a_v$.

\begin{lemma}[Deterministic Two-Level {SM} Pythagorean Identity]
\label{lem:sm_pythagorean}
For any $u_v \in \Sunit$, let $\hat{u}_v \in \R^D$ be the conditional-mean two-level vector $\hat{u}_{v,i} = \sgn(u_{v,i})(a_v + \delta_v m_{v,i})$ (setting $\sgn(0) = +1$).
Then, with zero distributional assumptions on $u_v$, the reconstructed vector $\hat u_v$ uniquely minimizes $\|u_v - \tilde{u}\|_2^2$ over all two-level vectors sharing bitplanes $(s_v,m_v)$. The populated-group means are unique, and the reconstruction satisfies
\begin{equation}
  \begin{aligned}
    \langle u_v, \hat{u}_v\rangle &= (D - s_v)a_v^2 + s_v b_v^2 \\
    &= \|\hat{u}_v\|_2^2 \triangleq \sigma_v^2 \in (0, 1], \\
    \|u_v - \hat{u}_v\|_2^2 &= 1 - \sigma_v^2.
  \end{aligned}
  \label{eq:sm_pythagorean}
\end{equation}
Moreover, for two two-level codes $\hat{u}_q, \hat{u}_v$, let $p_{\oplus} = \popcnt(\bar{x})$, $p_{10} = \popcnt(\bar{x} \wedge m_q)$, $p_{01} = \popcnt(\bar{x} \wedge m_v)$, $p_{11} = \popcnt(\bar{x} \wedge m_q \wedge m_v)$, and $n_{11} = \popcnt(m_q \wedge m_v)$ where $\bar{x} = \neg(s_q \oplus s_v)$ marks sign agreement.
Then $\langle \hat{u}_q, \hat{u}_v\rangle = a_q a_v(2 p_{\oplus} - D) + a_q \delta_v(2 p_{01} - s_v) + \delta_q a_v(2 p_{10} - s_q) + \delta_q \delta_v(2 p_{11} - n_{11})$ exactly\ifpreprint\ (proof in Appendix~\ref{app:sm_algebra})\fi.
\end{lemma}

\begin{proof}[Proof Sketch]
Because $\sgn(u_{v,i}) u_{v,i} = |u_{v,i}|$, splitting $\langle u_v, \hat{u}_v\rangle$ over $I_{0,v}$ and $I_{1,v}$ gives $a_v \sum_{i \in I_{0,v}} |u_{v,i}| + b_v \sum_{i \in I_{1,v}} |u_{v,i}| = (D - s_v)a_v^2 + s_v b_v^2 = \|\hat{u}_v\|_2^2$, whence $\langle u_v - \hat{u}_v, \hat{u}_v\rangle = 0$ and $\|u_v - \hat{u}_v\|_2^2 = 1 - \sigma_v^2$.
Expanding $\sgn_q \sgn_v = 2\bar{x}_i - 1$ over $\{1, m_q, m_v, m_q m_v\}$ yields the five-popcount identity.
\end{proof}

The main hop scorer uses the conditional-mean five-popcount product from Lemma~\ref{lem:sm_pythagorean}. With $t=\operatorname{clip}_{[-1,1]}(\langle\hat u_q,\hat u_v\rangle/(\sigma_q\sigma_v))$, it applies
\begin{equation}
  \hat\gamma_{\mathrm{sym}}=\frac{t}{\kappa+(1-\kappa)t^2},\qquad
  \kappa=\operatorname{clip}_{[0.5,1]}(\sigma_q\sigma_v).
  \label{eq:sm_calibration}
\end{equation}
The FP32 norms are precomputed at ingestion. For fixed $\kappa \in [0.5, 1]$, $g_\kappa(t) = t / (\kappa + (1 - \kappa)t^2)$ is strictly increasing on $[-1, 1]$ ($g'_\kappa(t) = [\kappa - (1 - \kappa)t^2] / [\kappa + (1 - \kappa)t^2]^2 > 0$ on $(-1, 1)$), preserving rank across equal-$\sigma_v$ candidates while calibrating hop scores ($\sigma_v$-dependent $\kappa$) ahead of \textsc{Stage-0.5}'s exact orthogonal projection. The canonical code $\operatorname{sgn}(u_i)(1+m_i)\in\{\pm1,\pm2\}$, with inverse norm $(D+3s)^{-1/2}$, supplies legacy scoring paths. Training-free encoding computes bitplanes and conditional means directly from each vector. Cone-axis fitting uses the deterministic sample procedure in \S\ref{subsec:polar_layout}.

\paragraph{Problem Statement and Discrete Comparison Operators.}
Given $N$ database points $\mathcal{X} \subset \Bball$ and query $q \in \Bball$, exact top-$k$ retrieval returns $\mathcal{N}_k(q)$ minimizing $d_{\mathbb{B}}(q, v)$, and {ANN} accuracy is evaluated via $\operatorname{Recall@}k = |\widehat{\mathcal{N}}_k(q) \cap \mathcal{N}_k(q)| / k$ against throughput ({QPS}) and p99 latency.
As established by comparison-decision stability theory for flat quantized search~\cite{xiao2026decisions}, proximity-graph construction and traversal depend not on point-wise metric reconstruction error $|\widehat{P}_{\mathbb{B}} - P_{\mathbb{B}}|$, but on sign preservation of two binary comparison operators:
\begin{enumerate}
  \item \emph{Shared-query ranking} during beam search: for query $q$ and candidates $x, y \in \mathcal{X}$, deciding the sign of
  \begin{equation}
    \Psi_{\mathrm{rank}}(q; x, y) \triangleq P_{\mathbb{B}}(q, x) - P_{\mathbb{B}}(q, y);
    \label{eq:psi_rank}
  \end{equation}
  \item \emph{Shared-target diversity pruning} during {Vamana} edge selection~\cite{subramanya2019diskann}: for vertex $u$, selected neighbor $w$, and candidate $v$ under geodesic slack $\alpha > 1$, deciding whether
  \begin{equation}
    \Psi_{\mathrm{prune}}(u, w, v; \alpha) \triangleq \alpha\,d_{\mathbb{B}}(w, v) - d_{\mathbb{B}}(u, v) \le 0.
    \label{eq:psi_prune}
  \end{equation}
\end{enumerate}
For any noisy estimator $\widehat{\Psi} = \Psi + \Delta\widehat{\Psi}$, the probability of a decision flip is controlled by its \emph{margin-to-noise ratio}~\cite{xiao2026decisions}:
\begin{equation}
  \MNR(\widehat{\Psi}) \triangleq \frac{|\E[\widehat{\Psi}]|}{\sqrt{\Var(\Delta\widehat{\Psi})}}.
  \label{eq:mnr_def}
\end{equation}
Section~\ref{sec:theory} characterizes how negative curvature reshapes both $\Var(\Delta\widehat{\Psi}_{\mathrm{rank}})$ and the geometry of $\Psi_{\mathrm{prune}}$.

\section{Conformal Polar Distortion and Hyperbolic Graph Pathologies}
\label{sec:theory}

Why do flat proximity-graph pruning rules~\cite{subramanya2019diskann} and compact directional codes~\cite{xiao2026quiver,xiao2026decisions} break when transplanted to negatively curved manifolds?
We analyze both the stochastic comparison error of the polar proxy $\widehat{P}_{\mathbb{B}}(q, v) = R(r_q, r_v) + W(r_q, r_v)(1 - \hat{\gamma}(u_q, u_v))$ and the deterministic geodesic geometry of {Vamana} pruning on $\Bball$.

\subsection{Conformal Flip Criterion, Query Depth Invariance, and Lorentz Blowup}
\label{subsec:non_cancellation}

Write the directional estimation error for any pair $(u_q, u_v)$ as $\varepsilon_\angle(q, v) \triangleq \hat{\gamma}(u_q, u_v) - \gamma_{qv}$.
For the normalized bilinear score $\langle\hat u_q/\sigma_q,\hat u_v/\sigma_v\rangle$, define directional residuals $\hat{u}_z / \sigma_z - u_z = e_z$, the bilinear error decomposition~\cite{bussgang1952crosscorrelation,xiao2026decisions} is
\begin{equation}
  \varepsilon_\angle(q, v)
  = \underbrace{\langle e_q, u_v\rangle}_{\eta_q(u_v)}
  + \underbrace{\langle u_q, e_v\rangle}_{\eta_v(u_q)}
  + \underbrace{\langle e_q, e_v\rangle}_{\xi_{qv}},
  \label{eq:bilinear_error}
\end{equation}
The calibrated hop score~\eqref{eq:sm_calibration} applies its rational map after this bilinear product (while \textsc{Stage-0.5} evaluates the uncalibrated linear projection directly at $e_q = 0$). Here $\eta_q(u_v)$ is the first-order query projection, $\eta_v(u_q)$ is the first-order candidate projection, and $\xi_{qv}$ is the second-order cross residual.
In flat cosine ranking ($W \equiv 1$), comparing angularly close candidates $u_x, u_y$ subtracts $\eta_q(u_x) - \eta_q(u_y) = \langle e_q, u_x - u_y\rangle$, with variance $(u_x-u_y)^\top\Sigma_q(u_x-u_y)$ for $\Sigma_q=\Cov(e_q)$. The scaling $\|\Sigma_q\|_2=O(1/D)$ gives standard deviation $O(\|u_x-u_y\|_2/\sqrt D)$~\cite{xiao2026decisions}.
On $\Bball$, conformal radial weights reshape this behavior in four distinct ways.

\begin{theorem}[Conformal Flip Criterion, Depth Invariance, and Lorentz-MIPS Blowup]
\label{thm:conformal_flip}
Let $q, x, y \in \Bball$ ($r_q > 0$) satisfy $P_{\mathbb{B}}(q, x) < P_{\mathbb{B}}(q, y)$ with true proxy margin $\Delta_P(q; x, y) \triangleq P_{\mathbb{B}}(q, y) - P_{\mathbb{B}}(q, x) > 0$, with real-valued radial scalars $(r_z,f_z)$. \ifpreprint Implementation rounding is quantified in Appendix~\ref{app:scalar_rounding}.\fi Then
\begin{enumerate}
  \item \textbf{Exact flip criterion and query boundary-depth invariance:}
  An approximate proxy inversion $\widehat{P}_{\mathbb{B}}(q, y) \le \widehat{P}_{\mathbb{B}}(q, x)$ occurs if and only if
  \begin{equation}
    r_y f_y\,\varepsilon_\angle(q, y) - r_x f_x\,\varepsilon_\angle(q, x)
    \ge \frac{\Delta_P(q; x, y)}{2 r_q f_q}.
    \label{eq:exact_flip_criterion}
  \end{equation}
  When $x, y$ lie on an \emph{isobathic shell} ($r_x = r_y = r > 0$),~\eqref{eq:exact_flip_criterion} simplifies to $\varepsilon_\angle(q, y) - \varepsilon_\angle(q, x) \ge \gamma_{qx} - \gamma_{qy}$: both $f_q$ and $f_x$ cancel identically, rendering same-shell flip probabilities completely invariant to query and candidate boundary proximity ($f_q, f_x \to \infty$).

  \item \textbf{Cross-shell query-noise non-cancellation:}
  For the bilinear error~\eqref{eq:bilinear_error}, the ranking perturbation $\Delta\widehat{\Psi}_{\mathrm{rank}}(q; x, y) = \widehat{P}_{\mathbb{B}}(q, x) - \widehat{P}_{\mathbb{B}}(q, y) - \Psi_{\mathrm{rank}}(q; x, y)$ contains an uncancelled query term $d_q(r_q; r_x, r_y)\,\eta_q(u_x)$ with $d_q(r_q; r_x, r_y) = W(r_q, r_y) - W(r_q, r_x)$.
  For arbitrary $\Sigma_q=\Cov(e_q)$, this first-order query term at $u_x=u_y$ has variance $d_q^2u_x^\top\Sigma_q u_x$. Specializing to $\Sigma_q=\frac{\kappa_q}{D}(I_D-u_qu_q^\top)$ gives
  \begin{equation}
    \begin{aligned}
      &\Var_{e_q}\!\bigl(d_q\eta_q(u_x)\mid u_x=u_y\bigr) \\
      &\quad= \frac{4\kappa_q}{c^2 D}\sinh^2\rho_q
        \bigl(\sinh\rho_y - \sinh\rho_x\bigr)^2(1 - \gamma_{qx}^2),
    \end{aligned}
    \label{eq:cross_shell_var}
  \end{equation}
  For fixed nonzero shell gap and $1-\gamma_{qx}^2>0$, this expression grows as $\Theta(e^{2\rho_q+2\max(\rho_x,\rho_y)}/D)$ at large depths. Inside a deep-shell band of width $\delta_\rho$, its weight mismatch is $O(\delta_\rho^2W^2)$. An unquantized query sets $e_q=0$ in pool re-scoring.

  \item \textbf{Spatial {Lorentz}-MIPS quantization scaling:}
  Under a spatial-error model $\Cov(\epsilon_v)=\Theta(\|\tilde z_v\|_2^2/D)I_D$, directly quantizing spatial coordinates $\tilde{z} = \frac{1}{\sqrt{c}}\sinh\rho_z\,u_z \in \R^D$ on $\Lmod$ to approximate the {Lorentz} inner product $-z_0 z'_0 + \langle \tilde{z}, \tilde{z}'\rangle$ without polar norm separation incurs coordinate error standard deviation $\Theta(\frac{1}{c\sqrt{D}}\sinh\rho_q\sinh\rho_v)$ even at $\theta_{qv} = 0$, giving nonzero spatial inner-product variance on aligned rays. Polar storage supplies the scalar radial lower bound independently of directional evaluation\ifpreprint\ (complete proof in Appendix~\ref{app:conformal_flips})\fi.
\end{enumerate}
\end{theorem}

\begin{proof}[Proof Sketch]
Because $R(r_q, r_v)$ and $W(r_q, r_v) = 2 r_q r_v f_q f_v$ are exact, $\widehat{P}_{\mathbb{B}}(q, v) - P_{\mathbb{B}}(q, v) = -2 r_q f_q r_v f_v \varepsilon_\angle(q, v)$.
Substituting this into $\widehat{P}_{\mathbb{B}}(q, y) - \widehat{P}_{\mathbb{B}}(q, x) \le 0$ and dividing by $2 r_q f_q > 0$ gives~\eqref{eq:exact_flip_criterion}; setting $r_x = r_y$ cancels $r_x f_x$ against $\Delta_P = 2 r_q r_x f_q f_x(\gamma_{qx} - \gamma_{qy})$.
Adding and subtracting $W(r_q, r_y)\eta_q(u_x)$ in~\eqref{eq:bilinear_error} isolates $d_q\,\eta_q(u_x)$ and yields~\eqref{eq:cross_shell_var}.
Under the stated spatial-error model on $\Lmod$, a quantizer applied to $\tilde{z}_v$ scales quantization step size by $\|\tilde{z}_v\|_2 = c^{-1/2}\sinh\rho_v$, injecting $\Theta(e^{\rho_q + \rho_v}/\sqrt{D})$ noise even along radial lines where polar $R(r_q, r_v)$ is exact.
\end{proof}

Theorem~\ref{thm:conformal_flip} establishes why \rquiver{} projects $\Lmod \to \Bball$ at ingestion, stores $(r_v,f_v)$ as unquantized FP32 scalars, isolates lateral graph edges inside isobathic shells $|\rho_v - \rho_u| \le \delta_\rho$, and applies an unquantized-query refinement pass (\textsc{Stage-0.5}) before cross-shell truncation.

\subsection{Hyperbolic Vamana Pathologies: Occlusion, Wells, and Bias}
\label{subsec:vamana_pathologies}

Even with exact geodesic distances, standard single-metric {Vamana} construction~\cite{subramanya2019diskann} exhibits severe topological degeneration in $\Bball$ because negative curvature bends geodesics toward the origin.

\begin{theorem}[Radial Occlusion Bound, Angular Escape Condition, and Intercept Bias]
\label{thm:vamana_pathologies}
Let $\Bball$ have curvature $-c < 0$ and {Vamana} pruning slack $\alpha > 1$.
\begin{enumerate}
  \item \textbf{Radial occlusion bound and exponential angular contraction:}
  Suppose outer vertex $u$ at geodesic radius $R_u = d_{\mathbb{B}}(0, u)$ admits an interior neighbor $p$ with $R_p < R_u / \alpha$.
  Then for \emph{every} vertex $v \in \Bball$ inside the closed geodesic ball
  \begin{equation}
    R_v \le R_{\mathrm{occ}}(R_u, R_p; \alpha) \triangleq \frac{R_u - \alpha R_p}{\alpha + 1},
    \label{eq:apollonius_occlusion}
  \end{equation}
  the pruning inequality $\alpha\,d_{\mathbb{B}}(p, v) \le d_{\mathbb{B}}(u, v)$ holds \emph{simultaneously for all spherical angles} $\theta_{pv} \in [0, \pi]$---including candidates $180^\circ$ opposite $p$.
  Whereas~\eqref{eq:apollonius_occlusion} is a generic metric ball, negative curvature drives severe same-shell angular collapse: for any outer shell $R_u = R_v = R > 0$ ($\rho = \sqrt{c}\,R$) and origin neighbor $p = 0$ under $\alpha \in (1, 2)$, $\alpha\,d_{\mathbb{B}}(0, v) \le d_{\mathbb{B}}(u, v)$ holds whenever $\sin(\theta_{uv}/2) \ge \sinh(\frac{\alpha}{2}\rho)/\sinh(\rho)$, shrinking the surviving unoccluded cone exponentially as $2 e^{-(2-\alpha)\rho/2}$ (from $73.7^\circ$ in the flat limit $c \to 0^+$ down to $23.1^\circ$ at $\rho = 4$ and $10.4^\circ$ at $\rho = 6$ for $\alpha = 1.2$).

  \item \textbf{Angular escape condition:}
  For any query $q$ at radius $R_q > 0$, a candidate $v$ at radius $R_v > 0$ is geodesically closer to $q$ than the origin $0$ ($d_{\mathbb{B}}(q, v) < R_q$) if and only if its spherical cosine alignment exceeds
  \begin{equation}
    \gamma_{qv} > \gamma_{\mathrm{esc}}(R_q, R_v) \triangleq \frac{\tanh(\sqrt{c}\,R_v / 2)}{\tanh(\sqrt{c}\,R_q)}.
    \label{eq:escape_cap}
  \end{equation}
  Outside this cap, $v$ fails to improve on the origin. Angular sensitivity vanishes along inward rays as $\sinh(\sqrt c R_v)\to0$.

  \item \textbf{Offset-intercept and proxy non-homogeneity bias:}
  Replacing geodesic distance with an affine surrogate $\hat{d}=a d+b$, $a>0$, gives pruning margin $a[d(u,v)-\alpha d(w,v)]+(1-\alpha)b$. Positive $b$ produces under-pruning and negative $b$ produces over-pruning. Direct linear proxy slack $\alpha P_{\mathbb{B}}(w,v)\le P_{\mathbb{B}}(u,v)$ over-prunes relative to geodesic slack, whose proxy threshold is $P_\alpha(P)=\frac{2}{c}[\cosh(\alpha\arccosh(1+cP/2))-1]>\alpha P$ for $P>0$\ifpreprint\ (complete proofs in Appendix~\ref{app:vamana_pathologies})\fi.
\end{enumerate}
\end{theorem}

\begin{proof}[Proof Sketch]
By the triangle inequality through $0$, $d_{\mathbb{B}}(p, v) \le R_p + R_v$ and $d_{\mathbb{B}}(u, v) \ge R_u - R_v$ for all $\theta_{pv}, \theta_{uv} \in [0, \pi]$; requiring $\alpha(R_p + R_v) \le R_u - R_v$ yields~\eqref{eq:apollonius_occlusion}.
Along any same-shell pair $R_u = R_v = R$, substituting $\frac{c}{4}P_{\mathbb{B}}(u, v) = \sinh^2(\frac{\sqrt{c}}{2}d_{\mathbb{B}}(u, v)) = \sinh^2\rho\,\sin^2(\theta_{uv}/2)$ reduces $\alpha\,d_{\mathbb{B}}(0, v) \le d_{\mathbb{B}}(u, v)$ directly to $\sin(\theta_{uv}/2) \ge \sinh(\frac{\alpha}{2}\rho)/\sinh\rho$.
For Claim~(2), $\cosh(\sqrt{c}\,d_{\mathbb{B}}(q, v)) < \cosh(\sqrt{c}R_q)$ expands via the hyperbolic law of cosines to $\cosh\rho_q(\cosh\rho_v - 1) < \sinh\rho_q\sinh\rho_v\,\gamma_{qv}$; dividing by $\sinh\rho_q\sinh\rho_v > 0$ and applying $(\cosh\rho_v - 1)/\sinh\rho_v = \tanh(\rho_v/2)$ gives~\eqref{eq:escape_cap}.
For Claim~(3), $\hat{d}(u, v) - \alpha\hat{d}(w, v) = a(\tilde{d}(u, v) - \alpha\tilde{d}(w, v)) + (1 - \alpha)b$.
\end{proof}

Theorem~\ref{thm:vamana_pathologies} directly motivates \rquiver{}'s Inside-Out insertion order, radially shielded isobathic ring, stem-excess geodesic dominance rule (\S\ref{subsec:compass}), and multi-entry polar router (\S\ref{subsec:router}).

\subsection{Boundary Margin Collapse, Tangent-Cone Recovery, and Pool Dilution}
\label{subsec:tangent_cone}

Even when $e_q = 0$, candidate-side quantization noise $W(r_q, r_v)\eta_v(u_q)$ persists.
When semantic subtrees concentrate inside a narrow cone around unit axis $\mu_c \in \Sunit$ near the boundary ($\rho_v \gg 1$), orthogonal decomposition writes
\begin{equation}
  \begin{gathered}
    u_v = \alpha_v \mu_c + \beta_v \tilde{r}_v, \qquad
    \alpha_v = \langle u_v, \mu_c\rangle, \\
    \beta_v = \sqrt{1 - \alpha_v^2}, \qquad
    \tilde{r}_v \in T_{\mu_c}\Sunit \cap \Sunit,
  \end{gathered}
  \label{eq:tangent_decomp}
\end{equation}
with $\langle \tilde{r}_v, \mu_c\rangle = 0$.
Decomposing $u_q = \alpha_{q,c}\mu_c + \beta_{q,c}\tilde{r}_{q,c}$ splits $\gamma_{qv}$ into macroscopic axial alignment and microscopic tangent variation:
\begin{equation}
  \gamma_{qv} = \langle u_q, u_v\rangle = \alpha_{q,c}\alpha_v + \beta_{q,c}\beta_v \langle \tilde{r}_{q,c}, \tilde{r}_v\rangle.
  \label{eq:tangent_inner}
\end{equation}

\begin{theorem}[Tangent-Cone Error Propagation and Fixed-Budget Admission]
\label{thm:tangent_cone}
Consider same-shell candidates $x,y$ with weight $W=W(r_q,r)>0$, common $\alpha=\sqrt{1-\beta^2}$, and $\beta,\beta_{q,c}>0$. Set $\Delta_{\tilde r}=\langle\tilde r_{q,c},\tilde r_y-\tilde r_x\rangle\ne0$. Variances below use a conditional error model with fixed displayed geometric quantities and zero-mean estimation errors.
\begin{enumerate}
  \item \textbf{Pairwise margin and covariance:}
  The true margin is $W\beta_{q,c}\beta|\Delta_{\tilde r}|$. For global cosine errors $g_x,g_y$, define $V_g=\Var(g_y)+\Var(g_x)-2\Cov(g_y,g_x)$. For $V_g>0$,
  \begin{equation}
    \MNR_{\mathrm{global}}=\frac{\beta_{q,c}\beta|\Delta_{\tilde r}|}{\sqrt{V_g}}.
    \label{eq:mnr_collapse}
  \end{equation}
  Thus margin collapse occurs when $\sqrt{V_g}/(\beta_{q,c}\beta|\Delta_{\tilde r}|)\to\infty$.

  \item \textbf{Fixed-budget admission under exchangeable ranks:}
  Suppose the retained pool contains $N_c(L)\ge B$ cone members, including $k_c(L)$ true neighbors, and all $B$ rerank slots are selected from these members. If their score ranks form a uniform permutation conditional on this pool, expected true-neighbor admission is $k_c(L)B/N_c(L)$. It decreases between two widths exactly when $k_c(L)/N_c(L)$ decreases.

  \item \textbf{Exact tangent-error attenuation:}
  With real-valued axial scalars, use
  \begin{equation}
    \hat{\gamma}_{\mathrm{cone}}(q, v) \triangleq \alpha_{q,c}\alpha_v + \beta_{q,c}\beta_v\,\hat{\gamma}_{\mathrm{asym}}(\tilde{r}_{q,c}, \tilde{r}_v)
    \label{eq:cone_estimator}
  \end{equation}
  and let $e_v=\hat\gamma_{\mathrm{asym}}(\tilde r_{q,c},\tilde r_v)-\langle\tilde r_{q,c},\tilde r_v\rangle$. Then cosine error equals $\beta_{q,c}\beta_v e_v$ pointwise. Set $V_r=\Var(e_y)+\Var(e_x)-2\Cov(e_y,e_x)$. For $V_r>0$,
  \begin{equation}
    \begin{aligned}
      \MNR_{\mathrm{cone}}&=\frac{|\Delta_{\tilde r}|}{\sqrt{V_r}},\\
      \frac{\MNR_{\mathrm{cone}}}{\MNR_{\mathrm{global}}}
      &=\frac{\sqrt{V_g}}{\beta_{q,c}\beta\sqrt{V_r}}.
    \end{aligned}
    \label{eq:mnr_recovery}
  \end{equation}
  Under equal residual variances $\kappa_{\mathrm{SM}}/D$ and pair correlation $\varrho<1$, $V_r=2\kappa_{\mathrm{SM}}(1-\varrho)/D$.
  \ifpreprint(Proof in Appendix~\ref{app:tangent_cone}.)\fi
\end{enumerate}
\end{theorem}

\begin{proof}[Proof Sketch]
Subtracting~\eqref{eq:tangent_inner} for $x,y$ cancels the axial term. Applying $\Var(X-Y)=\Var(X)+\Var(Y)-2\Cov(X,Y)$ gives $W^2V_g$ for global proxy-error variance and $W^2\beta_{q,c}^2\beta^2V_r$ for residual proxy-error variance. Uniform ranks give each cone member admission probability $B/N_c(L)$. Summing indicators over its $k_c(L)$ true neighbors proves Claim~(2).
\end{proof}

On synthetic {Clustered-Narrow}, increasing $L$ from $256$ to $2{,}048$ at $B=128$ reduces global-code recall from $34.22\%$ to $27.80\%$. The four-stage audit in \S\ref{subsec:eval_synthetic} measures beam retention and rerank admission directly. Tangent encoding raises recall to $99.385\%$ at $L=256$, $99.80\%$ at $L=384$, and $99.945\%$ peak. It maps residual normalization~\cite{jegou2011pq} into the existing 2-bit bitplanes. Equation~\eqref{eq:mnr_recovery} identifies the pairwise covariance and tangent scale governing this representation.

\subsection{Strong Convexity and Closed-Form Radial Minimization}
\label{subsec:geodesic_apex}

Next, we establish the exact radial lower bound and ray convexity properties governing shell filtering and entry-point selection.

\begin{proposition}[Radial Lower Bound, Strong Convexity, and Closed-Form Radial Minimizer]
\label{prop:geodesic_apex}
Let $q \in \Bball$ have radius $r_q$, conformal factor $f_q$, and depth $\rho_q = 2\atanh(\sqrt{c}\,r_q)$.
\begin{enumerate}
  \item \textbf{Lossless radial lower bound:}
  For every $v \in \Bball$,
  \begin{equation}
    P_{\mathbb{B}}(q, v) \ge P_{\mathrm{rad}}(r_q, r_v) \triangleq (r_q - r_v)^2 f_q f_v = \frac{2}{c}\bigl(\cosh|\rho_q - \rho_v| - 1\bigr),
    \label{eq:radial_lb}
  \end{equation}
  giving the radial lower bound used by the header gate. \ifpreprint Appendix~\ref{app:scalar_rounding} gives its FP32 perturbation bound.\fi

  \item \textbf{Strong convexity and closed-form radial minimizer:}
  Along any directional ray $\{v(\rho) : \rho \in \R\}$ with cosine alignment $\gamma = \langle u_q, \mu_a\rangle$, the hyperbolic cosine objective $g_\gamma(\rho) \triangleq \cosh(\sqrt{c}\,d_{\mathbb{B}}(q, v(\rho))) = \cosh\rho_q\cosh\rho - \gamma\sinh\rho_q\sinh\rho$ is strictly $1$-strongly convex on $\R$ ($g_\gamma''(\rho) = g_\gamma(\rho) \ge \cosh(\rho_q - |\rho|) \ge 1$) and attains its unique minimizer on $[0, \infty)$ at
  \begin{equation}
    \rho^*(\rho_q, \gamma) = \atanh\bigl(\tanh\rho_q \cdot \max(\gamma, 0)\bigr),
    \label{eq:geodesic_apex}
  \end{equation}
  and on any closed radial support $[\rho_{\min}, \rho_{\max}] \subset [0, \infty)$ at $\rho_{\mathrm{sup}}^* = \min\bigl(\max(\rho^*(\rho_q, \gamma), \rho_{\min}), \rho_{\max}\bigr)$, specializing the classical hyperbolic right-triangle orthogonal-projection identity and geodesic convexity of $\cosh(\sqrt{c}\,d_{\mathbb{B}})$ to bounded empirical support\ifpreprint\ (derivation in Appendix~\ref{app:warped_product})\fi.
\end{enumerate}
\end{proposition}

\begin{proof}[Proof Sketch]
Claim~(1) follows from $\gamma_{qv} \le 1$ in~\eqref{eq:polar_proxy}.
Differentiating $g_\gamma(\rho)$ twice yields $g_\gamma''(\rho) = g_\gamma(\rho) \ge \cosh\rho_q\cosh|\rho| - \sinh\rho_q\sinh|\rho| = \cosh(\rho_q - |\rho|) \ge 1$.
Setting $g_\gamma'(\rho) = \cosh\rho_q\sinh\rho - \gamma\sinh\rho_q\cosh\rho = 0$ gives $\tanh\rho = \gamma\tanh\rho_q$ for $\gamma > 0$ and boundary minimum $\rho^* = 0$ for $\gamma \le 0$; clamping follows by strict convexity.
\end{proof}

Unlike Euclidean space, for a coarse routing cone at $\gamma = 0.7$ and a deep query at $\rho_q = 4.0$, the closest point along ray $\mu_a$ bends inward to $\rho^* \approx 0.87$, while on hollow datasets such as {Cohere-1M} ($\rho_{0.01} = 1.44$), clamping to $[\rho_{\min}, \rho_{\max}]$ avoids targeting empty interior shells.

\subsection{Deterministic Asymmetric De-Biasing, CJ-Gate, and Sketch Variance}
\label{subsec:asym_cj}

Finally, we derive an algebraic estimator and two uncertainty models governing candidate-pool refinement, plateau expansion, and 256-bit sketch filtering.

\begin{proposition}[Deterministic Orthogonal-Projection Estimator, Off-Axis {CJ-Gate}, and Sketch Variance]
\label{prop:asym_cj_gate}
Let $\hat{u}_v$ be the two-level {SM} vector of $u_v \in \Sunit$ with Pythagorean norm $\sigma_v^2 = \|\hat{u}_v\|_2^2 = \langle u_v, \hat{u}_v\rangle \in (0, 1]$ from Lemma~\ref{lem:sm_pythagorean}.
\begin{enumerate}
  \item \textbf{Deterministic orthogonal-projection decomposition:}
  For \emph{every} deterministic pair $u_q, u_v \in \Sunit$, write $u_q = \gamma_{qv} u_v + w$ with $\langle w, u_v\rangle = 0$ ($\|w\|_2 = \sqrt{1 - \gamma_{qv}^2}$). Then without any distributional or rotational-symmetry assumption,
  \begin{equation}
    \frac{\langle u_q, \hat{u}_v\rangle}{\sigma_v^2}
    = \gamma_{qv} + \frac{\langle w, \hat{u}_v - \sigma_v^2 u_v\rangle}{\sigma_v^2}.
    \label{eq:orth_proj_asym}
  \end{equation}
  Dividing $\langle u_q, \hat{u}_v\rangle$ by $\sigma_v^2$ (rather than $\sigma_v$) recovers $\gamma_{qv}$ with zero multiplicative shrinkage along $u_v$, leaving only perpendicular residual $\langle w, \hat{u}_v\rangle / \sigma_v^2$.
  Algebraically analogous to {RaBitQ}'s $\langle \bar{o}, o\rangle$ inner-product divisor~\cite{gao2024rabitq}---which achieves unbiasedness in expectation via global random rotations---conditional-mean {SM} turns $\hat{u}_v$ into an orthogonal projection ($\langle u_v, \hat{u}_v\rangle = \sigma_v^2$, Lemma~\ref{lem:sm_pythagorean}) so~\eqref{eq:orth_proj_asym} holds deterministically without rotations or distributional hypotheses.

  \item \textbf{Off-axis Conformal Jacobian uncertainty (\textsc{CJ-Gate}):}
  Propagating directional standard deviation $\sigma_\gamma(q, v) = \sqrt{\kappa_{\mathrm{SM}}/D}\,\tilde{\beta}(q, v)$ through~\eqref{eq:polar_proxy}, where $\tilde{\beta}(q, v) = \beta_{q,c}\beta_v$ under cone encoding~\eqref{eq:cone_estimator} and $\tilde{\beta}(q, v) = 1$ globally, gives
  \begin{equation}
    \sigma_P(q, v) = \left|\frac{\partial P_{\mathbb{B}}(q, v)}{\partial \gamma_{qv}}\right|\sigma_\gamma(q, v)
    = 2 r_q r_v f_q f_v \sqrt{\frac{\kappa_{\mathrm{SM}}}{D}}\,\tilde{\beta}(q, v).
    \label{eq:cj_sigma}
  \end{equation}
  Using $\beta_v^2$ in place of $\beta_{q,c}\beta_v$ underestimates $\sigma_P(q, v)$ by factor $\beta_{q,c} / \beta_v$ on off-axis queries (up to $5.49\times$ when $\alpha_v = 0.995, \alpha_{q,c} = 0.84$).

  \item \textbf{Binomial sketch variance ratio:}
  For a $B_s$-bit sign sketch ($B_s = 256$) at observed Hamming distance $H = B_s p$ with arcsine cosine estimate $\hat{\gamma}_{B_s}(H) = \cos(\pi p)$, first-order delta propagation of $H \sim \operatorname{Binomial}(B_s, p)$ yields standard deviation $\sigma_{\mathrm{sk}}(p) = \frac{\pi}{\sqrt{B_s}}\sin(\pi p)\sqrt{p(1 - p)}$, whose ratio reaches $2.306\times$ at $H=128$ ($\hat\gamma=0$) and $2.265\times$ at $H=115$ ($\hat\gamma\approx0.16$) relative to high-similarity neighbors ($H_{\mathrm{ref}} = 48, \hat{\gamma} \approx 0.83$)\ifpreprint\ (derivations in Appendix~\ref{app:tangent_cone})\fi.
\end{enumerate}
\end{proposition}

\begin{proof}[Proof Sketch]
Substituting $u_q = \gamma_{qv} u_v + w$ into $\langle u_q, \hat{u}_v\rangle$ and applying $\langle u_v, \hat{u}_v\rangle = \sigma_v^2$ and $\langle w, u_v\rangle = 0$ gives $\langle u_q, \hat{u}_v\rangle = \gamma_{qv}\sigma_v^2 + \langle w, \hat{u}_v - \sigma_v^2 u_v\rangle$, proving~\eqref{eq:orth_proj_asym} deterministically.
Claim~(2) follows from $|\partial P_{\mathbb{B}} / \partial \gamma_{qv}| = W(r_q, r_v) = 2 r_q r_v f_q f_v$ and~\eqref{eq:cone_estimator}.
For Claim~(3), $|\frac{\mathrm{d}}{\mathrm{d}p}\cos(\pi p)| = \pi\sin(\pi p)$ multiplied by $\sqrt{\Var(H/B_s)} = \sqrt{p(1-p)/B_s}$ yields $\sigma_{\mathrm{sk}}(p)$.
\end{proof}

\section{Warped-Product Representation and Graph Topology}
\label{sec:design}

\begin{figure*}[t]
  \centering
  \includegraphics[width=\textwidth]{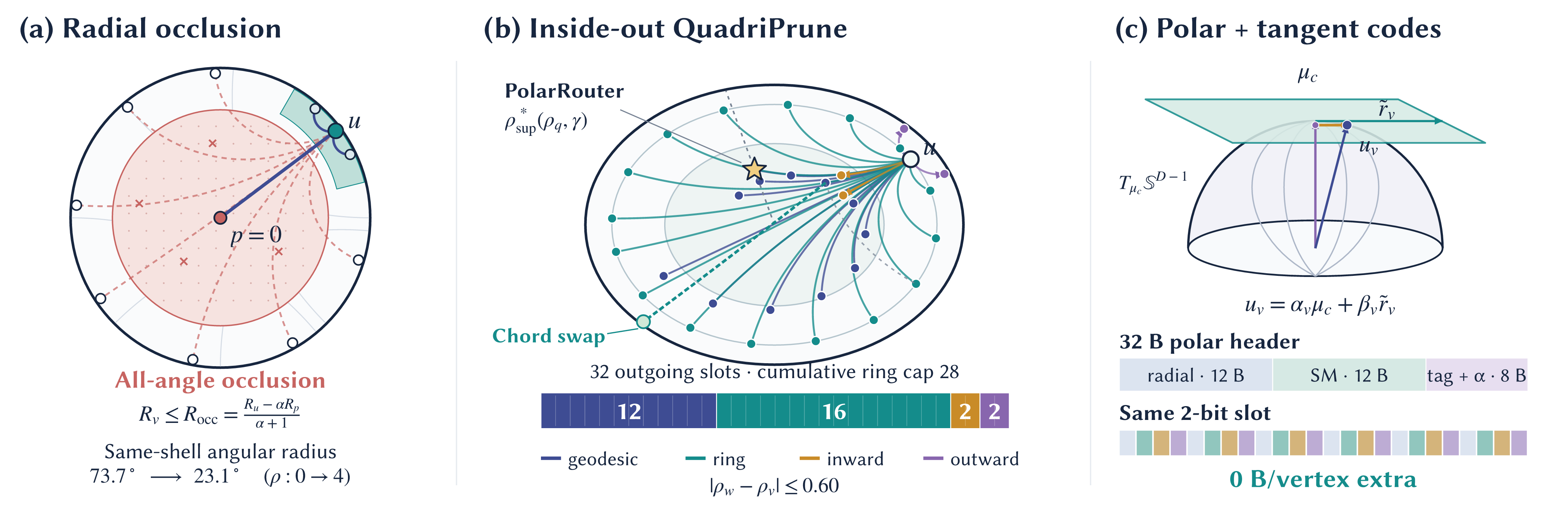}
  \caption{Geometric design of \rquiver{}. (a)~An origin hub exposes all-angle inner-ball occlusion and a shrinking same-shell angular cap ($\alpha=1.2$). (b)~A 32-slot example allocates 12 geodesic, 16 ring, two inward, and two outward links, with a cumulative ring cap of 28. \texttt{PolarRouter} selects ray-support entries, and chord swaps attach islands while preserving the BFS tree. (c)~Tangent encoding reuses the polar header and directional bitplanes, with cosine-error multiplier $\beta_{q,c}\beta_v$. On synthetic {Clustered-Narrow}, Recall@10 rises from $34.22\%$ to $99.385\%$ at $L=256,B=128$.}
  \label{fig:geometry_overview}
\end{figure*}

Figure~\ref{fig:geometry_overview} summarizes the geometry and representation. Guided by Theorems~\ref{thm:conformal_flip}--\ref{thm:tangent_cone} and Proposition~\ref{prop:geodesic_apex}, \rquiver{} models the index as a \emph{warped-product fiber-bundle graph} over $\R_{\ge 0} \times_{\sinh} \Sunit$, decoupling exact radial curvature from 2-bit spherical directions in memory and separating spherical diversity edges from radial transitions during construction.

\subsection{Zero-Overhead Polar Vertex Representation}
\label{subsec:polar_layout}

\paragraph{32-Byte \texttt{PolarNodeMeta} Header.}
For each vertex $v \in \{0, \dots, N-1\}$, \rquiver{} stores a fixed 32-byte polar header (half a 64-byte cache line) paired with packed 2-bit {SM} bitplanes occupying $2\lceil D / 64\rceil \times 8$ bytes:
\begin{itemize}
  \item \emph{Unquantized radial scalars (12\,B):} Euclidean radius $r_v \in [0, 1/\sqrt{c})$, boundary factor $F_v=(1-cr_v^2)^{-1}=f_v/2$, and dimensionless geodesic depth $\rho_v = 2\atanh(\sqrt{c}\,r_v)$, computed in FP64 at ingestion and stored in FP32.
  \item \emph{Conditional-mean two-level {SM} scalars (12\,B):} weak-coordinate mean $a_v$, level step $\delta_v = b_v - a_v$, and reconstructed two-level norm $\sigma_v = \|\hat{u}_v\|_2$ in FP32 (Lemma~\ref{lem:sm_pythagorean}).
  \item \emph{Bitpacked count/cone tag and axial/norm scalar (8\,B):} a 32-bit word packing the 20-bit strong-magnitude count $s_v$ with a 12-bit residual cone tag ($\text{cone\_id}$, with \texttt{0x0FFF} denoting global native {SM}), paired with an FP32 field storing axial alignment $\alpha_v = \langle u_v, \mu_c\rangle$ for residual-cone vertices or precomputed canonical inverse norm $(D+3s_v)^{-1/2}$ for global vertices. The triple $(a_v,\delta_v,\sigma_v)$ supports the main conditional-mean hop scorer and \textsc{Stage-0.5}. The global inverse-norm slot supports legacy canonical paths, while residual scoring reads $\alpha_v$ and computes $\beta_v$.
\end{itemize}
Table~\ref{tab:node_layout} gives the per-vertex logical payload at $D=768$, $M_0=32$ ($384$\,B with \texttt{SymSketch256} under \texttt{HotSmPage}, or $352$\,B without the sketch in split form), and Table~\ref{tab:memory_audit} reconciles this payload with the benchmark's navigation-byte counter. The measured allocation is $517.0$\,MB with page co-location ($2.44\times$ below the measured {Polar-SQ8} implementation at $1{,}261.0$\,MB and $6.90\times$ below {Polar-FP32} at $3{,}565.0$\,MB) or $485.0$\,MB in split form ($2.60\times$ and $7.35\times$ smaller).\ifpreprint\ Appendix~\ref{app:scalar_rounding} quantifies header rounding.\fi

\begin{table}[t]
  \centering
  \caption{Per-vertex logical representation payload at $D=768, M_0=32$ ($N=1\text{M}$). For the 2-bit and SQ8 paths, cold FP32 vectors are read during final reranking. Full allocation components appear in Table~\ref{tab:memory_audit}.}
  \label{tab:node_layout}
  \small
  \setlength{\tabcolsep}{3pt}
  \begin{tabularx}{\linewidth}{@{}>{\raggedright\arraybackslash}Xrrrr@{}}
    \toprule
    \textbf{Representation} & \textbf{\shortstack{Header\\(/Sketch)}} & \textbf{\shortstack{Dir.\\Code}} & \textbf{\shortstack{Active\\Edges}} & \textbf{\shortstack{Logical\\Total}} \\
    \midrule
    Full-Vector Nav (FP32) & 32\,B & 3,072\,B & 128\,B & 3,232\,B \\
    Polar-SQ8 (8-bit dir.) & 32\,B & 768\,B & 128\,B & 928\,B \\
    \rquiver{}-Split (no sketch) & 32\,B & 192\,B & 128\,B & 352\,B$^\dagger$ \\
    \textbf{\rquiver{} (\texttt{HotSmPage})} & \textbf{64\,B} & \textbf{192\,B} & \textbf{128\,B} & \textbf{384\,B$^\dagger$} \\
    \bottomrule
  \end{tabularx}
  \par\vspace{2pt}
  {\footnotesize\raggedright\noindent $^\dagger$\texttt{HotSmPage} adds $32$\,B \texttt{SymSketch256} to the $32$\,B polar header ($64$\,B \texttt{HotSmHeader}). Active edges count $M_0=32$ \texttt{u32} IDs ($128$\,B), while fixed-stride adjacency reserves $64$ slots/vertex ($256$\,MB; Table~\ref{tab:memory_audit}). Residual encoding reuses the same per-vertex slots ($4CD$\,B global dictionary for $C\le512$ axes, or $0.50$\,MiB at $C=256,D=512$).\par}
\end{table}

\begin{table*}[t]
 \centering
 \caption{Byte reconciliation on {Cohere-1M}, $N=10^6$, $D=768$, $M_0=32$. Entries are decimal MB. Fixed-stride adjacency reserves 64 neighbor slots and one degree word per vertex. The logical payload in Table~\ref{tab:node_layout} includes 32 edge slots.}
 \label{tab:memory_audit}
 \small
 \begin{tabular}{@{}lrrrr@{}}
 \toprule
 Component & 2-bit split & 2-bit page-co-located & Polar-SQ8 & Polar-FP32\\
 \midrule
 32-byte polar headers & 32.000000 & \multirow{3}{*}{256.000000} & 32.000000 & 32.000000\\
 2-bit directional storage & 192.000000 & & 192.000000 & 192.000000\\
 Sketch and record alignment & 0 & & 0 & 0\\
 32 logical edge slots, 32-bit IDs & 128.000000 & 128.000000 & 128.000000 & 128.000000\\
 32 reserved neighbor slots & 128.000000 & 128.000000 & 128.000000 & 128.000000\\
 Degree word & 4.000000 & 4.000000 & 4.000000 & 4.000000\\
 Router arrays & 0.974912 & 0.974912 & 0.974912 & 0.974912\\
 FP64 directional inverse norms & 0 & 0 & 8.000000 & 8.000000\\
 SQ8 directional coordinates & 0 & 0 & 768.000000 & 0\\
 FP32 coordinates used in traversal & 0 & 0 & 0 & 3,072.000000\\
 \midrule
 Navigation total & 484.974912 & 516.974912 & 1,260.974912 & 3,564.974912\\
 \bottomrule
 \end{tabular}
 \par\vspace{2pt}
 {\footnotesize The SQ8/FP32 controls retain their 2-bit store and add an 8-byte inverse norm per vertex. Router arrays comprise 393,216 B cone bitplanes, 32,768 B SM parameters, 8,192 B inverse norms, 524,288 B cell representatives, 16,384 B coarse shortlists, and 64 B origin hubs. Adjacency stores forward lists in one fixed-stride array. Reverse insertions update these lists in place. The counter sums navigation-array payloads. Process RSS includes the cold vector store, ID maps, and allocator state.}
\end{table*}

\paragraph{Purity-Gated \texttt{ResidualConeDict}.}
To realize Theorem~\ref{thm:tangent_cone}(3) without disturbing unclustered manifolds, \rquiver{} executes a deterministic cone-purity check at index build over a uniform sample of $\min(N, 8{,}192)$ vertices:
\begin{enumerate}
  \item \emph{Conformal depth gate:} If the 90th-percentile sample depth satisfies $p_{90}(\rho) < 1.0$, boundary conformal amplification is benign and residual encoding exits immediately with $0$ cones.
  \item \emph{Leader seeding and two-round purity refinement:} Otherwise, a deterministic spherical leader pass collects candidate unit axes separated by cosine $< \tau_{\mathrm{lead}} = 0.91$ (capped at $1{,}024$ candidates), followed by two rounds of spherical mean refinement assigning samples with $\langle u_s, \mu_c\rangle \ge \alpha_{\min} = 0.94$.
  A cone $\mu_c$ is retained if and only if it meets minimum sample support $\operatorname{clamp}(\lfloor N_{\mathrm{samp}}/1024\rfloor, 3, 12)$ and achieves high intra-cone mean alignment $\bar{\alpha}_c \ge 0.955$.
\end{enumerate}
Every database vertex $v$ achieving $\alpha_v = \langle u_v, \mu_c\rangle \ge 0.94$ against its nearest retained axis stores the 2-bit {SM} code of its normalized tangent residual $\tilde{r}_v = (u_v - \alpha_v\mu_c)/\sqrt{1 - \alpha_v^2}$ in the existing 192-byte bitplane slot and records $\text{cone\_id}$ in the 12-bit header field (with \texttt{0x0FFF} reserved for global {SM}, \textbf{0\,B/vertex} overhead), giving $\beta_v^2=1-\alpha_v^2\le0.1164$. The cosine-error multiplier is $\beta_{q,c}\beta_v$ in Theorem~\ref{thm:tangent_cone}(3).
Setting $\alpha_{\min} = 0.94$ and $\bar{\alpha}_c \ge 0.955$ activates residual coding only when tight concentration ($\beta_v^2 \le 0.1164$) outweighs axis-projection cost: all four diffuse real datasets pass through on the zero-overhead global path ($0$ cones), whereas both clustered benchmarks retain all $256$ axes ($0.50$\,MiB at $D=512$).

\subsection{Inside-Out Seeding and Stem-Excess QuadriPrune}
\label{subsec:compass}

By Theorem~\ref{thm:vamana_pathologies}(1), single-metric {Vamana} pruning allows interior vertices to occlude entire spherical neighborhoods of boundary vertices.
Algorithm~\ref{alg:compass_build} replaces isotropic pruning with an inside-out fiber-bundle construction.

\begin{algorithm}[t]
\caption{Warped-Product Construction \& Chord-Swap Repair}
\label{alg:compass_build}
\small
\begin{algorithmic}[1]
\Require Dataset $\mathcal{X} \subset \Bball$, max out-degree $M_0 = 2M$, build beam $L_c$, slack $\alpha > 1$
\Ensure Graph $G=(V,E)$ with maximum degree $M_0$ and measured entry-reachable set $\mathcal R$
\State Compute FP32 polar headers $(r_v, f_v, \rho_v)$, fit purity-gated \texttt{ResidualConeDict}, encode 2-bit {SM} payloads, and build \texttt{PolarRouter} (\S\ref{subsec:router})
\State Sort vertices $\pi = (v_1, \dots, v_N)$ in non-decreasing geodesic depth $\rho_{v_1} \le \dots \le \rho_{v_N}$ and initialize empty adjacency lists
\For{consecutive radial wave $B_t$ of up to 256 vertices in $\pi$}
  \For{each $u \in B_t$ in parallel}
    \State $\mathcal{C}_u \gets \textsc{PolarBeamSearch}(G, u, L_c)$ via 2-bit polar proxy $\widehat{P}_{\mathbb{B}}$
    \State Partition up to $M_0$ outgoing edges from $\mathcal{C}_u$ via \textbf{QuadriPrune}:
    \State \hspace{0.8em}(a) \emph{Geodesic fiber} (cumulative cap $\max(1,\lfloor3M_0/8\rfloor)$): stem-excess prune~\eqref{eq:dom_geo}
    \State \hspace{0.8em}(b) \emph{Shielded isobathic ring} (cumulative cap $\lfloor7M_0/8\rfloor$): rank by angle plus depth gap, test dominators with $|\rho_w-\rho_v|\le0.60$
    \State \hspace{0.8em}(c)--(d) \emph{Inward \& outward elevators} : split remaining capacity, select step-aligned transitions via~\eqref{eq:elevator_obj}, then apply fallback fill
    \State Insert reverse edges $v \to u$ for $v \in N^+(u)$, re-applying {QuadriPrune} when $\deg^+(v) > M_0$
  \EndFor
\EndFor
\State Final-prune every adjacency list to $M_0$ slots
\State $(\mathcal{R}, \texttt{parent}) \gets \textsc{DirectedBFSTree}(G, \text{entry\_hub})$ \Comment{\S\ref{subsec:repair}}
\For{each $u \in \pi$ with $u \notin \mathcal{R}$}
  \State Search $N^+(u)$, hubs, and the 256-vertex interior prefix for reachable $p$ with a free slot or removable non-tree chord
  \If{no qualifying anchor exists} \State Continue to the next vertex \EndIf
  \If{$\deg^+(p) < M_0$} append $p \to u$
  \Else\ replace peripheral chord $p \to w^\star$ s.t.\ $\texttt{parent}[w^\star] \ne p \land \deg^-(w^\star) \ge 2$ with $p \to u$
  \EndIf
  \State $\texttt{parent}[u] \gets p$; expand $(\mathcal{R}, \texttt{parent})$ via incremental {BFS} from $u$
\EndFor
\State \Return fixed-stride graph $G$, reachable set $\mathcal R$, and $4096$\,B \texttt{HotSmPage} layout
\end{algorithmic}
\end{algorithm}

\paragraph{Inside-Out Geodesic Seeding.}
Sorting vertices by non-decreasing depth schedules interior vertices before boundary waves. Router seeds are computed from the full database before these 256-vertex waves connect their adjacency lists.

\paragraph{Stem-Excess and Angle-Guarded Geodesic Dominance.}
To prevent interior vertices from occluding opposite subtrees via Theorem~\ref{thm:vamana_pathologies}(1)---a curvature-induced hub pathology distinct from flat angular redundancy~\cite{fu2019nsg,fu2022high} and cross-modal query shift~\cite{chen2024roargraph}---\rquiver{} replaces raw $\alpha$-pruning with \emph{stem-excess angle-guarded dominance}.
Let $\tilde{d}(x, y) = \sqrt{c}\,d_{\mathbb{B}}(x, y)$ denote dimensionless geodesic distance and $\ell_\angle(x, y) = 1 - \hat{\gamma}(u_x, u_y)$ denote angular loss.
Define the radial stem excess and effective non-stem distance of candidate $v$ relative to base $u$ and existing neighbor $w$:
\begin{align}
  \Delta_{\mathrm{stem}}(u, v; w) &\triangleq \max\!\bigl(0,\; |\rho_u - \rho_v| - |\rho_w - \rho_v|\bigr), \nonumber \\
  \tilde{d}_{\mathrm{eff}}(u, v; w) &\triangleq \max\!\bigl(0,\; \tilde{d}(u, v) - \Delta_{\mathrm{stem}}(u, v; w)\bigr).
  \label{eq:stem_excess}
\end{align}
Neighbor $w$ geodesically dominates candidate $v$ ($\operatorname{Dom}_{\mathrm{geo}}(w \mid u, v)$) if and only if
\begin{equation}
  \bigl[\ell_\angle(w, v) \le \ell_\angle(u, v)\bigr] \;\land\;
  \bigl[\alpha\,\tilde{d}(w, v) \le \tilde{d}_{\mathrm{eff}}(u, v; w)\bigr].
  \label{eq:dom_geo}
\end{equation}
Condition $\ell_\angle(w, v) \le \ell_\angle(u, v)$ forbids $w$ from pruning candidates at wider spherical angles than $u$ itself (preventing $180^\circ$ cross-origin occlusion), while subtracting $\Delta_{\mathrm{stem}}$ ensures $w$ shadows $v$ through genuine angular proximity rather than radial depth advantage alone.
Moreover, because $\tilde{d}(w, v) \ge |\rho_w - \rho_v|$ by the reverse triangle inequality, checking $\alpha|\rho_w - \rho_v| > \tilde{d}_{\mathrm{eff}}(u, v; w)$ rejects non-dominating pairs in $O(1)$ scalar arithmetic without evaluating $\arccosh$.

\paragraph{Shielded Isobathic Rings, Elevators, and Channel Allocation.}
Selection uses cumulative caps within each $M_0=32$-slot list.
(i)~The geodesic channel fills up to $\max(1,\lfloor3M_0/8\rfloor)=12$ slots via~\eqref{eq:dom_geo}.
(ii)~The ring channel extends the combined list to 28 entries. It orders candidates by $\ell_\angle(u,v)+0.85|\rho_u-\rho_v|$. A selected neighbor $w$ may dominate $v$ only if $|\rho_w-\rho_v|\le0.60$ and $\alpha\ell_\angle(w,v)\le\ell_\angle(u,v)$.
(iii)--(iv)~With $s$ entries selected and $h=M_0-s$ slots remaining, the inward pass targets $s+\lfloor h/2\rfloor$ total entries and the outward pass fills toward $M_0$. Thus a full 28-entry prefix leaves two slots for each radial direction. Their objective is
\begin{equation}
 J_{\mathrm{elev}}(u,v)=\ell_\angle(u,v)
 +0.15\bigl||\rho_v-\rho_u|-0.65\bigr|.
 \label{eq:elevator_obj}
\end{equation}
Candidates must satisfy $\rho_v<\rho_u-0.18$ inward or $\rho_v>\rho_u+0.18$ outward. Direction-specific dominators satisfy $\rho_w<\rho_u-0.10$ or $\rho_w>\rho_u+0.10$, respectively, and use the angular $\alpha$ test.
If these passes leave capacity, a diversity pass orders remaining candidates by $\ell_\angle(u,v)+0.25|\rho_u-\rho_v|$ and admits angular dominators within $|\rho_w-\rho_v|\le1.0$. A final ordered fill takes remaining distinct candidates up to $M_0$.
At constant depth the elevator sets are empty and the ring ranks by angle, using the fixed dataset-independent channel quotas and tolerances in Table~\ref{tab:fixed_parameters}. Appendix~\ref{app:vamana_pathologies} derives the channel-wise constant-depth and flat-curvature limits.

\subsection{Spanning-Tree Chord-Swap Island Repair}
\label{subsec:repair}

Parallel batch insertion and hub saturation can still leave $0.05\%\text{--}0.06\%$ of extreme boundary vertices with zero incoming edges ($\deg^-(u) = 0$: $518\text{--}626$ islands on {Cohere-1M}), capping asymptotic recall strictly below $100\%$.
The chord-swap pass in Algorithm~\ref{alg:compass_build} grows directed {BFS} entry reachability while preserving maximum out-degree $\deg^+(p)\le M_0$. A qualifying anchor has spare capacity or an eligible non-tree chord.
Directed {BFS} initializes a spanning tree $(\mathcal{R},\texttt{parent})$ from the router entry hub, each unreachable vertex $u \notin \mathcal{R}$ selects a reachable anchor $p \in \mathcal{R}$ (scanning $N^+(u)$, router hubs, and the interior core prefix) and either appends $p \to u$ (if $\deg^+(p) < M_0$) or replaces a peripheral non-tree chord $p \to w^\star$ satisfying both $\texttt{parent}[w^\star] \ne p$ and $\deg^-(w^\star) \ge 2$, decrementing $\deg^-(w^\star)$ in place and expanding $(\mathcal{R}, \texttt{parent})$ incrementally from $u$\ifpreprint\ (monotonic reachability invariant proved in Appendix~\ref{app:vamana_pathologies})\fi.
Because no spanning-tree edge is ever overwritten, $\mathcal{R}$ grows monotonically without cycle breakage; across all 21 million-scale {Zen~4} graphs, the pass reduces zero-in-degree islands to \textbf{0} ($1{,}000{,}000 / 1{,}000{,}000$ directed-{BFS} reachable from the entry hub) in under $0.3$\,s.

\subsection{Convex-Support Polar Router}
\label{subsec:router}

\rquiver{} builds $C=\min(N,\operatorname{clamp}(\lfloor N/100\rfloor,64,2048))$ directional landmarks in SM mode and crosses them with eight quantile-adaptive shells. At 1M scale this gives 2,048 fine cones and 16,384 cells, each storing up to eight 32-bit vertex IDs. For $C>256$, 64 coarse landmarks each retain a 128-cone query shortlist.
The query scores coarse landmarks, expands the leading shortlists, and uses the constrained radial minimizer $\rho_{\mathrm{sup}}^*(\rho_q,\gamma)$ from Proposition~\ref{prop:geodesic_apex}(2) to select shell representatives. The radial support endpoints are the smallest positive sampled database depth and the maximum depth. The variance-stabilized configuration activates a sampled isotropy guard that scans fine cones directly when coarse alignment falls below its dimension-scaled threshold. Table~\ref{tab:fixed_parameters} gives the fixed settings.

\begin{table*}[t]
 \centering
 \caption{Fixed construction and execution settings. Search operating points are scanned separately in \S\ref{subsec:eval_setup}.}
 \label{tab:fixed_parameters}
 \small
 \begin{tabularx}{\textwidth}{@{}l>{\raggedright\arraybackslash}X@{}}
 \toprule
 Mechanism & Fixed values and rules\\
 \midrule
 Construction & $M=16$, $M_0=32$, $L_c=128$, $\alpha=1.2$, radial wave size 256, repair fallback prefix 256\\
 Geodesic/ring caps & 12 geodesic slots, then 28 cumulative slots at $M_0=32$\\
 Ring & depth-gap weight 0.85, dominator--candidate tolerance 0.60, $\alpha_{\mathrm{iso}}=1.2$\\
 Elevators & step 0.65, step penalty 0.15, candidate depth offset 0.18, dominator depth offset 0.10\\
 Diversity fill & depth-gap weight 0.25, dominator--candidate tolerance 1.0\\
 Router & up to 2,048 fine landmarks, 64 coarse landmarks, 128-entry query shortlists, 8 shells, 8 IDs/cell, 16 origin hubs\\
 Shell boundaries & if $p_{01}(\rho)>0.35$, depth quantiles $[0.005,0.02,0.05,0.12,0.25,0.45,0.72]$\newline otherwise $[0.25,0.48,0.75,1.15,1.75,2.60,3.80]$\\
 Isotropy guard & coarse threshold $\operatorname{clip}_{[0.11,0.19]}(3.8/\sqrt{\max(D,64)})$\newline sampled median maximum fine alignment below $\operatorname{clip}_{[0.19,0.30]}(5.6/\sqrt{\max(D,64)})$\\
 Cone selection & sample $\min(N,8192)$, depth $p_{90}\ge1.0$, leader threshold 0.91, assignment 0.94, mean purity 0.955\newline support $\operatorname{clamp}(\lfloor N_{\mathrm{samp}}/1024\rfloor,3,12)$, 1,024 leader candidates, two refinement rounds, at most 512 retained axes\\
 Query cone cache & materialization $|\alpha_{q,c}|\ge0.20$, $\beta_{q,c}>10^{-6}$, at most 16 slots\\
 CJ-Gate & $\kappa_{\mathrm{SM}}=0.221$, relative gap 0.22, $z=2$, expansion offset 0.8, cap 768\\
 Sketch / prefetch & 256 bits, variance reference $H=48$, header prefetch window 8\\
 \bottomrule
 \end{tabularx}
\end{table*}

\section{Hardware-Conscious Polar Execution Engine}
\label{sec:system}

\begin{figure*}[t]
  \centering
  \includegraphics[width=\textwidth]{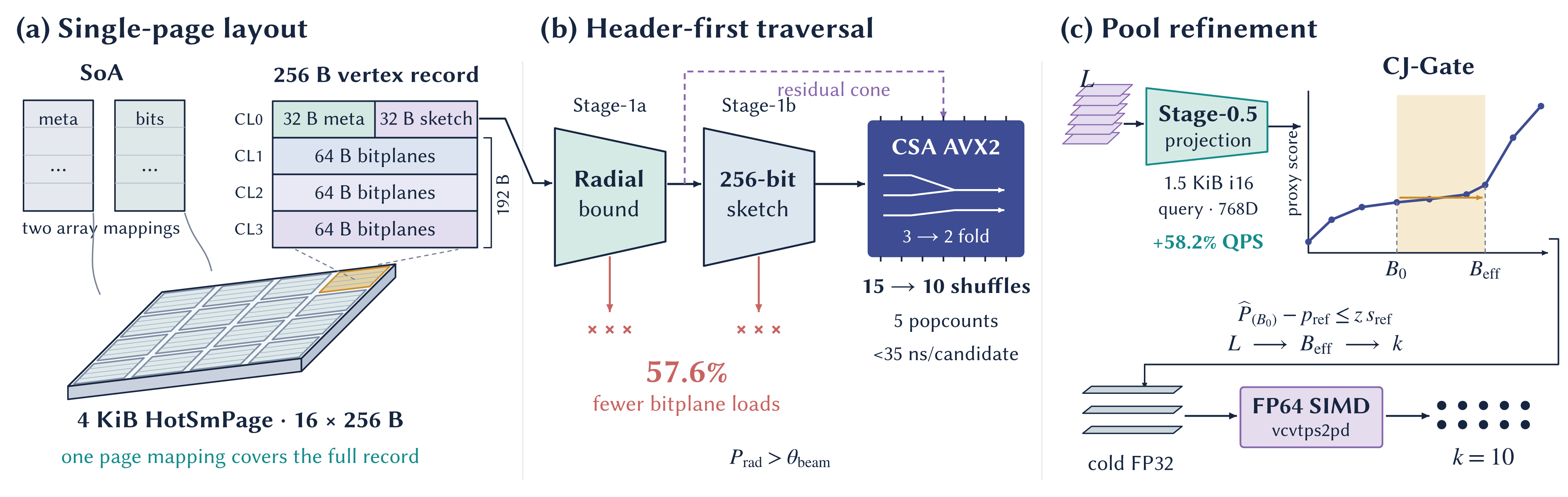}
  \caption{Header-first query execution at $D=768$. (a)~One $4$\,KiB page contains 16 records, each with a 64\,B header and 192\,B bitplanes. (b)~Radial and sketch gates precede full bitplane reads, with residual vertices bypassing the sketch gate. The skip rate and kernel annotations use the {Cohere-1M} measurements. (c)~Asymmetric projection precedes adaptive rerank admission and cold-vector FP64 evaluation. The score curve is schematic. \textsc{Stage-0.5} improves $99.5\%$-recall {Cohere} QPS by $58.2\%$ in the controlled ablation. The measured $+63.7\%$ ({Cohere}, $99.5\%$ recall) and $+76.5\%$ ({MERU}, $99\%$ recall) gains combine \textsc{CJ-Gate} with stabilized sketching.}
  \label{fig:execution_overview}
\end{figure*}

Figure~\ref{fig:execution_overview} follows a query from page translation to final reranking. On million-scale hyperbolic graphs, achieving $99\%$ Recall@10 at beam width $L \in [128, 256]$ expands $300\text{--}350$ vertices and evaluates $4{,}500\text{--}5{,}800$ candidate neighbors per query, while reading only $B \in [64, 128]$ cold FP32 vectors during final reranking.
End-to-end latency is therefore governed by random-access {DTLB} walks, cache-line traffic during neighbor expansion, and 2-bit bitplane population counts.
Query-private scratch buffers are reset by generation stamps and reused across requests. Candidate scoring depends on the current query and index state.

\subsection{Single-Page Co-Location Avoiding Cross-Array DTLB Walks}
\label{subsec:colocated_page}

In conventional Structure-of-Arrays ({SoA}) engines, polar headers ($32$\,MB at $N=10^6$) and 2-bit bitplanes ($192$\,MB at $D=768$) occupy separate contiguous arrays hundreds of megabytes apart in virtual memory (whereas disk-block reorderers~\cite{wang2024starling} and neighbor-embedded quantizers~\cite{gou2025symphonyqg} either require graph-isomorphism renumbering or duplicate codes across edges).
Because x86-64 cores provide only $64\text{--}96$ L1 {DTLB} entries and $1{,}024\text{--}3{,}072$ L2 {DTLB} entries for $4$\,KiB pages, $5{,}500$ random neighbor hops across a $224$\,MB working set miss the {DTLB} on nearly every hop---forcing two independent page-table walks per candidate under {SoA} (one for metadata, a second for bitplanes).

\begin{table}[t]
  \centering
  \caption{Hardware-page-aligned $4096$\,B \texttt{HotSmPage} layout with power-of-two per-vertex record stride $S_{\mathrm{rec}} = \operatorname{next\_pow2}(64 + 16\lceil D/64\rceil)$\,B (shown at $D=768$: $S_{\mathrm{rec}} = 256$\,B, $K = 16$ vertices/page). Resolving the $64$\,B \texttt{HotSmHeader} translates the entire vertex record on the identical $4$\,KiB OS page.}
  \label{tab:hot_sm_page}
  \small
  \setlength{\tabcolsep}{5pt}
  \begin{tabularx}{\linewidth}{@{}l>{\raggedright\arraybackslash}Xr@{}}
    \toprule
    \textbf{Record Offset} & \textbf{Per-Vertex Slot in $4096$\,B \texttt{HotSmPage} ($D=768$)} & \textbf{Size} \\
    \midrule
    \texttt{0x00..0x1F} & \texttt{PolarNodeMeta} (radial, {SM}, \& cone scalars) & $32$\,B \\
    \texttt{0x20..0x3F} & \texttt{SymSketch256} ($4 \times \texttt{u64}$ strided sign sketch) & $32$\,B \\
    \texttt{0x40..0x9F} & Positive-sign {SM} bitplane (\texttt{pos}, $12 \times \texttt{u64}$) & $96$\,B \\
    \texttt{0xA0..0xFF} & Strong-magnitude {SM} bitplane (\texttt{strong}, $12 \times \texttt{u64}$) & $96$\,B \\
    \bottomrule
  \end{tabularx}
\end{table}

Table~\ref{tab:hot_sm_page} shows our page-co-located \texttt{HotSmPage} layout:
each vertex occupies a power-of-two record of $S_{\mathrm{rec}} = \operatorname{next\_pow2}(64 + 16\lceil D/64\rceil)$ bytes ($256$\,B at $D=512, 768$ packing $K = 16$ vertices/page; $512$\,B at $D=1536$ packing $K = 8$ vertices/page; $128$\,B at $D=20$ packing $K = 32$ vertices/page, supporting up to $D = 16{,}128$ within one $4$\,KiB page).
Because $S_{\mathrm{rec}}$ divides $4096$ evenly ($\text{page\_id} = \text{vid} / K$, $\text{slot} = \text{vid} \bmod K$), every vertex's 64-byte \texttt{HotSmHeader} (\texttt{PolarNodeMeta} + \texttt{SymSketch256} in cache line~0) and its complete 2-bit {SM} bitplanes reside contiguously inside a single $4096$-byte OS page.
Once hardware resolves the virtual-to-physical translation for \texttt{HotSmHeader} at $\text{base} + S_{\mathrm{rec}} \cdot \text{slot}$ (supported up to $D \le 16{,}128$), any subsequent bitplane reads (\texttt{0x40..0xFF} at $D=768$) share the \emph{exact same $4$\,KiB virtual-page translation}, avoiding cross-array secondary page-table walks per accessed vertex record.

\subsection{Variance-Stabilized Two-Level Beam Filtering (\texttt{SymSketch256})}
\label{subsec:symsketch}

Even within one $4$\,KiB page, fetching full 2-bit {SM} bitplanes ($192$\,B = $3$ cache lines at $D=768$; $384$\,B = $6$ cache lines at $D=1536$) for every unvisited neighbor incurs unnecessary memory traffic once the beam tightens around $q$.
\rquiver{} first prefetches only the single 64-byte \texttt{HotSmHeader} cache line (consuming just one hardware Line Fill Buffer entry per neighbor) and executes a two-stage gate against the current $L$-th beam bound $\theta_{\mathrm{beam}}$:
\begin{enumerate}
  \item \textbf{Stage-1a (Exact Radial Lower-Bound Gate, $0$ extra cache lines):}
  Using $(r_v, f_v)$ from \texttt{HotSmHeader} (\texttt{0x00..0x1F}), if $P_{\mathrm{rad}}(r_q, r_v) = (r_q - r_v)^2 f_q f_v > \theta_{\mathrm{beam}}$ (Proposition~\ref{prop:geodesic_apex}(1)), $v$ is discarded by the lossless radial bound. \ifpreprint FP32 header perturbations follow Appendix~\ref{app:scalar_rounding}.\fi

  \item \textbf{Stage-1b (Binomial-Stabilized \texttt{SymSketch256} Gate, same $64$\,B header line):}
  The upper half of \texttt{HotSmHeader} (\texttt{0x20..0x3F}) stores a $256$-bit ($32$\,B) strided sign sketch.
  Computing the 256-bit Hamming distance $H \in \{0, \dots, 256\}$ against $q$'s sketch via four \texttt{u64} \texttt{XOR}+\texttt{POPCNT} instructions indexes two compile-time lookup tables ($2$\,KB total in L1): the exact arcsine table $\hat{\gamma}_{256} = \texttt{COS\_PI\_256}[H] = \cos(\pi H / 256)$ and the binomial-model standard-deviation ratio table $\rho_{\mathrm{std}}(H) = \texttt{SKETCH256\_STD\_RATIO}[H] = \max(1, \sigma_{\mathrm{sk}}(H/256) / \sigma_{\mathrm{sk}}(48/256))$ derived in Proposition~\ref{prop:asym_cj_gate}(3).
  Forming the variance-stabilized optimistic cosine bound $\hat{\gamma}_{\mathrm{upper}} = \min(1, \hat{\gamma}_{256} + \Delta_0 \cdot \rho_{\mathrm{std}}(H))$ yields optimistic proxy $\widehat{P}_{\mathbb{B}}^{\mathrm{opt}}(q, v) = R(r_q, r_v) + W(r_q, r_v)(1 - \hat{\gamma}_{\mathrm{upper}})$; whenever $\widehat{P}_{\mathbb{B}}^{\mathrm{opt}}(q, v) > \theta_{\mathrm{beam}}$ (for non-residual vertices once the beam fills), $v$ is pruned without prefetching or reading its trailing bitplane cache lines.
\end{enumerate}
Scaling the sketch margin by $\rho_{\mathrm{std}}(H)$ keeps tight filtering ($1.0\times$) for high-similarity neighbors ($H \le 48$) while automatically widening to $2.265\times$ at $H=115$ and a maximum of $2.306\times$ at $H=128$ to prevent premature pruning on moderate-angle manifolds.
In our controlled three-seed ablation on {Cohere-1M}, Stage-1a and \texttt{SymSketch256} skip $57.6\%$ of full bitplane loads, raising single-thread $99\%$-recall QPS from $1{,}752$ ({SoA}) to $1{,}909$ (\texttt{HotSmPage}) and $2{,}071$ (\texttt{HotSmPage} + \texttt{SymSketch256}, $+18.2\%$ total).

\subsection{3-to-2 Carry-Save AVX2 Bitplane Kernel}
\label{subsec:csa_kernel}

During hop traversal, the engine evaluates the conditional-mean product in Lemma~\ref{lem:sm_pythagorean}, followed by~\eqref{eq:sm_calibration}. For disagreement mask $x=s_q\oplus s_v$, the five counts are $\popcnt(x)$, $\popcnt(x\wedge m_q)$, $\popcnt(x\wedge m_v)$, $\popcnt(m_q\wedge m_v)$, and $\popcnt(x\wedge m_q\wedge m_v)$. Together with the two headers' $(a,\delta,\sigma,s)$, these counts reconstruct the bilinear product without square roots in the main scorer.

\paragraph{Residual-Aware Hop Dispatch.}
Let $K(a,b)$ denote this calibrated symmetric scorer and $A(q,b)$ the asymmetric scorer of \S\ref{subsec:stage05_rerank}. A global vertex uses $K(u_q,u_v)$. A residual vertex uses its cone tag to read axis $\mu_c$ and cache $\alpha_{q,c}=\langle u_q,\mu_c\rangle$, $\beta_{q,c}=\sqrt{1-\alpha_{q,c}^2}$. Its header stores $\alpha_v$ in the shared axial/norm slot and the conditional means of $\tilde r_v$.
\begin{algorithm}[t]
 \caption{Residual-aware candidate cosine scoring}
 \label{alg:residual_score}
 \begin{algorithmic}[1]
 \Require Header $h_v$, candidate bitplanes, query cache, phase
 \If{$h_v$ has a global cone tag}
  \State \Return $K(u_q,u_v)$ for hops, $A(u_q,u_v)$ for pool refinement
 \EndIf
 \State Read cached $(\alpha_{q,c},\beta_{q,c})$, computing one axis dot product on a miss
 \State $\alpha_v\gets h_v.\alpha$, $\beta_v\gets\sqrt{\max(0,1-\alpha_v^2)}$
 \If{candidate residual is degenerate}
  \State \Return $\alpha_{q,c}\alpha_v$
 \EndIf
 \If{a residual slot is cached, or $|\alpha_{q,c}|\ge0.20$, $\beta_{q,c}>10^{-6}$ and a free slot exists}
  \State Reuse or materialize unit $\tilde r_{q,c}$, its bitplanes, and SM parameters
  \State $t_{\mathrm{res}}\gets K(\tilde r_{q,c},\tilde r_v)$ for hops, $A(\tilde r_{q,c},\tilde r_v)$ for refinement
  \State \Return $\operatorname{clip}_{[-1,1]}(\alpha_{q,c}\alpha_v+\beta_{q,c}\beta_v t_{\mathrm{res}})$
 \Else
  \State $t_{\mathrm{uq}}\gets K(u_q,\tilde r_v)$ for hops, $A(u_q,\tilde r_v)$ for refinement \Comment{$\tilde r_v\perp\mu_c\Rightarrow\langle u_q,\tilde r_v\rangle=\beta_{q,c}\langle\tilde r_{q,c},\tilde r_v\rangle$}
  \State \Return $\operatorname{clip}_{[-1,1]}(\alpha_{q,c}\alpha_v+\beta_v t_{\mathrm{uq}})$
 \EndIf
 \end{algorithmic}
\end{algorithm}
Each query caches projections for at most 512 dictionary axes and materializes at most 16 residual queries. A first cone visit reads one $D$-coordinate axis and performs $O(D)$ work. Materialization adds $O(D)$ re-orthogonalization and encoding, with $O(16D+512)$ thread-local storage. A query visiting every dictionary axis performs at most 512 axis projections and 16 residual encodings. Repeated visits reuse these transforms. Upper-layer scoring without a query cache uses the global-query/residual-candidate branch. The radial gate applies to both tags, and residual vertices proceed directly to the full bitplane scorer after that gate.

Invoking {AVX2} \texttt{vpshufb} nibble-lookup popcounts~\cite{mula2018faster} independently on every bitplane would execute $15$ \texttt{vpshufb} shuffle instructions per $768$D candidate.
Adapting bitwise carry-save reduction~\cite{mula2018faster} to {SM} bitplanes, \rquiver{} compresses triplets $(A, B, C)$ across 256-bit \texttt{YMM} registers via a \emph{3-to-2 Carry-Save Adder} ({CSA}):
\begin{equation}
  \mathrm{Sum} = A \oplus B \oplus C, \qquad
  \mathrm{Carry} = (A \wedge B) \vee \bigl((A \oplus B) \wedge C\bigr),
  \label{eq:csa_identity}
\end{equation}
which satisfies $\popcnt(A) + \popcnt(B) + \popcnt(C) = \popcnt(\mathrm{Sum}) + 2\,\popcnt(\mathrm{Carry})$ identically\ifpreprint\ (verified bit-by-bit in Appendix~\ref{app:sm_algebra})\fi, cutting \texttt{vpshufb} shuffle instructions by $33\%$ ($15 \to 10$ per $768$D candidate) and evaluating each surviving candidate in $<35$\,ns alongside distance-2 software prefetching.

\subsection{L1-Bounded \textsc{Stage-0.5} Asymmetric Projection, \textsc{CJ-Gate}, and FP64 Reranking}
\label{subsec:stage05_rerank}

Once beam search converges to a pool $\mathcal{P}$ of size $L$, fetching all $L$ cold FP32 vectors ($L \times 3$\,KB) would dominate latency at large $L$.
\rquiver{} executes a three-step tail pipeline:
\begin{enumerate}
  \item \textbf{L1-Bounded \textsc{Stage-0.5} Asymmetric Re-Scoring:}
  By Theorem~\ref{thm:conformal_flip}(2) and Proposition~\ref{prop:asym_cj_gate}(1), evaluating $\hat{\gamma}_{\mathrm{asym}}(u_q, u_v) = \operatorname{clip}_{[-1,1]}(\langle u_q, \hat{u}_v\rangle / \sigma_v^2)$ with unquantized $u_q$ and $(a_v, \delta_v, \sigma_v)$ already resident in \texttt{PolarNodeMeta} eliminates both cross-shell query-noise non-cancellation ($e_q = 0$) and norm-inflation bias.
  Let $S_{\mathrm{tot}} = \sum_{i=1}^D u_{q,i}$ and define three masked signed-query subset sums over candidate bitplanes $(s_v, m_v)$: positive-sign sum $S_+ = \sum_{i : s_{v,i} = 1} u_{q,i}$, strong-candidate sum $S_h = \sum_{i : m_{v,i} = 1} u_{q,i}$, and joint positive-strong sum $S_{+h} = \sum_{i : s_{v,i} = 1 \land m_{v,i} = 1} u_{q,i}$ (equivalently expressed via unsigned $|u_{q,i}|$ and sign-agreement masks $\mathbb{I}[s_{q,i}=s_{v,i}]$).
  Then $\langle u_q, \hat{u}_v\rangle$ reduces algebraically to
  \begin{equation}
    \langle u_q, \hat{u}_v\rangle = a_v\bigl(2 S_+ - S_{\mathrm{tot}}\bigr) + \delta_v\bigl(2 S_{+h} - S_h\bigr).
    \label{eq:asym_subset_sums}
  \end{equation}
  Unlike 4-bit nibble {SIMD} lookup tables~\cite{andre2015cache,andre2017accelerated,trent2024bbq}, \texttt{AsymQuerySm} encodes $u_q$ once per query into a $1.5$\,KiB L1-resident \texttt{i16} fixed-point array ($q_{\mathrm{i16},i} = \operatorname{round}(8100\,u_{q,i})$ at $D=768$; Cauchy--Schwarz guarantees 16-lane sums remain $\le32{,}408<2^{15}-1$ without overflow) and accumulates $(S_+, S_h, S_{+h})$ via register-resident {AVX2} 16-lane bit-mask selection and \texttt{vpmaddwd} (or an optional $2$\,KiB byte-mask {LUT}) exclusively over the $L$ pool members.
  For residual-cone pool members, \textsc{Stage-0.5} reuses the cone cache from Algorithm~\ref{alg:residual_score} and lazily builds a fixed-point query array for each materialized residual. It applies $A$ in the same materialized or global-query branch.

  \item \textbf{\textsc{CJ-Gate} Observable-Margin Plateau Expansion:}
  Sort $\mathcal P$ by asymmetric proxy and let $B_0=\min(B,L)$, $j_{\mathrm{ref}}=\min(k,B_0)$, $p_{\mathrm{ref}}=\widehat P_{(j_{\mathrm{ref}})}$. Compute one reference uncertainty $s_{\mathrm{ref}}$ from~\eqref{eq:cj_sigma}, using empirical $\kappa_{\mathrm{SM}}=0.221$ to target near-tie plateaus in the $\ge 99.5\%$ high-recall regime. A materialized cone uses attenuation $\beta_{q,c}\beta_v$, a global-query/residual branch uses $\beta_v$, and a global code uses 1.
  If $B_0<L$ and $\widehat P_{(B_0)}-p_{\mathrm{ref}}\le z s_{\mathrm{ref}}$, extend the rerank prefix while
  \begin{equation}
   \widehat P_{(j)}\le p_{\mathrm{ref}}+(z+0.8)s_{\mathrm{ref}},\qquad
   j\le\min(L,\max(B_0,B_{\max})).
  \end{equation}
  The measured full configuration fixes the relative-gap parameter at $0.22$, giving $z=\operatorname{clip}_{[0.5,4.5]}(0.22\cdot2/0.22)=2$ and $B_{\max}=768$. With no router metadata, the reference scale is $\max(p_{\mathrm{ref}},10^{-9})\cdot0.22$.

  \item \textbf{FP64 {SIMD} Geodesic Reranking:}
  Near $\|v\|_2 \to 1/\sqrt{c}$, single-precision evaluation of $1 - c\|v\|_2^2$ loses mantissa precision.
  \rquiver{} prefetches cold vector $v_{(j+2)}$ while widening $q$ and $v_{(j)}$ via \texttt{vcvtps2pd} into 8-wide {AVX-512F} (or 4-wide {AVX2+FMA}) FP64 accumulators, computing exact $P_{\mathbb{B}}(q, v_{(j)})$ in double precision and invoking $\asinh$~\eqref{eq:proxy_to_geodesic} only on the final $k = 10$ results.
\end{enumerate}

\section{Experimental Evaluation}
\label{sec:eval}

Our evaluation examines how warped-product graph indexing in \rquiver{} performs across real and synthetic curved manifolds.
We first measure construction speed, hot navigation memory, and matched-recall throughput relative to full-precision and 8-bit polar graphs (\S\ref{subsec:eval_repr}), compare against exact hyperbolic graph indices and flat-proxy vector databases (\S\ref{subsec:eval_external}), and test multi-threaded scaling from $1$ to $24$ hardware threads (\S\ref{subsec:eval_scaling}).
We then isolate each topological and execution mechanism via controlled ablations (\S\ref{subsec:eval_ablation}) and evaluate external baselines on four synthetic distributions, using a four-stage recall-loss audit to characterize the measured traversal, beam-retention, and rerank-admission losses (\S\ref{subsec:eval_synthetic}).

\subsection{Experimental Setup}
\label{subsec:eval_setup}

\paragraph{Hardware and Protocol.}
Experiments execute in two isolated Linux containers on {AMD EPYC 9654} ({Zen~4}) hosts, each restricted via cgroups to 32 logical CPUs and $60$\,GiB RAM. The CPU sets are \texttt{64-95} and \texttt{128-159}. Swap is disabled and transparent huge pages ({THP}) are set to \texttt{never}, using standard $4$\,KiB pages for {DTLB} translation. Each instance runs one timed benchmark at a time, and measurement manifests record instance placement and source/binary fingerprints.
All Rust binaries are compiled in release mode targeting native {AVX-512F} and {AVX2+FMA}.
Unless otherwise stated, graph indices use base degree $M = 16$ (maximum out-degree $M_0 = 32$), construction beam $L_c = 128$, geodesic slack $\alpha = 1.2$, and $32$ build threads.
Following community {ANN} evaluation protocols~\cite{aumuller2020annbenchmarks,simhadri2022results}, controlled {Cohere-1M} representation comparisons and {\rquiver{}-Core} component ablations (alongside multi-seed {\rquiver{}-Core} stability checks on {MERU-1M} and {OpenAI-1536}) are built across \textbf{three independent random seeds} (\texttt{12345}, \texttt{12446}, \texttt{12547}) and report three-graph means.
Across all $210$ three-graph operating points, Recall@10 differs by at most $0.22$ percentage points across seeds, and $99\%$-recall {\rquiver{}-Core} QPS varies within $\pm 2.64\%$ on {Cohere-1M} ($\pm 0.20\%$ on {MERU-1M}).
Relying on this cross-seed stability to keep full multi-system parameter sweeps tractable, end-to-end \rquiver{}, {Global-SM}, synthetic stress tests, and external baselines evaluate one build per $(M, L_c)$ configuration (seed \texttt{12345} for configurable build randomness; {HDB} retains its upstream graph-builder randomization). All use the same {Zen~4} resource quotas and IEEE-754 FP64 geodesic ground truth across $1, 2, 4, 8, 16, 24$ query threads.

\paragraph{Fixed Parameters and Operating-Point Selection.}
Table~\ref{tab:fixed_parameters} gives the shared construction, cone-purity, and uncertainty settings. The full-method sweep uses
\[
 \begin{aligned}
 L\in\{&32,64,96,128,192,200,208,\\
       &256,384,512,768,1024,1536,2048\},
 \end{aligned}
\]
rerank budgets $B\in\{32,64,128\}$ subject to $B\le L$, and sketch margins $\Delta_0\in\{0.08,0.12,0.20\}$. Controlled engine sweeps additionally include full-pool reranking and sketch-disabled paths. Each query-thread count uses its measured qualifying operating points.
Curves and recall-threshold columns report per-configuration empirical frontiers, with each recall column independently selecting its fastest qualifying measured build/search point per thread count. {MERU-1M}, {OpenAI-1536}, {WordNet-Hazy}, and the four synthetic datasets tune on disjoint $2{,}000$-query splits and report archived test-curve frontiers, whereas {Cohere-1M} evaluates directly on its public $1{,}000$-query set. Cone-purity thresholds and \textsc{CJ-Gate} constants stay fixed across datasets.

\begin{table}[t]
  \centering
  \caption{Hyperbolic benchmarks in $\Bunit$ ($c=1$).}
  \label{tab:datasets}
  \small
  \setlength{\tabcolsep}{3.5pt}
  \begin{tabularx}{\linewidth}{@{}lrrr>{\raggedright\arraybackslash}X@{}}
    \toprule
    \textbf{Dataset} & $N$ & $D$ & $|Q|$ & \textbf{Manifold Geometry} \\
    \midrule
    Cohere-1M & 1,000K & 768 & 1,000 & Mapped Wikipedia ($\exp_0^1$, $\rho_{0.01}=1.44$) \\
    MERU-1M~\cite{desai2023meru} & 1,000K & 512 & 10,000 & Native {Lorentz}$\to\Bunit$ multimodal \\
    OpenAI-1536 & 990K & 1,536 & 8,000 & Mapped DBpedia ($\exp_0^1$, $\text{med}\,\rho=2.0$) \\
    WordNet-Hazy & 83.9K & 20 & 2,000 & Native {Lorentz}$\to\mathbb{B}_1^{20}$ taxonomy \\
    \midrule
    Uniform-Narrow & 1,000K & 512 & 2,000 & Isotropic $\mathbb{S}^{511}$, $\rho \in [3.9, 4.1]$ \\
    Uniform-Wide & 1,000K & 512 & 2,000 & Isotropic $\mathbb{S}^{511}$, $\rho \in [0.4, 6.0]$ \\
    Clustered-Narrow & 1,000K & 512 & 2,000 & $256$ tight cones, $\rho \in [3.9, 4.1]$ \\
    Clustered-Wide & 1,000K & 512 & 2,000 & $256$ tight cones, $\rho \in [0.4, 6.0]$ \\
    \bottomrule
  \end{tabularx}
\end{table}

\paragraph{Benchmarks.}
Table~\ref{tab:datasets} summarizes the eight hyperbolic benchmarks in $\Bunit$ ($c=1$).
For Euclidean embeddings mapped into $\Bunit$, both \textbf{Cohere-1M} and \textbf{OpenAI-1536} use the origin exponential map from centered coordinates: given database mean $\bar{x}$ and global radial scale $s = 2 / \operatorname{median}_i(\|x_i - \bar{x}\|_2)$, each vector is mapped as
\[
\Phi(x) = \tanh\!\left(\frac{s}{2}\|x - \bar{x}\|_2\right)\frac{x - \bar{x}}{\|x - \bar{x}\|_2},
\]
clamped at Euclidean radius $1 - 10^{-6}$ so that the database median geodesic depth is fixed at $\rho_{0.50} = 2.0$.
On \textbf{Cohere-1M} (1M 768D Wikipedia embeddings, $1{,}000$ test queries, $s = 0.2644$, zero clipped vectors), norm concentration after mean-centering narrows base depths to $[0.98, 3.39]$ ($\rho_{0.01} = 1.45$, $\rho_{0.99} = 2.69$) and query depths to $[1.25, 2.98]$ ($\rho_{0.01} = 1.44$, $\rho_{0.99} = 2.67$), forming a hollow boundary-concentrated shell with no vectors below $\rho = 0.98$.
\textbf{OpenAI-1536} applies unit $\ell_2$-normalization before mean-centering and $\Phi(x)$ ($s = 2.093$) across 990K 1536D \texttt{text-embedding-3-large} {DBpedia} embeddings with $8{,}000$ held-out test queries and $2{,}000$ tuning queries.
In contrast, \textbf{MERU-1M} contains 1M 512D native {Lorentz} hyperboloid {MERU--ImageNet} vision-language embeddings~\cite{desai2023meru} mapped isometrically via~\eqref{eq:lorentz_isometry} and evaluated on $10{,}000$ held-out test queries alongside $2{,}000$ tuning queries, and \textbf{WordNet-Hazy} adapts the public $20$D native {Lorentz} {WordNet} weights (\texttt{wordnet\_full.emb}, $83{,}943$ base nodes and $2{,}000$ disjoint test queries) isometrically into $\mathbb{B}_1^{20}$ ($\le 2.53\times 10^{-15}$ max numerical error).
Four 1M-vector 512D synthetic benchmarks cross directional geometry (isotropic $\mathbb{S}^{511}$ versus $256$ tight spherical cones) with radial depth support (narrow boundary shell $\rho \in [3.9, 4.1]$ versus wide hierarchy $\rho \in [0.4, 6.0]$).

\paragraph{Baselines.}
Within our engine family, we evaluate two orthogonal groups of controlled variants:
(1)~\emph{Representation and frozen-graph controls:} \textbf{Polar-FP32} and \textbf{Polar-SQ8} share \rquiver{}'s {QuadriPrune} topology and router while substituting unquantized FP32 or 8-bit scalar-quantized unit directions during both construction and search, whereas \textbf{SameGraph-FP32} and \textbf{SameGraph-SQ8} freeze the 2-bit-constructed graph and replace only the search-time directional representation to decouple graph quality from traversal speed (plotted as \textbf{FP32 (same graph)} and \textbf{SQ8 (same graph)} alongside 2-bit {SM} in Figure~\ref{fig:abl_repr}).
(2)~\emph{Cumulative 2-bit engine progression:} \textbf{\rquiver{}-Split} (split-array 2-bit {QuadriPrune} without chord-swap island repair, \texttt{HotSmPage} co-location, or \texttt{SymSketch256}) $\to$ \textbf{\rquiver{}-Core} (adding chord-swap repair, $4096$\,B single-page co-location, and \texttt{SymSketch256} under global 2-bit {SM}) $\to$ full \textbf{\rquiver{}} (further enabling purity-gated \texttt{ResidualConeDict} and adaptive \textsc{CJ-Gate} plateau expansion; in \S\ref{subsec:eval_synthetic}, \textbf{Global-SM} denotes the unrescued global-code baseline without \texttt{ResidualConeDict}, \textsc{CJ-Gate}, or variance-stabilized sketch margins).
Under the same hardware and FP64 ground-truth protocol, we compare against four external systems using the exact variant names reported in Figures~\ref{fig:real_pareto}--\ref{fig:scaling} and Tables~\ref{tab:external_baselines} and~\ref{tab:synthetic_external}:
\textbf{Exact-RobustPrune}~\cite{subramanya2019diskann,prokhorenkova2022hyperbolic} ($M/L_c = 16/128, L \le 2{,}048$) using exact geodesic {Vamana} pruning without polar decoupling;
\textbf{HyperspaceDB} (\textbf{HDB}, \texttt{YARlabs/hyperspace-db} commit \texttt{89d0406}, native {AVX-512F} {Poincar\'e} {HNSW})~\cite{glukhota2026hyperspacedb} across FP32 and FP64 hyperbolic modes and dynamic/flat storage layouts (\textbf{HDB FP32 dynamic}, \textbf{HDB FP32 flat}, \textbf{HDB FP64 dynamic}, \textbf{HDB FP64 flat});
and flat inner-product {HNSW} indices in \textbf{Faiss} (\texttt{faiss-cpu 1.16.0} \texttt{IndexHNSWFlat})~\cite{douze2024faiss}---tested with mean-centered {Lorentz} inner product (\textbf{Faiss centered IP}), unit-direction search with exact FP64 reranking (\textbf{Faiss direction}), and uncentered {Lorentz} inner product (\textbf{Faiss raw Lorentz-IP})---and \textbf{Milvus} ({Milvus Standalone 3.0.5} {HNSW})~\cite{wang2021milvus} (\textbf{Milvus centered IP}), building {HDB}/{Faiss}/{Milvus} at both $(M=16, L_c=128)$ and $(M=32, L_c=256)$ and scanning $18$ $ef$ values up to $4{,}096$, all paired with exact FP64 geodesic reranking.

\subsection{Index Construction and Polar Representation Comparison}
\label{subsec:eval_repr}

\begin{table*}[t]
  \centering
  \caption{Controlled polar representation comparisons on {Cohere-1M} and measured \rquiver{} coverage across four real workloads (Zen~4, 32 build threads). {Cohere-1M} representation and progression controls report three-graph means; end-to-end full \rquiver{} rows follow the single-build protocol of \S\ref{subsec:eval_setup}. Internal representation controls were measured on {Cohere-1M} only.}
  \label{tab:zen4_repr}
  \small
  \setlength{\tabcolsep}{3pt}
  \begin{tabular}{@{}llrrrrrrrrrrrr@{}}
    \toprule
    & & \textbf{Build} & \textbf{Hot Nav} & \textbf{Zero-In} & \multicolumn{3}{c}{\textbf{Recall@10 $\ge 98\%$}} & \multicolumn{3}{c}{\textbf{Recall@10 $\ge 99\%$}} & \multicolumn{3}{c}{\textbf{Recall@10 $\ge 99.5\%$}} \\
    \cmidrule(lr){6-8}\cmidrule(lr){9-11}\cmidrule(lr){12-14}
    \textbf{Dataset} & \textbf{Method} & \textbf{32T (s)} & \textbf{(MB)} & \textbf{Islands} & \textbf{1T QPS} & \textbf{p99} & \textbf{24T QPS} & \textbf{1T QPS} & \textbf{p99} & \textbf{24T QPS} & \textbf{1T QPS} & \textbf{p99} & \textbf{24T QPS} \\
    \midrule
    \multirow{7}{*}{Cohere-1M}
    & Polar-FP32 & 190.2 & 3,565.0$^\ddagger$ & 0 & 1,027 & 1.19 & 14,925 & 448 & 2.69 & 6,735 & 258 & 4.67 & 3,937 \\
    & Polar-SQ8 & 66.9 & 1,261.0 & 0 & 1,887 & 0.68 & 34,639 & 878 & 1.44 & 16,844 & 372 & 3.39 & 7,016 \\
    & SameGraph-FP32 & 32.5 & 3,565.0$^\ddagger$ & 0 & 1,031 & 1.27 & \missing & 454 & 2.60 & \missing & 113 & 10.54 & \missing \\
    & SameGraph-SQ8 & 32.5 & 1,261.0 & 0 & 1,979 & 0.62 & \missing & 894 & 1.52 & \missing & 222 & 5.57 & \missing \\
    & \rquiver{}-Split & \textbf{32.3} & \textbf{485.0} & 626 & 2,760 & 0.45 & 56,365 & 1,749 & 0.67 & 38,203 & 618 & 1.88 & 14,439 \\
    & \rquiver{}-Core & 32.5 & \textbf{485.0} & \textbf{0} & \textbf{3,841} & \textbf{0.31} & \textbf{79,047} & \textbf{2,071} & \textbf{0.56} & \textbf{45,833} & 688 & \textbf{1.63} & 16,059 \\
    & \textbf{\rquiver{}} & 35.0 & 517.0$^*$ & \textbf{0} & 2,907 & 0.41 & 58,162 & 1,930 & 0.62 & 42,466 & \textbf{738} & 1.65 & \textbf{17,137} \\
    \midrule
    MERU-1M & \textbf{\rquiver{}} & \textbf{25.7} & \textbf{516.8}$^*$ & \textbf{0} & \textbf{2,547}$^\S$ & \textbf{0.50} & \textbf{53,483}$^\S$ & \textbf{2,547} & \textbf{0.50} & \textbf{53,483} & \textbf{1,669} & \textbf{0.81} & \textbf{36,092} \\
    OpenAI-1536 & \textbf{\rquiver{}} & \textbf{43.6} & \textbf{765.6}$^*$ & \textbf{0} & \textbf{1,378} & \textbf{1.03} & \textbf{25,725} & \textbf{792} & \textbf{1.81} & \textbf{14,731} & \textbf{481} & \textbf{3.05} & \textbf{9,094} \\
    WordNet-Hazy & \textbf{\rquiver{}} & \textbf{1.5} & \textbf{32.8}$^*$ & \textbf{0} & \textbf{4,130} & \textbf{0.29} & \textbf{89,847} & \textbf{3,089} & \textbf{0.38} & \textbf{65,405} & \textbf{2,341} & \textbf{0.49} & \textbf{52,371} \\
    \bottomrule
  \end{tabular}
  \par\vspace{2pt}
  {\footnotesize \textit{Note:} Latency p99 is in milliseconds (1T). $^*$Full $4096$\,B \texttt{HotSmPage} allocation including $32$\,B/vertex \texttt{SymSketch256} and page alignment; split allocations without the sketch are $485.0$, $420.8$, $670.6$, and $26.1$\,MB (Table~\ref{tab:memory_audit}). {\rquiver{}-Core} in row~6 also uses the $517.0$\,MB \texttt{HotSmPage} layout at runtime ($485.0$\,MB base split counter). $^\ddagger$Includes $3{,}072$\,MB FP32 vectors accessed during traversal; the remaining $493.0$\,MB comprises $256$\,MB 64-slot adjacency + $4$\,MB degree words, $32$\,MB polar headers, $192$\,MB retained 2-bit storage, $8$\,MB inverse norms, and $0.975$\,MB router arrays (Table~\ref{tab:memory_audit}). $^\S$Attains $99.25\%$ Recall@10 at the same operating point ($4{,}521$\,QPS 1T and $94{,}846$\,QPS 24T at $\ge 95\%$ Recall@10).}
\end{table*}

Table~\ref{tab:zen4_repr} evaluates controlled polar directional representations on {Cohere-1M} and summarizes end-to-end \rquiver{} performance across all four real hyperbolic workloads.

\paragraph{Construction Speed and Memory Footprint.}
On {Cohere-1M}, \rquiver{} builds a $100\%$ {BFS}-reachable 1M-vertex hyperbolic index in \textbf{$35.0$\,s} on 32 threads---$5.44\times$ faster than {Polar-FP32} ($190.2$\,s) and $1.91\times$ faster than {Polar-SQ8} ($66.9$\,s)---while using $517.0$\,MB hot navigation memory under $4096$\,B \texttt{HotSmPage} co-location ($2.44\times$ smaller than {Polar-SQ8} and $6.90\times$ smaller than full-vector traversal; $485.0$\,MB, or $2.60\times$ and $7.35\times$ smaller, in split form without the sketch).
On native hyperbolic {MERU-1M} ($512$D), {OpenAI-1536} ($1536$D), and {WordNet-Hazy} ($20$D), 32-thread construction completes in $25.7$\,s, $43.6$\,s, and $1.5$\,s, respectively, with zero unreachable islands and $0$ retained residual cones across every real graph.

\begin{figure}[t]
  \centering
  \includegraphics[width=0.8\linewidth]{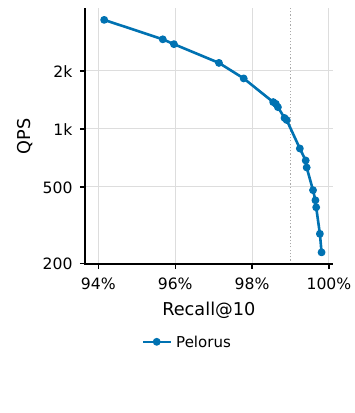}
  \caption{Measured \rquiver{} Recall@10--QPS frontier on {OpenAI-1536} ($990$K vectors, $1536$D, Zen~4, one query thread).}
  \label{fig:pareto_openai}
\end{figure}

\paragraph{Matched-Recall Throughput and Graph-Quality Crossover.}
At $\ge 99\%$ Recall@10 on {Cohere-1M}, \rquiver{} reaches \textbf{$1{,}930$\,QPS} ($0.62$\,ms p99) on 1 thread and \textbf{$42{,}466$\,QPS} ($1.20$\,ms p99) on 24 threads, improving over {Polar-FP32} by \textbf{$4.30\times$ (1T)} and \textbf{$6.31\times$ (24T)}, and over {Polar-SQ8} by \textbf{$2.20\times$ (1T)} and \textbf{$2.52\times$ (24T)}.
Comparing {\rquiver{}-Core} and full \rquiver{} highlights the exact precision--throughput trade-off of adaptive plateau gating: on diffuse {Cohere-1M} ($0$ active residual cones), {\rquiver{}-Core} attains slightly higher throughput at moderate $98\%\text{--}99\%$ targets ($3{,}841$ / $2{,}071$\,QPS 1T vs.\ $2{,}907$ / $1{,}930$\,QPS) because \textsc{CJ-Gate} conservatively promotes near-tie boundary plateau candidates into FP64 reranking, whereas at $\ge 99.5\%$ Recall@10 that adaptive expansion pushes full \rquiver{} ahead to \textbf{$738$\,QPS} (1T) and \textbf{$17{,}137$\,QPS} (24T)---outperforming {\rquiver{}-Core} ($688$ / $16{,}059$\,QPS) as well as {Polar-FP32} by \textbf{$2.86\times$ (1T)} and \textbf{$4.35\times$ (24T)} while safeguarding concentrated boundary clusters (\S\ref{subsec:eval_synthetic}).
The frozen-graph crossover variants isolate why 2-bit construction works:
running unquantized FP32 directional navigation over the 2-bit-constructed graph ({SameGraph-FP32}) achieves $454$\,QPS at $99\%$ recall versus $448$\,QPS on the FP32-constructed graph, confirming that polar 2-bit {SM} comparisons preserve {Vamana} topology quality at $>5\times$ lower build cost.
Meanwhile, replacing search-time FP32 reads with the 2-bit page-co-located engine accelerates traversal on that identical graph by $4.25\times$ under \rquiver{} ($4.56\times$ under {\rquiver{}-Core}) at $99\%$ recall and $6.53\times$ at $99.5\%$ recall ($738$ vs.\ $113$\,QPS).
On native hyperbolic {MERU-1M}, \rquiver{} achieves $2{,}547$\,QPS (1T) and $53{,}483$\,QPS (24T) at $99.25\%$ Recall@10 ($1{,}669$ and $36{,}092$\,QPS at $99.76\%$ Recall@10); on 1536D {OpenAI-1536} (Figure~\ref{fig:pareto_openai}), it sustains $792$\,QPS (1T) and $14{,}731$\,QPS (24T) at $99.24\%$ Recall@10; and on 20D {WordNet-Hazy}, it reaches $3{,}089$\,QPS (1T) and $65{,}405$\,QPS (24T) at $99.36\%$ Recall@10 ($2{,}341$ and $52{,}371$\,QPS at $99.79\%$).

\subsection{Comparison with Hyperbolic and Flat-Proxy Engines}
\label{subsec:eval_external}

\begin{table*}[t]
  \centering
  \caption{Measured comparison of \rquiver{} and baselines on {Cohere-1M} and {MERU-1M} (Zen~4, 32 build threads) at Recall@10 targets of $90\%$, $95\%$, $98\%$, $99\%$, and $99.5\%$. Panels (a) and (b) report 1-thread and 24-thread QPS, respectively.}
  \label{tab:external_baselines}
  \footnotesize
  \setlength{\tabcolsep}{2pt}
  \begin{tabular}{@{}llrrrrrrrrr@{}}
\multicolumn{11}{@{}l}{\textbf{(a) Single-thread QPS}} \\
\toprule
\textbf{Dataset} & \textbf{Method} & \textbf{$M/L_c$} & \textbf{\shortstack{Build\\32T (s)}} & \textbf{\shortstack{Peak\\Rec@10}} & \textbf{$\ge90\%$} & \textbf{$\ge95\%$} & \textbf{$\ge98\%$} & \textbf{$\ge99\%$} & \textbf{$\ge99.5\%$} & \textbf{\shortstack{p99 (ms)\\at $99\%$}} \\
\midrule
Cohere-1M & Pelorus & 16/128 & 35.0 & 99.700\% & 4,939 & 4,939 & 2,907 & 1,930 & 738 & 0.62 \\
 & Faiss centered IP & 32/256 & 167.3 & 98.910\% & 1,652 & 830 & 127 & \missing & \missing & \missing \\
 & Faiss direction & 16/128 & 77.6 & 65.170\% & \missing & \missing & \missing & \missing & \missing & \missing \\
 & Faiss raw Lorentz-IP & 16/128 & 7.4 & 0.120\% & \missing & \missing & \missing & \missing & \missing & \missing \\
 & Milvus centered IP & 32/256 & 361.7 & 98.690\% & 252 & 238 & 76 & \missing & \missing & \missing \\
 & HDB FP32 dynamic & 32/256 & 414.8 & 96.780\% & 839 & 290 & \missing & \missing & \missing & \missing \\
 & HDB FP32 flat & 32/256 & 416.3 & 96.670\% & 946 & 279 & \missing & \missing & \missing & \missing \\
 & HDB FP64 dynamic & 32/256 & 866.5 & 94.430\% & 67 & \missing & \missing & \missing & \missing & \missing \\
 & HDB FP64 flat & 32/256 & 869.3 & 94.330\% & 74 & \missing & \missing & \missing & \missing & \missing \\
 & Exact-RobustPrune & 16/128 & 75.5 & 93.940\% & 629 & \missing & \missing & \missing & \missing & \missing \\
 & Polar-SQ8 & 16/128 & 66.9 & 99.780\% & 2,428 & 2,428 & 1,887 & 878 & 372 & 1.44 \\
 & Polar-FP32 & 16/128 & 190.2 & 99.783\% & 1,355 & 1,355 & 1,027 & 448 & 258 & 2.69 \\
\midrule
MERU-1M & Pelorus & 16/128 & 25.7 & 99.983\% & 5,846 & 4,521 & 2,547 & 2,547 & 1,669 & 0.50 \\
 & Faiss centered IP & 32/256 & 99.2 & 99.988\% & 4,312 & 2,800 & 1,903 & 1,059 & 872 & 1.43 \\
 & Faiss direction & 16/128 & 48.8 & 100.000\% & 3,460 & 1,814 & 1,480 & 1,078 & 1,078 & 1.30 \\
 & Faiss raw Lorentz-IP & 16/128 & 36.3 & 93.858\% & 841 & \missing & \missing & \missing & \missing & \missing \\
 & Milvus centered IP & 32/256 & 271.6 & 99.991\% & 297 & 297 & 297 & 260 & 233 & 5.30 \\
 & HDB FP32 dynamic & 16/128 & 86.5 & 99.947\% & 3,136 & 2,103 & 1,493 & 1,188 & 875 & 1.27 \\
 & HDB FP32 flat & 16/128 & 87.6 & 99.947\% & 3,187 & 2,184 & 1,579 & 1,296 & 966 & 1.15 \\
 & HDB FP64 dynamic & 32/256 & 631.4 & 99.955\% & 615 & 428 & 281 & 196 & 165 & 8.83 \\
 & HDB FP64 flat & 32/256 & 633.9 & 99.707\% & 480 & 320 & 186 & 93 & 46 & 21.68 \\
\bottomrule
\end{tabular}
\par\vspace{5pt}
\begin{tabular*}{\linewidth}{@{\extracolsep{\fill}}llrrrrr@{}}
\multicolumn{7}{@{}l}{\textbf{(b) 24-thread QPS (queries/s)}} \\
\toprule
\textbf{Dataset} & \textbf{Method} & \textbf{$\ge90\%$} & \textbf{$\ge95\%$} & \textbf{$\ge98\%$} & \textbf{$\ge99\%$} & \textbf{$\ge99.5\%$} \\
\midrule
Cohere-1M & Pelorus & 96,584 & 96,584 & 58,162 & 42,466 & 17,137 \\
 & Faiss centered IP & 10,717 & 10,717 & 1,660 & \missing & \missing \\
 & Faiss direction & \missing & \missing & \missing & \missing & \missing \\
 & Faiss raw Lorentz-IP & \missing & \missing & \missing & \missing & \missing \\
 & Milvus centered IP & 1,115 & 1,115 & 999 & \missing & \missing \\
 & HDB FP32 dynamic & 3,416 & 3,416 & \missing & \missing & \missing \\
 & HDB FP32 flat & 3,656 & 3,656 & \missing & \missing & \missing \\
 & HDB FP64 dynamic & \missing & \missing & \missing & \missing & \missing \\
 & HDB FP64 flat & \missing & \missing & \missing & \missing & \missing \\
 & Exact-RobustPrune & \missing & \missing & \missing & \missing & \missing \\
 & Polar-SQ8 & 46,034 & 46,034 & 34,639 & 16,844 & 7,016 \\
 & Polar-FP32 & 19,932 & 19,932 & 14,925 & 6,735 & 3,937 \\
\midrule
MERU-1M & Pelorus & 94,846 & 94,846 & 53,483 & 53,483 & 36,092 \\
 & Faiss centered IP & 33,189 & 33,189 & 22,488 & 13,005 & 9,349 \\
 & Faiss direction & 43,798 & 43,798 & 19,883 & 14,967 & 9,866 \\
 & Faiss raw Lorentz-IP & \missing & \missing & \missing & \missing & \missing \\
 & Milvus centered IP & 1,208 & 1,208 & 1,208 & 1,130 & 1,088 \\
 & HDB FP32 dynamic & 30,015 & 30,015 & 18,413 & 13,780 & 11,038 \\
 & HDB FP32 flat & 26,235 & 26,235 & 19,434 & 15,890 & 11,699 \\
 & HDB FP64 dynamic & 9,588 & 9,588 & 6,121 & 4,098 & 3,414 \\
 & HDB FP64 flat & 6,362 & 6,362 & 4,039 & 1,933 & 925 \\
\bottomrule
\end{tabular*}

  \par\vspace{2pt}
  {\footnotesize \textit{Note:} HDB denotes HyperspaceDB. A dash denotes the absence of a qualifying measured point at that thread count. {Polar-SQ8} and {Polar-FP32} report three-graph means; \rquiver{} and external systems use the single-build protocol of \S\ref{subsec:eval_setup}. Build time and $M/L_c$ correspond to the fastest qualifying 1T configuration at $99\%$ recall, or the peak-recall configuration when $99\%$ is unreached. Each QPS column independently selects its fastest measured qualifying configuration. The $90\%$ and $95\%$ columns are selected from the existing measured configurations. Peak recall is the largest measured 1T recall; p99 is in milliseconds at the selected $99\%$ 1T point.}
\end{table*}

\begin{figure*}[t]
  \centering
  \begin{subfigure}[t]{0.485\textwidth}
    \centering
    \includegraphics[width=\linewidth]{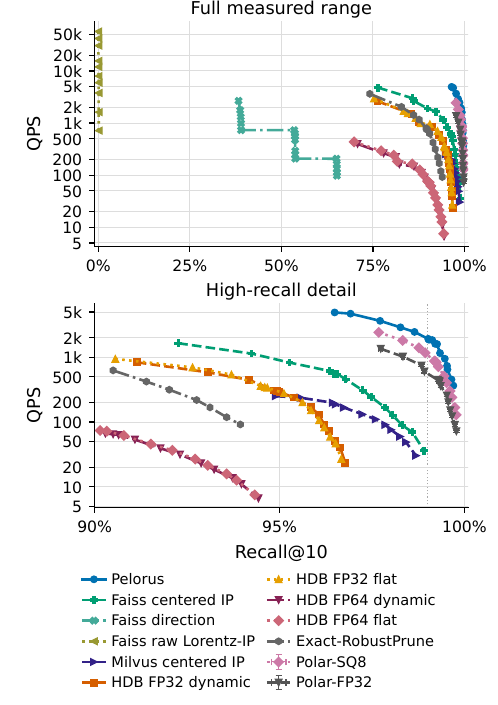}
    \caption{Cohere-1M ($768$D, 1 thread)}
    \label{fig:pareto_cohere}
  \end{subfigure}\hfill
  \begin{subfigure}[t]{0.485\textwidth}
    \centering
    \includegraphics[width=\linewidth]{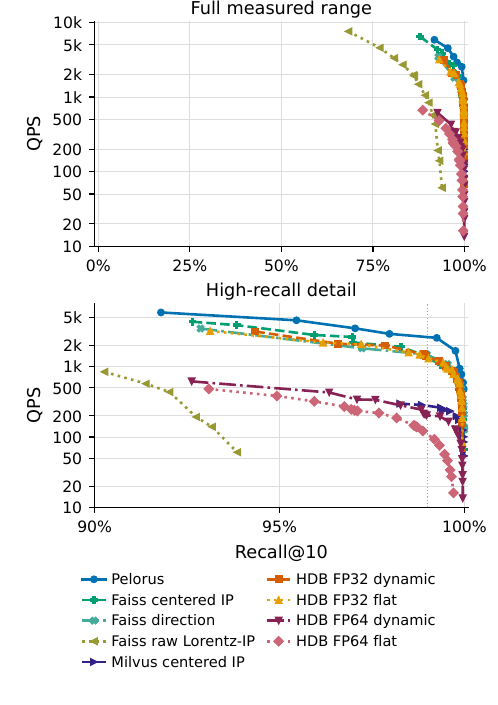}
    \caption{MERU-1M ($512$D, 1 thread)}
    \label{fig:pareto_meru}
  \end{subfigure}
  \caption{Matched-dataset Recall@10--QPS frontiers of \rquiver{} and all measured external baselines on Zen~4. Each panel shows the full measured recall range and a high-recall detail view. HDB denotes HyperspaceDB. {Polar-FP32} and {Polar-SQ8} report three-graph means with min--max error bars, while \rquiver{} and external baselines follow the single-build protocol of \S\ref{subsec:eval_setup}. Each method's frontier includes its measured build and search configurations; lines connect nondominated measurements.}
  \label{fig:real_pareto}
\end{figure*}

Table~\ref{tab:external_baselines} and Figure~\ref{fig:real_pareto} place \rquiver{} alongside {Exact-RobustPrune}~\cite{prokhorenkova2022hyperbolic}, {HDB} ({HyperspaceDB}~\cite{glukhota2026hyperspacedb}), and flat inner-product {HNSW} proxies in {Faiss}~\cite{douze2024faiss} ({Faiss centered IP}, {Faiss direction}, {Faiss raw Lorentz-IP}) and {Milvus}~\cite{wang2021milvus} ({Milvus centered IP}).
On the mapped {Cohere-1M} benchmark (Figure~\ref{fig:pareto_cohere}), \rquiver{} builds in $35.0$\,s, reaches $99.70\%$ peak Recall@10, and sustains $1{,}930$\,QPS (1T) and $42{,}466$\,QPS (24T) at $\ge 99\%$ recall.
Because external baselines miss $99\%$ even at $ef = 4{,}096$, Table~\ref{tab:external_baselines} reports their higher-recall $32/256$ builds on {Cohere-1M}: {Exact-RobustPrune} ($16/128$) reaches $93.94\%$ peak Recall@10 ($75.5$\,s build), {HDB FP32 dynamic} and {HDB FP32 flat} ($32/256$) reach $96.78\%$ and $96.67\%$ ($414.8$\,s and $416.3$\,s build), and {HDB FP64 dynamic} and {HDB FP64 flat} ($32/256$) reach $94.43\%$ and $94.33\%$ ($866.5$\,s and $869.3$\,s); at matched $95\%$ recall, \rquiver{} delivers $4{,}939$\,QPS (1T) and $96{,}584$\,QPS (24T) compared with $290$\,QPS (1T, $17.0\times$) and $3{,}416$\,QPS (24T, $28.3\times$) for {HDB FP32 dynamic}.
Among flat {HNSW} proxies with exact FP64 reranking on {Cohere-1M}, {Faiss centered IP} and {Milvus centered IP} ($32/256$) peak at $98.91\%$ and $98.69\%$ Recall@10 ($167.3$\,s and $361.7$\,s build; at $98\%$ recall, \rquiver{} attains $2{,}907$\,QPS on 1T and $58{,}162$\,QPS on 24T vs.\ $127$ / $1{,}660$\,QPS for {Faiss centered IP} and $76$ / $999$\,QPS for {Milvus centered IP}), while {Faiss direction} and {Faiss raw Lorentz-IP} reach $65.17\%$ and $0.12\%$. Under matched $M/L_c = 16/128$ ($ef \le 4{,}096$), external {Cohere-1M} peak recalls drop further to $93.84\%$/$93.64\%$ for {HDB FP32 dynamic/flat} ($106.1$\,s/$107.5$\,s build; $558$/$530$\,QPS 1T at $\ge 90\%$, missing $95\%$), $89.94\%$/$89.74\%$ for {HDB FP64}, and $97.93\%$/$97.99\%$ for {Faiss/Milvus centered IP} ($517$/$176$\,QPS 1T at $\ge 95\%$, missing $98\%$).

On native hyperbolic {MERU-1M} (Figure~\ref{fig:pareto_meru}), where radial depth variance is wider and interior connectivity is denser, all external systems except {Faiss raw Lorentz-IP} ($93.86\%$ peak) reach $99\%$ recall.
At $\ge 99\%$ Recall@10, \rquiver{} builds in $25.7$\,s and achieves \textbf{$2{,}547$\,QPS} (1T, $0.50$\,ms p99), outperforming {HDB FP32 flat} ($1{,}296$\,QPS, $1.97\times$), {HDB FP32 dynamic} ($1{,}188$\,QPS, $2.14\times$), {Faiss direction} ($1{,}078$\,QPS, $2.36\times$), {Faiss centered IP} ($1{,}059$\,QPS, $2.41\times$), {Milvus centered IP} ($260$\,QPS, $9.80\times$), and {HDB FP64 dynamic/flat} ($196$/$93$\,QPS, $13.0\times$/$27.4\times$).
On 24 threads, \rquiver{}'s compact working set widens its advantage to \textbf{$53{,}483$\,QPS} versus $15{,}890$\,QPS for {HDB FP32 flat} ($3.37\times$), $14{,}967$\,QPS for {Faiss direction} ($3.57\times$), $13{,}780$\,QPS for {HDB FP32 dynamic} ($3.88\times$), $13{,}005$\,QPS for {Faiss centered IP} ($4.11\times$), $4{,}098$\,QPS for {HDB FP64 dynamic} ($13.1\times$), $1{,}933$\,QPS for {HDB FP64 flat} ($27.7\times$), and $1{,}130$\,QPS for {Milvus centered IP} ($47.3\times$).
The full-range plots in Figure~\ref{fig:real_pareto} also retain low-recall operating points, while the high-recall detail views expose the $90\%\text{--}100\%$ trade-offs directly.

\subsection{Multi-Threaded Scaling Efficiency}
\label{subsec:eval_scaling}

\begin{figure*}[t]
  \centering
  \begin{subfigure}[t]{0.485\linewidth}
    \centering
    \includegraphics[width=\linewidth]{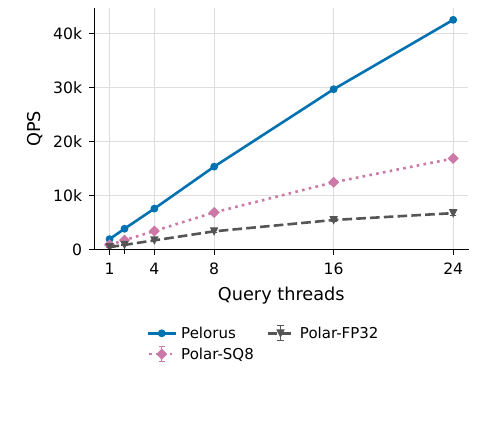}
    \caption{Cohere-1M ($\ge 99\%$ Recall)}
    \label{fig:scale_cohere}
  \end{subfigure}\hfill
  \begin{subfigure}[t]{0.485\linewidth}
    \centering
    \includegraphics[width=\linewidth]{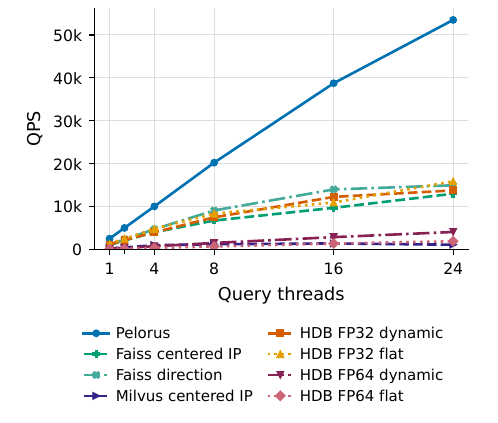}
    \caption{MERU-1M ($\ge 99\%$ Recall)}
    \label{fig:scale_meru}
  \end{subfigure}
  \caption{Multi-threaded query scaling from 1 to 24 threads on Zen~4 at Recall@10 $\ge 99\%$. On {Cohere-1M}, \rquiver{} is compared against {Polar-SQ8} and {Polar-FP32} (three-graph means with min--max error bars; external baseline peak recalls are reported in Table~\ref{tab:external_baselines}). On {MERU-1M}, \rquiver{} is compared against all external systems reaching $99\%$ recall; each method selects its fastest qualifying configuration per thread count.}
  \label{fig:scaling}
\end{figure*}

Figure~\ref{fig:scaling} reports throughput scaling from $1$ to $24$ threads at $\ge 99\%$ Recall@10.
With stateless, read-only traversal over a compact hot working set ($485.0$\,MB logical and $517.0$\,MB page-co-located on {Cohere-1M}, reading $64\text{--}256$\,B per visited neighbor at $D=768$ with header, sketch, and complete bitplanes co-located on a single $4$\,KiB page), \rquiver{} scales by \textbf{$22.0\times$} at 24 threads on {Cohere-1M} ($1{,}930 \to 42{,}466$\,QPS, $91.7\%$ parallel efficiency; Figure~\ref{fig:scale_cohere}), \textbf{$21.0\times$} on {MERU-1M} ($2{,}547 \to 53{,}483$\,QPS), \textbf{$18.6\times$} on {OpenAI-1536} ($792 \to 14{,}731$\,QPS), and \textbf{$21.2\times$} on {WordNet-Hazy} ($3{,}089 \to 65{,}405$\,QPS).
By contrast, {Polar-FP32} reads $3{,}072$\,B ($48$ cache lines) per visited neighbor across a $3.56$\,GB footprint and scales by $15.0\times$ at 24 threads on {Cohere-1M} ($448 \to 6{,}735$\,QPS), widening \rquiver{}'s speedup from $4.30\times$ (1T) to \textbf{$6.31\times$ (24T)}; similarly, on {MERU-1M} (Figure~\ref{fig:scale_meru}), in-process FP32 systems scale by $11.6\times\text{--}12.3\times$ ({HDB FP32 dynamic/flat}: $1{,}188 \to 13{,}780$\,QPS and $1{,}296 \to 15{,}890$\,QPS; {Faiss centered IP}: $1{,}059 \to 13{,}005$\,QPS) and {Milvus centered IP} scales by $4.3\times$ ($260 \to 1{,}130$\,QPS), further widening \rquiver{}'s multi-threaded lead.

\subsection{Geometry and Execution Ablation Study}
\label{subsec:eval_ablation}

\begin{table}[t]
  \centering
  \caption{Controlled component ablations around {\rquiver{}-Core} on {Cohere-1M} (Zen~4, 1 thread, three-graph means).}
  \label{tab:ablation_cohere}
  \small
  \setlength{\tabcolsep}{3pt}
  \begin{tabular}{@{}lrrrr@{}}
    \toprule
    & \textbf{Zero-In} & \textbf{Build} & \multicolumn{2}{c}{\textbf{1T QPS (p99 ms)}} \\
    \cmidrule(lr){4-5}
    \textbf{Variant} & \textbf{Islands} & \textbf{32T (s)} & \textbf{$\ge 99\%$} & \textbf{$\ge 99.5\%$} \\
    \midrule
    \textbf{\rquiver{}-Core} & \textbf{0} & 32.5 & \textbf{2,071} (0.56) & \textbf{688} (1.63) \\
    w/o \texttt{SymSketch256} & 0 & 32.5 & 1,909 (0.60) & 593 (1.90) \\
    w/o \texttt{HotSmPage} ({SoA}) & 0 & 32.5 & 1,752 (0.65) & 537 (2.11) \\
    w/o \textsc{Stage-0.5} Asym Gate & 0 & 32.5 & 1,659 (0.70) & 435 (2.66) \\
    w/o Chord-Swap Repair & 518 & 32.4 & 1,723 (0.66) & 540 (2.08) \\
    w/o QuadriPrune & 0 & \textbf{29.2} & 394 (2.92) & \emph{Unreached} \\
    \bottomrule
  \end{tabular}
\end{table}

\begin{figure*}[t]
  \centering
  \begin{subfigure}[t]{0.485\linewidth}
    \centering
    \includegraphics[width=\linewidth]{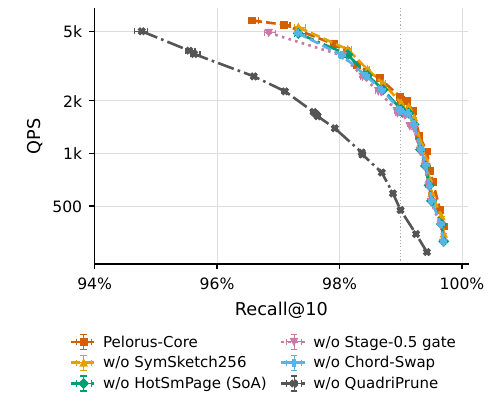}
    \caption{Topology \& decision ablations}
    \label{fig:abl_geom}
  \end{subfigure}\hfill
  \begin{subfigure}[t]{0.485\linewidth}
    \centering
    \includegraphics[width=\linewidth]{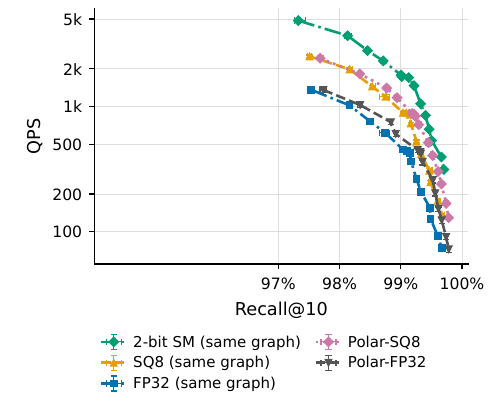}
    \caption{Representation crossover}
    \label{fig:abl_repr}
  \end{subfigure}
  \caption{Single-thread ablation Pareto curves around {\rquiver{}-Core} on {Cohere-1M} (Zen~4, three-graph means; error bars show graph min--max). In the geometry panel, ``w/o Stage-0.5 gate'' and ``w/o Chord-Swap'' correspond to {w/o \textsc{Stage-0.5} Asym Gate} and {w/o Chord-Swap Repair} in Table~\ref{tab:ablation_cohere}; in the representation crossover (matched {SoA} search policy), ``2-bit SM (same graph)'' denotes {w/o \texttt{HotSmPage} ({SoA})}, while ``SQ8 (same graph)'' and ``FP32 (same graph)'' denote {SameGraph-SQ8} and {SameGraph-FP32} in Table~\ref{tab:zen4_repr}.}
  \label{fig:ablation_pdf}
\end{figure*}

Table~\ref{tab:ablation_cohere} and Figure~\ref{fig:ablation_pdf} (Figures~\ref{fig:abl_geom}--\ref{fig:abl_repr}) quantify each topological and execution component on {Cohere-1M}:
\begin{itemize}
  \item \emph{QuadriPrune ($5.26\times$ throughput gain):}
  Replacing {QuadriPrune}'s four-channel quotas and stem-excess dominance with global hyperbolic $\alpha$-pruning produces the radial occlusion characterized by Theorem~\ref{thm:vamana_pathologies}(1), dropping $99\%$-recall QPS from $2{,}071$ to $394$ ($2.92$\,ms p99) and failing to reach $99.5\%$ recall even at $L = 2{,}048$.
  \item \emph{\textsc{Stage-0.5} Asymmetric Refinement ({w/o \textsc{Stage-0.5} Asym Gate} in Table~\ref{tab:ablation_cohere}, $+58.2\%$ at $99.5\%$ recall):}
  Removing lazy orthogonal-projection re-scoring leaves cross-shell query error $d_q\,\eta_q(u_x)$ uncancelled during pool truncation (Theorem~\ref{thm:conformal_flip}(2)), reducing $99.5\%$-recall throughput from $688$ to $435$\,QPS ($1{,}659$\,QPS at $99\%$ recall).
  \item \emph{Spanning-Tree Chord-Swap Repair:}
  Omitting chord swaps leaves $518$ zero-in-degree boundary islands per million vertices and lowers {SoA} $99\%$-recall QPS from $1{,}752$ to $1{,}723$.
  \item \emph{\texttt{HotSmPage} and \texttt{SymSketch256} ($+18.2\%$ at $99\%$, $+28.1\%$ at $99.5\%$ recall):}
  Single-page $4096$\,B co-location lifts $99\%$-recall QPS from $1{,}752$ to $1{,}909$ ($+9.0\%$) by avoiding cross-array secondary {DTLB} walks; adding binomial-stabilized \texttt{SymSketch256} ($57.6\%$ header-line bitplane skip rate) reaches $2{,}071$\,QPS ($688$\,QPS vs.\ $537$\,QPS at $99.5\%$ recall).
\end{itemize}

\subsection{Synthetic Stress Tests, Recall-Loss Audit, and Applicability Boundaries}
\label{subsec:eval_synthetic}

\begin{table}[t]
  \centering
  \caption{Synthetic extreme manifolds and real-manifold trade-offs on Zen~4 (complete \rquiver{} with tangent-cone rescue and adaptive \textsc{CJ-Gate} vs.\ \textbf{Global-SM} without cone rescue or \textsc{CJ-Gate}, $0$\,B/vertex overhead).}
  \label{tab:synthetic_rescue}
  \footnotesize
  \setlength{\tabcolsep}{1.7pt}
  \begin{tabular}{@{}lrrrrrr@{}}
    \toprule
    & \multicolumn{2}{c}{\textbf{Max Rec@10}} & & \multicolumn{3}{c}{\textbf{$\ge 99\%$ QPS (Zen~4)}} \\
    \cmidrule(lr){2-3}\cmidrule(lr){5-7}
    \textbf{Dataset} & \textbf{\shortstack{Global\\SM}} & \textbf{\rquiver{}} & \textbf{Cones} & \textbf{\shortstack{Global\\SM 1T}} & \textbf{\rquiver{} 1T} & \textbf{\rquiver{} 24T} \\
    \midrule
    Clustered-Narrow & 34.22\% & \textbf{99.95\%} & 256 & \emph{Unr.} & \textbf{1,840} & \textbf{38,876} \\
    Clustered-Wide & 73.87\% & \textbf{100.00\%} & 256 & \emph{Unr.} & \textbf{2,704} & \textbf{57,144} \\
    Uniform-Wide & 97.93\% & \textbf{98.88\%} & 0 & \emph{Unr.} & $^\dagger$\emph{98\%:300} & $^\dagger$\emph{6,739} \\
    Uniform-Narrow & 62.41\% & \textbf{64.74\%}$^\ddagger$ & 0 & \emph{Unr.} & \emph{Unr.} & \emph{Unr.} \\
    \midrule
    Cohere-1M ($99\%$) & 99.70\% & \textbf{99.70\%} & 0 & \textbf{1,932} & 1,930 & 42,466 \\
    Cohere-1M ($99.5\%$) & 99.70\% & \textbf{99.70\%} & 0 & 451 & \textbf{738} & \textbf{17,137} \\
    MERU-1M ($99\%$) & 99.95\% & \textbf{99.98\%} & 0 & 1,443 & \textbf{2,547} & \textbf{53,483} \\
    MERU-1M ($99.5\%$) & 99.95\% & \textbf{99.98\%} & 0 & 1,443 & \textbf{1,669} & \textbf{36,092} \\
    OpenAI-1536 ($99\%$) & 99.81\% & \textbf{99.81\%} & 0 & \textbf{820} & 792 & 14,731 \\
    OpenAI-1536 ($99.5\%$) & 99.81\% & \textbf{99.81\%} & 0 & \textbf{487} & 481 & 9,094 \\
    \bottomrule
  \end{tabular}
  \par\vspace{2pt}
  {\footnotesize $^\dagger$At $\ge 98\%$ Recall@10 ($98.51\%$ at $L=1{,}536, B=128$; {Global-SM} caps at $97.93\%$). $^\ddagger$On the 32-query diagnostic probe, FP64 geodesic traversal over FP32 coordinates over the same degree-32 graph visits and returns $62.2\%$ ($\mathcal{L}_{\mathrm{reach}}=37.8\%,\ \mathcal{L}_{\mathrm{beam}}=\mathcal{L}_{\mathrm{gate}}=0\%$), characterizing traversal loss on this constructed graph.}
\end{table}

\begin{table*}[t]
  \centering
  \caption{Measured comparison on all four synthetic distributions (1M vectors, 512D, Zen~4). Each QPS column selects the fastest measured configuration reaching its Recall@10 threshold. Panels (a) and (b) report one and 24 query threads.}
  \label{tab:synthetic_external}
  \footnotesize
  \setlength{\tabcolsep}{2pt}
  \begin{tabular}{@{}llrrrrrrrrr@{}}
\multicolumn{11}{@{}l}{\textbf{(a) Single-thread QPS}} \\
\toprule
\textbf{Dataset} & \textbf{Method} & \textbf{$M/L_c$} & \textbf{\shortstack{Build\\32T (s)}} & \textbf{\shortstack{Peak\\Rec@10}} & \textbf{$\ge90\%$} & \textbf{$\ge95\%$} & \textbf{$\ge98\%$} & \textbf{$\ge99\%$} & \textbf{$\ge99.5\%$} & \textbf{\shortstack{p99 (ms)\\at $99\%$}} \\
\midrule
Uniform-Narrow & Pelorus & 16/128 & 46.8 & 64.735\% & \missing & \missing & \missing & \missing & \missing & \missing \\
 & HDB FP32 flat & 32/256 & 478.2 & 92.170\% & 18 & \missing & \missing & \missing & \missing & \missing \\
 & Faiss centered IP & 32/256 & 311.3 & 29.860\% & \missing & \missing & \missing & \missing & \missing & \missing \\
 & Milvus centered IP & 32/256 & 573.6 & 46.445\% & \missing & \missing & \missing & \missing & \missing & \missing \\
\midrule
Uniform-Wide & Pelorus & 16/128 & 31.9 & 98.880\% & 639 & 397 & 300 & \missing & \missing & \missing \\
 & HDB FP32 flat & 32/256 & 299.6 & 97.860\% & 241 & 103 & \missing & \missing & \missing & \missing \\
 & Faiss centered IP & 32/256 & 140.1 & 48.640\% & \missing & \missing & \missing & \missing & \missing & \missing \\
 & Milvus centered IP & 32/256 & 255.1 & 54.040\% & \missing & \missing & \missing & \missing & \missing & \missing \\
\midrule
Clustered-Narrow & Pelorus & 16/128 & 25.3 & 99.945\% & 3,371 & 2,875 & 2,092 & 1,840 & 1,551 & 0.67 \\
 & HDB FP32 flat & 32/256 & 106.4 & 99.950\% & 1,568 & 1,315 & 1,133 & 998 & 998 & 1.33 \\
 & Faiss centered IP & 32/256 & 35.2 & 21.245\% & \missing & \missing & \missing & \missing & \missing & \missing \\
 & Milvus centered IP & 32/256 & 146.8 & 19.525\% & \missing & \missing & \missing & \missing & \missing & \missing \\
\midrule
Clustered-Wide & Pelorus & 16/128 & 19.4 & 99.995\% & 4,113 & 3,572 & 3,141 & 2,704 & 2,704 & 0.44 \\
 & HDB FP32 flat & 32/256 & 73.7 & 99.500\% & 1,037 & 908 & 735 & 735 & 645 & 2.14 \\
 & Faiss centered IP & 32/256 & 38.3 & 4.015\% & \missing & \missing & \missing & \missing & \missing & \missing \\
 & Milvus centered IP & 32/256 & 273.7 & 9.330\% & \missing & \missing & \missing & \missing & \missing & \missing \\
\bottomrule
\end{tabular}
\par\vspace{5pt}
\begin{tabular*}{\linewidth}{@{\extracolsep{\fill}}llrrrrr@{}}
\multicolumn{7}{@{}l}{\textbf{(b) 24-thread QPS (queries/s)}} \\
\toprule
\textbf{Dataset} & \textbf{Method} & \textbf{$\ge90\%$} & \textbf{$\ge95\%$} & \textbf{$\ge98\%$} & \textbf{$\ge99\%$} & \textbf{$\ge99.5\%$} \\
\midrule
Uniform-Narrow & Pelorus & \missing & \missing & \missing & \missing & \missing \\
 & HDB FP32 flat & 288 & \missing & \missing & \missing & \missing \\
 & Faiss centered IP & \missing & \missing & \missing & \missing & \missing \\
 & Milvus centered IP & \missing & \missing & \missing & \missing & \missing \\
\midrule
Uniform-Wide & Pelorus & 9,363 & 9,363 & 6,739 & \missing & \missing \\
 & HDB FP32 flat & 1,727 & 1,072 & \missing & \missing & \missing \\
 & Faiss centered IP & \missing & \missing & \missing & \missing & \missing \\
 & Milvus centered IP & \missing & \missing & \missing & \missing & \missing \\
\midrule
Clustered-Narrow & Pelorus & 55,231 & 55,231 & 44,175 & 38,876 & 33,132 \\
 & HDB FP32 flat & 17,024 & 17,024 & 14,483 & 13,311 & 13,311 \\
 & Faiss centered IP & \missing & \missing & \missing & \missing & \missing \\
 & Milvus centered IP & \missing & \missing & \missing & \missing & \missing \\
\midrule
Clustered-Wide & Pelorus & 72,780 & 72,780 & 70,801 & 57,144 & 57,144 \\
 & HDB FP32 flat & 15,090 & 15,090 & 12,519 & 12,519 & 12,011 \\
 & Faiss centered IP & \missing & \missing & \missing & \missing & \missing \\
 & Milvus centered IP & \missing & \missing & \missing & \missing & \missing \\
\bottomrule
\end{tabular*}

  \par\vspace{2pt}
  {\footnotesize \textit{Note:} Each row uses one graph and 32 build threads. \rquiver{} uses $M/L_c=16/128$ and search widths through $2{,}048$; the three external baselines use $32/256$ and widths through $4{,}096$. Every timed test condition processes $2{,}000$ queries over five rounds. Peak recall is the largest measured 1T value. The p99 column uses the selected $99\%$-qualifying 1T point. A dash denotes the absence of a qualifying measured point for that thread count. Existing measured configurations supply the $90\%$ and $95\%$ selections.}
\end{table*}

\paragraph{External Comparisons Across Distributions.}
Table~\ref{tab:synthetic_external} reports all 12 external-baseline builds alongside \rquiver{} on the four synthetic distributions. On {Clustered-Narrow}, at Recall@10 $\ge99\%$, \rquiver{} reaches $1{,}840$\,QPS (1T) and $38{,}876$\,QPS (24T), compared with HDB FP32 flat at $998$ and $13{,}311$\,QPS ($1.84\times$ and $2.92\times$). Their selected points attain $99.385\%$ and $99.5\%$ recall, respectively. On {Clustered-Wide}, the corresponding rates are $2{,}704$/$57{,}144$ for \rquiver{} and $735$/$12{,}519$ for HDB ($3.68\times$/$4.56\times$), at $99.61\%$ and $99.495\%$ recall. These are common-threshold throughput comparisons across the reported construction budgets.
Centered Faiss and Milvus attain peak recalls of $21.245\%$/$19.525\%$ on {Clustered-Narrow} and $4.015\%$/$9.33\%$ on {Clustered-Wide}. On {Uniform-Wide}, peak recalls are $98.88\%$ for \rquiver{}, $97.86\%$ for HDB, $48.64\%$ for Faiss, and $54.04\%$ for Milvus. The {Uniform-Narrow} results expose a separate recall weakness of the current \rquiver{} configuration, analyzed below.

\newpage
\paragraph{Four-Stage Telescoping Recall-Loss Audit.}
To pinpoint where candidates are lost on pathological manifolds, let $G_k(q)$ be the exact top-$k$ set, $V(q)$ the vertices visited during traversal, $C_L(q) \subseteq V(q)$ the retained beam pool of size $L$, $F_B(q) \subseteq C_L(q)$ the candidates admitted to exact reranking, and $R_k(q) \subseteq F_B(q)$ the returned top-$k$.
Because $R_k(q) \subseteq F_B(q) \subseteq C_L(q) \subseteq V(q)$, recall miss decomposes telescopically with zero overlap\ifpreprint\ (Appendix~\ref{app:loss_decomposition})\fi:
\begingroup
\predisplaypenalty=0
\begin{equation}
  \begin{aligned}
    &1 - \operatorname{Recall@}k(q) \\
    &= \underbrace{1 - \frac{|G_k \cap V|}{k}}_{\mathcal{L}_{\mathrm{reach}}(q)}
      + \underbrace{\frac{|G_k \cap (V \setminus C_L)|}{k}}_{\mathcal{L}_{\mathrm{beam}}(q)} \\
    &\quad+ \underbrace{\frac{|G_k \cap (C_L \setminus F_B)|}{k}}_{\mathcal{L}_{\mathrm{gate}}(q)}
      + \underbrace{\frac{|G_k \cap (F_B \setminus R_k)|}{k}}_{\mathcal{L}_{\mathrm{rerank}}(q) \equiv 0}.
  \end{aligned}
  \label{eq:four_stage_loss}
\end{equation}
\endgroup
On {Clustered-Narrow} ($N=1\text{M}$, $256$ tight cones at $\rho \in [3.9, 4.1]$) at $L=256, B=128$, auditing {Global-SM} without tangent-cone rescue or \textsc{CJ-Gate} on the diagnostic subset reveals $|G_{10} \cap V| / 10 = 93.9\%$ ($\mathcal{L}_{\mathrm{reach}} = 6.1\%$), yet pool retention drops to $49.3\%$ ($\mathcal{L}_{\mathrm{beam}} = 44.6\%$) and rerank admission drops to $33.0\%$ ($\mathcal{L}_{\mathrm{gate}} = 16.3\%, \mathcal{L}_{\mathrm{rerank}} = 0\%$; $34.22\%$ across all $2{,}000$ queries).
Increasing $L$ from $256$ to $2{,}048$ at fixed $B = 128$ monotonically \emph{lowers} {Global-SM} recall from $34.22\%$ ($2{,}071$\,QPS) through $31.78\%$ ($L=512$) and $29.43\%$ ($L=1{,}024$) down to $27.80\%$ ($L=2{,}048$) via intra-cluster pool dilution.

\paragraph{Synthetic Tangent-Cone Rescue and Real-Data Gate Gains.}
Table~\ref{tab:synthetic_rescue} demonstrates that \rquiver{}'s purity-gated \texttt{ResidualConeDict} automatically detects all $256$ tight cones ($0$\,B/vertex) on both synthetic clustered benchmarks, lifting {Clustered-Narrow} at fixed $L=256, B=128$ from $34.22\%$ to \textbf{$99.39\%$} Recall@10 ($1{,}840$\,QPS 1T and $38{,}876$\,QPS 24T; climbing to $99.80\%$ at $L=384, B=128$ and $99.95\%$ peak) and lifting {Clustered-Wide} from $73.87\%$ to \textbf{$99.61\%$} at $L=192, B=64$ ($2{,}704$\,QPS 1T and $57{,}144$\,QPS 24T; $99.995\%$ peak).
On the unclustered real benchmarks ({Cohere-1M}, {MERU-1M}, {OpenAI-1536}, and {WordNet-Hazy}), the strict purity gate retains \textbf{$0$ cones}, avoiding unnecessary residual dictionary construction while adaptive \textsc{CJ-Gate} and variance-stabilized sketching raise {Cohere-1M} $99.5\%$-recall QPS by \textbf{$+63.7\%$} ($451 \to 738$\,QPS 1T, $17{,}137$\,QPS 24T) and {MERU-1M} $99\%$-recall QPS by \textbf{$+76.5\%$} ($1{,}443 \to 2{,}547$\,QPS 1T, $53{,}483$\,QPS 24T) at $<3.5\%$ overhead on 1536D {OpenAI-1536} ($792$ vs.\ $820$\,QPS).

\paragraph{Degree-and-Budget Limits on Isotropic Narrow Shells.}
On {Uniform-Narrow} (1M points with uniform directions on $\mathbb{S}^{511}$, $\rho \in [3.9, 4.1]$, and local intrinsic dimensionality $\mathrm{LID} \approx 511$~\cite{houle2017lid}), \rquiver{} ($16/128, L\le2{,}048$) reaches $64.735\%$ recall ($62.41\%$ under {Global-SM}), whereas {HDB FP32 flat} ($32/256$) reaches $85.08\%$ at $ef=2{,}048$ and $92.17\%$ at $ef=4{,}096$ ($18$\,QPS 1T and $288$\,QPS 24T at $\ge 90\%$) on the full $2{,}000$-query test set, and {Faiss}/{Milvus} peak at $29.86\%$/$46.445\%$.
On a separate $32$-query probe over that same degree-32 \rquiver{} graph, exact FP64 geodesic search over FP32 coordinates achieves $62.2\%$ Recall@10 ($\mathcal{L}_{\mathrm{reach}}=37.8\%,\ \mathcal{L}_{\mathrm{beam}}=\mathcal{L}_{\mathrm{gate}}=0\%$), isolating graph reachability rather than 2-bit quantization on the $16/128$ build; separating single-layer {QuadriPrune} structure from {HDB}'s larger $M/L_c=32/256, ef\le4{,}096$ budget requires matched-degree builds.
Widening radial depth to $\rho \in [0.4, 6.0]$ ({Uniform-Wide}) enables exact polar radial lower-bound pruning (Proposition~\ref{prop:geodesic_apex}(1)) across shells and restores Recall@10 above $98\%$ (\textbf{$98.88\%$} peak, $300$\,QPS 1T and $6{,}739$\,QPS 24T at $98.51\%$).

\section{Related Work}
\label{sec:related}

We position \rquiver{} relative to academic and industrial work across four areas: hyperbolic representation learning and non-Euclidean indexing, proximity-graph topology design, low-bit vector quantization, and production vector database systems.

\subsection{Hyperbolic Representations and Non-Euclidean Indexing}
\label{subsec:rel_hyperbolic}

\paragraph{Curved Representation Models.}
Negatively curved manifolds embed trees and asymmetric hierarchies with exponentially lower distortion than flat Euclidean spaces~\cite{nickel2017poincare,sala2018representation}.
Foundational models formulate optimization directly on the {Poincar\'e} ball for taxonomies, neural layers, and lexical semantics~\cite{ganea2018hyperbolic,tifrea2019poincare}, or parameterize partial orders via {Poincar\'e} entailment cones~\cite{ganea2018entailment}.
Parallel lines adopt the {Lorentz} hyperboloid and mixed-curvature product manifolds to stabilize Riemannian gradients across graph neural networks~\cite{nickel2018lorentz,chami2019hgcn,gu2019mixed}, low-dimensional knowledge graphs~\cite{chami2020low}, and fully hyperbolic Transformer backbones~\cite{chen2022fully}.
In visual and multimodal retrieval, hyperbolic metric learning captures semantic uncertainty and part-whole hierarchies in image encoders~\cite{khrulkov2020hyperbolic,ermolov2022hyperbolic}, scaling to million-item contrastive vision-language alignment in Meta's {MERU}~\cite{desai2023meru} and hierarchical passage retrieval for generation in {HypRAG}~\cite{madhu2026hyprag,lewis2020rag}.
\rquiver{} unifies both {Poincar\'e} and {Lorentz} workloads at ingestion via exact stereographic projection~\eqref{eq:lorentz_isometry}, avoiding unbounded spatial-norm quantization blowup on $\Lmod$ (Theorem~\ref{thm:conformal_flip}(3)).

\paragraph{Metric, Hashing, and Graph Search in Curved Spaces.}
Classical metric indexes such as vantage-point trees ({vp-tree})~\cite{yianilos1993data} and navigating nets in doubling or negatively curved spaces~\cite{krauthgamer2006navigating} prune subtrees via exact triangle inequalities, but shell concentration in $D \ge 512$ dimensions causes metric ball bounds to overlap nearly the entire dataset.
On the hashing and Euclidean-adapter side, Deng et al.~\cite{deng2026hyperboliclsh} recently constructed native locality-sensitive hashing ({LSH}) families in hyperbolic space, though data-oblivious hashing requires many hash tables or wide candidate probes to sustain $99\%+$ recall in high dimensions, while {HyEm}~\cite{deng2026hyem} stores radius-bounded origin log-maps inside Euclidean {HNSW} before query-adaptive hyperbolic reranking rather than traversing a native curved graph.
Empirical non-Euclidean graph libraries led by {NMSLIB}~\cite{boytsov2013engineering} (deployed in early OpenSearch engines), the theoretical hyperbolic graph analysis of Prokhorenkova et al.~\cite{prokhorenkova2022hyperbolic}, recent low-dimensional navigable-graph bounds proving $\Omega(n)$ worst-case out-degree without Steiner points~\cite{deng2026navigable}, and the dedicated hyperbolic engine {HyperspaceDB}~\cite{glukhota2026hyperspacedb} demonstrate that greedy proximity-graph routing adapts to intrinsic manifold structure.
Nevertheless, existing hyperbolic graph indices either evaluate geodesic distances over full-precision coordinates at every hop---fetching $2\text{--}6$\,KB per visited vertex and requiring FP64 arithmetic near $\|x\|_2 \to 1/\sqrt{c}$---or quantize ambient Cartesian/{Lorentz} coordinates without isolating conformal weights while pruning edges under a single isotropic predicate.
\rquiver{} instead factors the warped-product metric into exact 32-byte radial scalars plus 2-bit spherical bitplanes and replaces isotropic pruning with stem-excess {QuadriPrune} four-channel diversity.

\subsection{Proximity-Graph Topology and Skewed Routing}
\label{subsec:rel_graphs}

\paragraph{Flat Proximity Graphs and Benchmarks.}
Navigable small-world graphs ({NSW}~\cite{malkov2014approximate} and hierarchical {HNSW}~\cite{malkov2020hnsw}), approximate relative-neighborhood and satellite-system graphs ({NSG}~\cite{fu2019nsg} and {NSSG}~\cite{fu2022high}), and Microsoft's $\alpha$-slack {Vamana} index ({DiskANN})~\cite{subramanya2019diskann} define the standard in-memory and SSD-resident {ANN} paradigms, as established across standardized community benchmarks~\cite{aumuller2020annbenchmarks,wang2021comprehensive,simhadri2022results,azizi2025graph}.
Production extensions of {DiskANN} support streaming updates in {FreshDiskANN}~\cite{singh2021freshdiskann}, centroid-partitioned disk posting lists in {SPANN}~\cite{chen2021spann}, and structured predicate filtering in {Filtered-DiskANN}~\cite{gollapudi2023filtered}.
Transplanting these pruning rules from their assumed flat ambient metrics to $\Bball$ permits radial occlusion of entire spherical neighborhoods by interior vertices. The angular escape condition characterizes when a candidate improves on the origin (Theorem~\ref{thm:vamana_pathologies}).

\paragraph{Skewed Distributions, Hub Occlusion, and Block Locality.}
Recent database indices examine routing pathologies and block layout from complementary angles.
{RoarGraph}~\cite{chen2024roargraph} shows that cross-modal out-of-distribution ({OOD}) queries suffer hub occlusion and long detours on flat inner-product graphs, mitigating shift by building a bipartite graph over historical query projections.
In negatively curved spaces, hub occlusion arises even for \emph{in-distribution} queries as an intrinsic geometric consequence of $\sinh(\sqrt{c}R)$ fiber expansion rather than query-modality shift.
\rquiver{} controls radial occlusion through Inside-Out depth seeding, stem-excess angle-guarded dominance, shielded isobathic rings, spanning-tree chord-swap island repair, and routing via the closed-form radial minimizer, with angular ring pruning on constant-depth data.
For storage locality, {Starling}~\cite{wang2024starling} reorders disk-resident graph neighborhoods into $4$\,KiB SSD segments to reduce physical block reads per hop.
Addressing the in-memory {DTLB} translation bottleneck rather than SSD block I/O, \rquiver{}'s \texttt{HotSmPage} co-locates polar headers, 256-bit sketches, and complete 2-bit bitplanes of $K$ consecutive vertex IDs inside one $4096$\,B OS page ($K=16$ at $D=768$, Table~\ref{tab:hot_sm_page}), avoiding cross-array secondary page-table walks without graph-isomorphism reordering.

\subsection{Low-Bit Vector Quantization in Academia and Industry}
\label{subsec:rel_quantization}

\paragraph{Subspace and Anisotropic Codebooks.}
Product quantization ({PQ}~\cite{jegou2011pq} and {OPQ}~\cite{ge2013opq}) partitions flat vectors into learned subspaces, while {PQ Fast Scan}~\cite{andre2015cache} and {Quick ADC}~\cite{andre2017accelerated} pack 4-bit lookup tables into {SIMD} registers via \texttt{vpshufb} shuffles.
In industrial maximum inner-product search ({MIPS}), Google's {ScaNN}~\cite{guo2020accelerating} observes that quantization error parallel to a candidate vector perturbs rankings more severely than orthogonal error, motivating anisotropic codebook losses and spilled orthogonal-residual assignment in {SOAR}~\cite{sun2023soar}.
Subspace codebooks, however, require offline iterative clustering and assume fixed flat inner-product weights; similarly, {HiHPQ}~\cite{qiu2024hihpq} learns end-to-end trainable hyperbolic product quantization codebooks during representation training rather than serving a training-free post-hoc graph index.
On $\Bball$, the directional weight $W(r_q, r_v) = \frac{2}{c}\sinh\rho_q\sinh\rho_v$ varies exponentially across candidate pairs and diverges near $r_v \to 1/\sqrt{c}$, which \rquiver{} isolates analytically in unquantized polar headers before touching directional codes.

\paragraph{Industrial Locally-Adaptive Scalar and Binary Quantization.}
To avoid codebook training during graph traversal, Intel's Scalable Vector Search ({SVS}) introduces Locally-Adaptive Vector Quantization ({LVQ})~\cite{aguerrebere2023similarity}, centering and scaling each vector individually into 4-bit or 8-bit integers, and {LeanVec}~\cite{tepper2024leanvec}, which pairs learned linear dimensionality reduction with {LVQ} to shrink memory bandwidth.
In enterprise search engines, Apache {Lucene} and {Elasticsearch} deploy Better Binary Quantization ({BBQ})~\cite{trent2024bbq}, combining centered 1-bit database codes with asymmetric 4-bit query lookup tables.
Two technical boundaries separate these flat production quantizers from \rquiver{}.
On coordinate adaptation, {LVQ}'s per-vector scalar centering and scaling operate on ambient coordinates and cannot factor out the nonlinear {Poincar\'e} denominator $(1 - c\|x\|_2^2)$ or prevent tight semantic cones ($\alpha_v = \langle u_v, \mu_c\rangle \to 1$) from expending dynamic range on the shared cluster axis $\mu_c$.
\rquiver{}'s purity-gated \texttt{ResidualConeDict} explicitly subtracts $\alpha_v\mu_c$ in the spherical tangent space $T_{\mu_c}\Sunit$ at $0$\,B/vertex overhead, attenuating residual cosine error by $\beta_{q,c}\beta_v$ and preserving the pairwise covariance in its {$\MNR$} expression (Theorem~\ref{thm:tangent_cone}).
On bit-width fidelity, compared with 1-bit {BBQ}~\cite{trent2024bbq}, two-level 2-bit {SM} retains strong-magnitude coordinates and pairs an L1-resident fixed-point masked subset-sum kernel (\textsc{Stage-0.5}, contrasting with 4-bit nibble {SIMD} shuffle tables~\cite{andre2015cache,andre2017accelerated}) with a deterministic Pythagorean orthogonal-projection divisor $\sigma_v^2$ (Proposition~\ref{prop:asym_cj_gate}(1)) that removes multiplicative shrinkage deterministically.

\paragraph{Randomized Bit Quantization, Quantized Graphs, and Decision Theory.}
Locality-sensitive hashing and random hyperplane rounding ({SimHash})~\cite{indyk1998approximate,charikar2002similarity} initiated 1-bit angular estimation.
Extending 1-bit hyperplane rounding, {PolarQuant}~\cite{han2026polarquant} applies random preconditioning and recursive 2D polar coordinate transforms in flat {Euclidean} space to eliminate block-scale metadata in {KV}-cache quantization (orthogonal to our warped-product separation of singular hyperbolic conformal factors on $\Bball$).
Recent database work achieves strong theoretical guarantees in flat Euclidean space via {RaBitQ}~\cite{gao2024rabitq}, asymptotically optimal extended {RaBitQ}~\cite{gao2025extendedrabitq}, and {SymphonyQG}~\cite{gou2025symphonyqg}, which tightly co-layouts randomized quantized codes inside graph neighbor blocks.
Both {RaBitQ} and {SymphonyQG} rely on a \emph{global random orthogonal rotation} to equalize coordinate variances.
On contrastive neural embeddings---whose representations exhibit structured coordinate anisotropy and local Gaussian concentration~\cite{wang2020understanding,betser2026infonce}---prior flat two-bit graph indexing~\cite{xiao2026quiver} and flat comparison-decision stability theory~\cite{xiao2026decisions} establish how shared-endpoint residual covariance governs ranking and pruning flips and why global random rotations can weaken native cancellation.
Taking that flat decision framework as its foundation, \rquiver{} addresses negatively curved warped-product manifolds by decoupling singular conformal factors into unquantized polar headers, isolating cross-shell weight mismatch, rescuing concentrated boundary cones at zero byte overhead, and gating off-axis conformal uncertainty via \textsc{CJ-Gate}.

\subsection{Production Vector Database Systems}
\label{subsec:rel_systems}

Modern vector data management spans specialized vector engines and relational-hybrid databases~\cite{pan2024survey}.
Meta's {Faiss} provides CPU and GPU kernels for flat, IVF-PQ, and {HNSW} search~\cite{johnson2019billion,douze2024faiss}, while Zilliz's {Milvus}~\cite{wang2021milvus} and cloud-native {Manu}~\cite{guo2022manu} disaggregate storage, streaming ingestion, and multi-replica index execution.
On the hybrid analytical and transactional front, Alibaba's {PASE}~\cite{yang2020pase} and {AnalyticDB-V}~\cite{wei2020analyticdb} integrate quantized and graph vector indexes into {PostgreSQL} and cloud analytical optimizers.
Microsoft's {VBASE}~\cite{zhang2023vbase} unifies online vector traversal and relational predicates by exploiting relaxed distance monotonicity along proximity-graph iterators, and {SingleStore-V}~\cite{chen2024singlestore} embeds segmented {HNSW} and subspace compression inside a distributed SQL engine.
All of these industrial engines assume flat $L_2$, cosine, or inner-product distance kernels.
Hosting hyperbolic embeddings ({MERU}, {HypRAG}) on flat inner-product indices requires fetching wide candidate pools across mismatched radial shells before exact geodesic reranking.
Because \rquiver{}'s polar proxy $P_{\mathbb{B}}(q, v)$ is strictly order-isomorphic to $d_{\mathbb{B}}(q, v)$ (satisfying {VBASE}-style relaxed monotonicity~\cite{zhang2023vbase}) and encapsulates its state in stateless $4096$\,B pages plus fixed-stride adjacency arrays, our polar 2-bit operator can serve directly as a native curved-manifold index engine inside modular vector DBMS architectures.

\section{Conclusion}
\label{sec:conclusion}

We presented \textbf{\rquiver{}}, a warped-product proximity-graph index for hyperbolic vector search in the {Poincar\'e} ball and {Lorentz} hyperboloid.
By factoring the warped-product manifold into unquantized FP32 radial and conformal scalars paired with scale-invariant 2-bit Sign-Magnitude spherical codes, \rquiver{} prevents quantization noise from entering singular boundary denominators while compressing per-vertex logical hot navigation state to $384$\,B ($517.0$\,MB page-co-located at $N=10^6, D=768$, $2.44\times$ below 8-bit {Polar-SQ8}; $352$\,B / $485.0$\,MB / $2.60\times$ in split form).
Our conformal polar distortion and hyperbolic graph-pathology analysis characterized query boundary-depth invariance, cross-shell query-noise non-cancellation, radial occlusion and angular-cone collapse, and boundary cluster {$\MNR$} collapse, directly motivating stem-excess {QuadriPrune} construction, spanning-tree chord-swap {BFS} island repair, routing via the constrained radial minimizer, deterministic orthogonal-projection de-biasing, purity-gated zero-byte-per-node tangent-cone residuals, and $4096$\,B single-page co-location.
Across eight real and synthetic hyperbolic benchmarks on {AMD Zen~4}, \rquiver{} accelerates 32-thread index construction by $5.44\times$ over {Polar-FP32} ($1.91\times$ vs.\ {Polar-SQ8}), improves $99\%$-recall throughput by $4.30\times$ (1T) to $6.31\times$ (24T) over full-precision polar search on {Cohere-1M}, and reaches $99.70\%$ peak Recall@10 on {Cohere-1M} (vs.\ $96.78\%$ for {HyperspaceDB} and $98.91\%$ for centered {Faiss}) alongside $53{,}483$\,QPS at $99\%$ recall on native hyperbolic {MERU-1M} ($3.37\times$--$3.88\times$ over {HyperspaceDB} FP32) on the global 2-bit path ($0$ cones across diffuse real sets), while activating $256$ cones on clustered shells to raise {Clustered-Narrow} recall from $34.22\%$ to $99.385\%$ at $L=256$ ($99.80\%$ at $L=384$).
The synthetic stress tests further delineate how manifold geometry shapes these gains: at $\ge 99\%$ recall, \rquiver{} exceeds {HyperspaceDB} FP32 flat throughput by $1.84\times$ / $2.92\times$ (1T / 24T) on {Clustered-Narrow} and $3.68\times$ / $4.56\times$ on {Clustered-Wide}, whereas on isotropic {Uniform-Narrow} \rquiver{} ($M/L_c = 16/128, L \le 2{,}048$) peaks at $64.735\%$ Recall@10 compared with $92.17\%$ for {HyperspaceDB} ($32/256, ef \le 4{,}096$), delineating degree-and-budget limits on unclustered narrow shells.

\bibliographystyle{ACM-Reference-Format}
\bibliography{references}

\ifpreprint
\clearpage
\appendix

\section{Riemannian Geometry, Stereographic Isometry, and Warped-Product Proofs}
\label{app:warped_product}

\subsection{Algebraic Proof of the Lorentz-to-Poincar\'e Stereographic Isometry}
Let $z = (z_0, \tilde{z}), z' = (z'_0, \tilde{z}') \in \Lmod$ satisfy the upper-sheet hyperboloid constraint $-z_0^2 + \|\tilde{z}\|_2^2 = -1/c$ with $z_0 > 0$, so $c\|\tilde{z}\|_2^2 = c z_0^2 - 1$.
For $x = \Pi(z) = \frac{\tilde{z}}{1 + \sqrt{c}\,z_0}$, we first verify $x \in \Bball$:
\begin{equation}
  \begin{aligned}
    1 - c\|x\|_2^2
    &= 1 - \frac{c\|\tilde{z}\|_2^2}{(1 + \sqrt{c}\,z_0)^2} \\
    &= 1 - \frac{c z_0^2 - 1}{(1 + \sqrt{c}\,z_0)^2} \\
    &= \frac{2(1 + \sqrt{c}\,z_0)}{(1 + \sqrt{c}\,z_0)^2} \\
    &= \frac{2}{1 + \sqrt{c}\,z_0} > 0,
  \end{aligned}
  \label{eq:app_conformal_lorentz}
\end{equation}
which simultaneously yields the exact conformal factor identity $f_{\Pi(z)} = \frac{2}{1 - c\|\Pi(z)\|_2^2} = 1 + \sqrt{c}\,z_0$.
Next, denote the scaled {Lorentz} inner product by $L(z, z') \triangleq c(z_0 z'_0 - \langle \tilde{z}, \tilde{z}'\rangle) \ge 1$.
Expanding $\|x - x'\|_2^2$ for $x = \frac{\tilde{z}}{f_x}$ and $x' = \frac{\tilde{z}'}{f_{x'}}$ where $f_x = 1 + \sqrt{c}\,z_0$ and $f_{x'} = 1 + \sqrt{c}\,z'_0$:
\begin{equation}
  \begin{aligned}
    &c f_x f_{x'}\|x - x'\|_2^2 \\
    &= \frac{c\|\tilde{z}\|_2^2 f_{x'}}{f_x} + \frac{c\|\tilde{z}'\|_2^2 f_x}{f_{x'}}
      - 2c\langle \tilde{z}, \tilde{z}'\rangle \\
    &= (\sqrt{c}\,z_0 - 1)(1 + \sqrt{c}\,z'_0) \\
    &\quad+ (\sqrt{c}\,z'_0 - 1)(1 + \sqrt{c}\,z_0)
      - 2c\langle \tilde{z}, \tilde{z}'\rangle \\
    &= 2\bigl(c z_0 z'_0 - 1 - c\langle \tilde{z}, \tilde{z}'\rangle\bigr) \\
    &= 2\bigl(L(z, z') - 1\bigr).
  \end{aligned}
\end{equation}
Substituting $f_x f_{x'} = \frac{4}{(1 - c\|x\|_2^2)(1 - c\|x'\|_2^2)}$ gives
\begin{equation}
  1 + \frac{2c\|x - x'\|_2^2}{(1 - c\|x\|_2^2)(1 - c\|x'\|_2^2)}
  = 1 + \frac{c}{2} f_x f_{x'}\|x - x'\|_2^2
  = L(z, z').
\end{equation}
Applying $\frac{1}{\sqrt{c}}\arccosh(\cdot)$ to both sides proves $d_{\mathbb{B}}(\Pi(z), \Pi(z')) = d_{\mathbb{L}}(z, z')$ identically for all $z, z' \in \Lmod$. \qed

\subsection{Warped-Product Metric Tensor, Hyperbolic Law of Cosines, and Flat Limit}
Write $x = r u \in \Bball \setminus \{0\}$ with $r \in (0, 1/\sqrt{c})$ and $u \in \Sunit$.
In Euclidean polar coordinates, $\mathrm{d}x^\top \mathrm{d}x = \mathrm{d}r^2 + r^2 \mathrm{d}\Omega_{D-1}^2$.
From $\rho = \sqrt{c}\,R = 2\atanh(\sqrt{c}\,r)$, we have $\sqrt{c}\,r = \tanh(\frac{\sqrt{c}}{2}R)$, whence
\begin{equation}
  \begin{aligned}
    \mathrm{d}r &= \frac{1}{2}\operatorname{sech}^2\!\left(\frac{\sqrt{c}}{2}R\right)\mathrm{d}R, \\
    1 - c r^2 &= \operatorname{sech}^2\!\left(\frac{\sqrt{c}}{2}R\right), \\
    f_x &= \frac{2}{1 - c r^2} = 2\cosh^2\!\left(\frac{\sqrt{c}}{2}R\right).
  \end{aligned}
\end{equation}
Multiplying $\mathrm{d}x^\top \mathrm{d}x$ by $f_x^2$ gives the Riemannian line element:
\begin{equation}
  \begin{aligned}
    \mathrm{d}s^2 &= f_x^2\bigl(\mathrm{d}r^2 + r^2\mathrm{d}\Omega_{D-1}^2\bigr) \\
    &= \mathrm{d}R^2 + \frac{4}{c}\cosh^4\!\left(\frac{\sqrt{c}}{2}R\right) \\
    &\quad\times\tanh^2\!\left(\frac{\sqrt{c}}{2}R\right)\mathrm{d}\Omega_{D-1}^2 \\
    &= \mathrm{d}R^2 + \frac{1}{c}\sinh^2(\sqrt{c}\,R)\,\mathrm{d}\Omega_{D-1}^2,
  \end{aligned}
\end{equation}
establishing~\eqref{eq:warped_metric}.
Furthermore, using $\sinh\rho_z = \sqrt{c}\,r_z f_z$ and $\cosh\rho_z = f_z - 1 = c r_z^2 f_z + 1$, direct expansion of $\frac{c}{4}P_{\mathbb{B}}(q, v)$ gives
\begin{equation}
  \begin{aligned}
    &\frac{c}{4}P_{\mathbb{B}}(q, v) \\
    &= \frac{c}{4}(r_q - r_v)^2 f_q f_v
      + \frac{c}{2}r_q r_v f_q f_v (1 - \cos\theta_{qv}) \\
    &= \frac{1}{2}\bigl(\cosh\rho_q\cosh\rho_v - \sinh\rho_q\sinh\rho_v - 1\bigr) \\
    &\quad+ \sinh\rho_q\sinh\rho_v\sin^2\!\left(\frac{\theta_{qv}}{2}\right) \\
    &= \sinh^2\!\left(\frac{\rho_q - \rho_v}{2}\right)
      + \sinh\rho_q\sinh\rho_v\sin^2\!\left(\frac{\theta_{qv}}{2}\right) \\
    &= \sinh^2\!\left(\frac{\sqrt{c}}{2}d_{\mathbb{B}}(q, v)\right),
  \end{aligned}
\end{equation}
where the final equality uses $\sinh^2(t/2) = \frac{1}{2}(\cosh t - 1)$.
Taking $c \to 0^+$ with fixed $q, v \in \R^D$, $\lim_{c \to 0^+} f_z = 2$, so $\lim_{c \to 0^+} \frac{1}{4}P_{\mathbb{B}}(q, v) = \|q - v\|_2^2$ and $\lim_{c \to 0^+} d_{\mathbb{B}}(q, v) = 2\|q - v\|_2$. \qed

\subsection{Complete Proof of Proposition~\ref{prop:geodesic_apex} (Strong Convexity and Constrained Radial Minimization)}
\begin{proof}
For Claim~(1), since $u_q, u_v \in \Sunit$, Cauchy--Schwarz gives $\gamma_{qv} = \langle u_q, u_v\rangle \le 1$.
Since $W(r_q, r_v) = 2 r_q r_v f_q f_v \ge 0$, $P_{\mathbb{B}}(q, v) = R(r_q, r_v) + W(r_q, r_v)(1 - \gamma_{qv}) \ge R(r_q, r_v) = P_{\mathrm{rad}}(r_q, r_v)$, with equality at $u_v = u_q$, where $d_{\mathbb{B}}(q, v) = |\rho_q - \rho_v|/\sqrt{c}$ yields $P_{\mathrm{rad}}(r_q, r_v) = \frac{2}{c}(\cosh|\rho_q - \rho_v| - 1)$.

For Claim~(2), consider $g_\gamma(\rho) = \cosh\rho_q\cosh\rho - \gamma\sinh\rho_q\sinh\rho$ for $\rho \in \R$ and $\gamma \in [-1, 1]$.
Differentiating twice with respect to $\rho$:
\begin{align}
  g_\gamma'(\rho) &= \cosh\rho_q\sinh\rho - \gamma\sinh\rho_q\cosh\rho, \\
  g_\gamma''(\rho) &= \cosh\rho_q\cosh\rho - \gamma\sinh\rho_q\sinh\rho = g_\gamma(\rho).
\end{align}
Since $|\gamma| \le 1$ and $\sinh\rho_q \ge 0$, $|\gamma\sinh\rho_q\sinh\rho| \le \sinh\rho_q|\sinh\rho|$, hence for all $\rho \in \R$:
\begin{equation}
  \begin{aligned}
    g_\gamma''(\rho) &= g_\gamma(\rho) \\
    &\ge \cosh\rho_q\cosh|\rho| - \sinh\rho_q\sinh|\rho| \\
    &= \cosh(\rho_q - |\rho|) \ge 1 > 0.
  \end{aligned}
\end{equation}
Thus $g_\gamma(\rho)$ is globally $1$-strongly convex on $\R$.
Evaluating the first-order optimality condition $g_\gamma'(\rho) = 0$ divides by $\cosh\rho_q\cosh\rho \ge 1$ to obtain $\tanh\rho = \gamma\tanh\rho_q \in (-1, 1)$, so the unique unconstrained minimizer on $\R$ is $\rho_{\R}^* = \atanh(\gamma\tanh\rho_q)$.
Restricting $\rho$ to $[0, \infty)$: if $\gamma \le 0$, $\rho_{\R}^* \le 0$ and $g_\gamma'(\rho) > 0$ for all $\rho > 0$, so the unique minimizer on $[0, \infty)$ is $0 = \atanh(\max(\gamma, 0)\tanh\rho_q)$; if $\gamma > 0$, $\rho_{\R}^* > 0$ lies in the interior.
Similarly, for any closed interval $[\rho_{\min}, \rho_{\max}] \subset [0, \infty)$, a $1$-strongly convex univariate function decreases strictly on $(-\infty, \rho_{\R}^*)$ and increases strictly on $(\rho_{\R}^*, \infty)$, so its unique minimizer on $[\rho_{\min}, \rho_{\max}]$ is $\min(\max(\rho_{\R}^*, \rho_{\min}), \rho_{\max}) = \rho_{\mathrm{sup}}^*$, which also minimizes $d_{\mathbb{B}}(q, v(\rho)) = \frac{1}{\sqrt{c}}\arccosh(g_\gamma(\rho))$ by strict monotonicity of $\arccosh$ on $[1, \infty)$.
\end{proof}

\section{Two-Level Spherical SM Algebra, Bitplane Kernels, and Deterministic De-Biasing}
\label{app:sm_algebra}

\subsection{Complete Proof of Lemma~\ref{lem:sm_pythagorean} (Conditional-Mean Pythagorean Identity)}
\begin{proof}
Let $u_v \in \Sunit$ have bitplanes $s_{v,i} = \mathbb{I}[u_{v,i} \ge 0]$ and $m_{v,i} = \mathbb{I}[|u_{v,i}| > \tau_v]$, partitioning $\{1,\dots,D\}$ into $I_{0,v}$ ($|I_{0,v}| = D - s_v$) and $I_{1,v}$ ($|I_{1,v}| = s_v$).
For any candidate two-level vector $\tilde{u}_i = \sgn(u_{v,i})(a + (b - a)m_{v,i})$ parameterized by $(a, b) \in \R^2$, because $\sgn(u_{v,i})u_{v,i} = |u_{v,i}|$ for every $i$ (including $u_{v,i} = 0$ where $|0| = 0$), the squared Euclidean reconstruction error splits additively across $I_{0,v}$ and $I_{1,v}$:
\begin{equation}
  \|u_v - \tilde{u}\|_2^2
  = \sum_{i \in I_{0,v}}\bigl(|u_{v,i}| - a\bigr)^2 + \sum_{i \in I_{1,v}}\bigl(|u_{v,i}| - b\bigr)^2.
\end{equation}
When $0 < s_v < D$, both sums are strictly convex quadratics minimized uniquely at the empirical conditional means $a_v = \frac{1}{D - s_v}\sum_{i \in I_{0,v}} |u_{v,i}|$ and $b_v = \frac{1}{s_v}\sum_{i \in I_{1,v}} |u_{v,i}|$ For $s_v=0$, the weak mean is uniquely $a_v=\tau_v>0$ and the strong mean is unused. The implementation chooses $b_v=0$, giving $\delta_v=-a_v$. Every reconstructed coordinate uses $a_v$, so the reconstructed vector remains unique. The strict mean threshold implies $s_v<D$ in real arithmetic.
At $(a_v, b_v)$, setting $\hat{u}_{v,i} = \sgn(u_{v,i})(a_v + \delta_v m_{v,i})$ with $\delta_v = b_v - a_v$ gives:
\begin{align}
  \langle u_v, \hat{u}_v\rangle
  &= \sum_{i \in I_{0,v}} |u_{v,i}| a_v + \sum_{i \in I_{1,v}} |u_{v,i}| b_v
   = (D - s_v)a_v^2 + s_v b_v^2, \\
  \|\hat{u}_v\|_2^2
  &= \sum_{i \in I_{0,v}} a_v^2 + \sum_{i \in I_{1,v}} b_v^2
   = (D - s_v)a_v^2 + s_v b_v^2 \triangleq \sigma_v^2.
\end{align}
Hence $\langle u_v - \hat{u}_v, \hat{u}_v\rangle = \langle u_v, \hat{u}_v\rangle - \|\hat{u}_v\|_2^2 = 0$, and the Pythagorean identity yields
\begin{equation}
  \|u_v - \hat{u}_v\|_2^2 = \|u_v\|_2^2 - 2\langle u_v, \hat{u}_v\rangle + \|\hat{u}_v\|_2^2 = 1 - \sigma_v^2 \ge 0,
\end{equation}
which implies $\sigma_v^2 \le 1$; meanwhile $\sigma_v^2 \ge \frac{1}{D}(\sum_{i=1}^D |u_{v,i}|)^2 \ge \frac{1}{D}\|u_v\|_2^2 = \frac{1}{D} > 0$.

For the five-popcount bilinear formula, let $\sigma_{q,i} = \sgn(u_{q,i}) \in \{\pm 1\}$, $\sigma_{v,i} = \sgn(u_{v,i}) \in \{\pm 1\}$, and $\bar{x}_i = \mathbb{I}[s_{q,i} = s_{v,i}] \in \{0, 1\}$, so $\sigma_{q,i}\sigma_{v,i} = 2\bar{x}_i - 1$.
Expanding $\langle \hat{u}_q, \hat{u}_v\rangle = \sum_{i=1}^D (2\bar{x}_i - 1)(a_q + \delta_q m_{q,i})(a_v + \delta_v m_{v,i})$ term by term:
\begin{equation}
  \begin{aligned}
    \sum_{i=1}^D (2\bar{x}_i - 1) &= 2 p_\oplus - D, \\
    \sum_{i=1}^D (2\bar{x}_i - 1)m_{v,i} &= 2 p_{01} - s_v, \\
    \sum_{i=1}^D (2\bar{x}_i - 1)m_{q,i} &= 2 p_{10} - s_q, \\
    \sum_{i=1}^D (2\bar{x}_i - 1)m_{q,i}m_{v,i} &= 2 p_{11} - n_{11},
  \end{aligned}
\end{equation}
proving Lemma~\ref{lem:sm_pythagorean}.
\end{proof}

\subsection{Bit-by-Bit Proof of the 3-to-2 Carry-Save Adder ({CSA}) Identity}
For any single bit position $i$ across three binary words $A, B, C \in \{0, 1\}^W$, integer addition of bits $A_i + B_i + C_i \in \{0, 1, 2, 3\}$ equals the two-bit binary representation $\mathrm{Sum}_i + 2\,\mathrm{Carry}_i$, where the full-adder truth table gives
\begin{equation}
  \mathrm{Sum}_i = A_i \oplus B_i \oplus C_i, \qquad
  \mathrm{Carry}_i = (A_i \wedge B_i) \vee \bigl((A_i \oplus B_i) \wedge C_i\bigr).
\end{equation}
Checking all $8$ combinations $(A_i, B_i, C_i) \in \{0, 1\}^3$:
for weight $0$, $(0, 0) \mapsto 0$; for weight $1$, $(\mathrm{Sum}_i, \mathrm{Carry}_i) = (1, 0) \mapsto 1$; for weight $2$, $(0, 1) \mapsto 2$; and for weight $3$, $(1, 1) \mapsto 1 + 2 = 3$.
Summing over $i = 1, \dots, W$ yields $\popcnt(A) + \popcnt(B) + \popcnt(C) = \popcnt(\mathrm{Sum}) + 2\,\popcnt(\mathrm{Carry})$ identically. \qed

\subsection{Three-Subset-Sum Identity, Fixed-Point Overflow Bound, and Proof of Proposition~\ref{prop:asym_cj_gate}(1)}
\paragraph{Derivation of~\eqref{eq:asym_subset_sums} and Overflow-Free \texttt{i16} Accumulation.}
Write each candidate sign as $\sigma_{v,i} = 2 s_{v,i} - 1 \in \{\pm 1\}$ with $s_{v,i} \in \{0, 1\}$.
Expanding $\hat{u}_{v,i} = (2 s_{v,i} - 1)(a_v + \delta_v m_{v,i})$ directly against signed query coordinates $u_{q,i}$ gives
\begin{align}
  \langle u_q, \hat{u}_v\rangle
  &= \sum_{i=1}^D u_{q,i}(2 s_{v,i} - 1)\bigl(a_v + \delta_v m_{v,i}\bigr) \nonumber \\
  &= a_v\!\left(2\sum_{i=1}^D u_{q,i} s_{v,i} - \sum_{i=1}^D u_{q,i}\right) \nonumber \\
  &\quad+ \delta_v\!\left(2\sum_{i=1}^D u_{q,i} s_{v,i} m_{v,i} - \sum_{i=1}^D u_{q,i} m_{v,i}\right) \nonumber \\
  &= a_v\bigl(2 S_+ - S_{\mathrm{tot}}\bigr) + \delta_v\bigl(2 S_{+h} - S_h\bigr),
\end{align}
and substituting $u_{q,i} = (2 s_{q,i} - 1)|u_{q,i}|$ yields the equivalent unsigned sign-agreement form with $\bar{x}_i = \mathbb{I}[s_{q,i} = s_{v,i}]$.
When \texttt{AsymQuerySm} encodes $u_q \in \Sunit$ as $q_{\mathrm{i16},i} = \operatorname{round}(8100\,u_{q,i})$ and accumulates up to four $64$-coordinate chunks ($16$ SIMD steps per \texttt{i16} lane) before widening via \texttt{vpmaddwd} to \texttt{i32}, Cauchy--Schwarz bounds every 16-term lane sum by
\[
  \begin{aligned}
    \sum_{t=1}^{16} |q_{\mathrm{i16},i_t}|
    &\le 8100\sum_{t=1}^{16} |u_{q,i_t}| + 8 \\
    &\le 8100\sqrt{16}\,\|u_q\|_2 + 8 \\
    &= 32{,}408 < 32{,}767 = 2^{15} - 1,
  \end{aligned}
\]
guaranteeing strictly overflow-free \texttt{i16} SIMD accumulation for all unit directions $u_q \in \Sunit$.

\paragraph{Complete Proof of Proposition~\ref{prop:asym_cj_gate}(1).}
\begin{proof}
Let $u_q, u_v \in \Sunit$ be arbitrary deterministic unit vectors with $\gamma_{qv} = \langle u_q, u_v\rangle$.
Define $w \triangleq u_q - \gamma_{qv} u_v$; then $\langle w, u_v\rangle = \langle u_q, u_v\rangle - \gamma_{qv}\|u_v\|_2^2 = 0$ and $\|w\|_2^2 = 1 - \gamma_{qv}^2$.
Taking the inner product of $u_q = \gamma_{qv} u_v + w$ with $\hat{u}_v$ and applying $\langle u_v, \hat{u}_v\rangle = \sigma_v^2 > 0$ from Lemma~\ref{lem:sm_pythagorean}:
\begin{equation}
  \langle u_q, \hat{u}_v\rangle
  = \gamma_{qv}\langle u_v, \hat{u}_v\rangle + \langle w, \hat{u}_v\rangle
  = \gamma_{qv}\sigma_v^2 + \langle w, \hat{u}_v - \sigma_v^2 u_v\rangle,
\end{equation}
where the last step uses $\langle w, u_v\rangle = 0$.
Dividing both sides by $\sigma_v^2$ yields~\eqref{eq:orth_proj_asym} deterministically for every pair $(u_q, u_v)$ without invoking any distributional or rotational-symmetry hypothesis.
In particular, along $u_q = \pm u_v$ ($w = 0$), $\langle \pm u_v, \hat{u}_v\rangle / \sigma_v^2 = \pm 1$ is exact, whereas standard cosine normalization $\langle \pm u_v, \hat{u}_v\rangle / \sigma_v = \pm \sigma_v$ suffers deterministic multiplicative shrinkage $1 - \sigma_v > 0$.
\end{proof}

\section{Conformal Comparison Flip Theory and Suboptimality of Direct Lorentz-MIPS Quantization}
\label{app:conformal_flips}

\begin{proof}[Complete Proof of Theorem~\ref{thm:conformal_flip}]
\textbf{Part (1): Exact flip criterion and depth invariance.}
For any $v \in \Bball$, since $(r_q, f_q, r_v, f_v)$ are stored without quantization error, $R(r_q, r_v)$ and $W(r_q, r_v) = 2 r_q r_v f_q f_v$ are exact, so
\begin{equation}
  \begin{aligned}
    &\widehat{P}_{\mathbb{B}}(q, v) - P_{\mathbb{B}}(q, v) \\
    &= W(r_q, r_v)\bigl(\gamma_{qv} - \hat{\gamma}(u_q, u_v)\bigr) \\
    &= -2 r_q f_q r_v f_v\,\varepsilon_\angle(q, v).
  \end{aligned}
\end{equation}
Given $P_{\mathbb{B}}(q, x) < P_{\mathbb{B}}(q, y)$ with $\Delta_P(q; x, y) = P_{\mathbb{B}}(q, y) - P_{\mathbb{B}}(q, x) > 0$, a ranking flip $\widehat{P}_{\mathbb{B}}(q, y) \le \widehat{P}_{\mathbb{B}}(q, x)$ holds if and only if
\begin{equation}
  \begin{aligned}
    &\Delta_P(q; x, y) + \bigl(\widehat{P}_{\mathbb{B}}(q, y) - P_{\mathbb{B}}(q, y)\bigr) \\
    &\qquad- \bigl(\widehat{P}_{\mathbb{B}}(q, x) - P_{\mathbb{B}}(q, x)\bigr) \le 0 \\
    &\iff 2 r_q f_q\bigl(r_y f_y\,\varepsilon_\angle(q, y)
      - r_x f_x\,\varepsilon_\angle(q, x)\bigr) \\
    &\qquad\ge \Delta_P(q; x, y).
  \end{aligned}
\end{equation}
Since $r_q > 0$ implies $2 r_q f_q > 0$, dividing by $2 r_q f_q$ yields~\eqref{eq:exact_flip_criterion}.
Notice that $\frac{\Delta_P(q; x, y)}{2 r_q f_q} = \frac{(r_q - r_y)^2 f_y - (r_q - r_x)^2 f_x}{2 r_q} + r_y f_y(1 - \gamma_{qy}) - r_x f_x(1 - \gamma_{qx})$ remains completely bounded as $r_q \to 1/\sqrt{c}$ ($f_q \to \infty$ cancels out of both numerator and denominator).
Furthermore, on an isobathic shell $r_x = r_y = r > 0$ ($f_x = f_y = f$), $R(r_q, r_y) = R(r_q, r_x)$ cancels in $\Delta_P$, leaving $\Delta_P(q; x, y) = 2 r_q f_q r f (\gamma_{qx} - \gamma_{qy})$; dividing~\eqref{eq:exact_flip_criterion} by $r f > 0$ yields $\varepsilon_\angle(q, y) - \varepsilon_\angle(q, x) \ge \gamma_{qx} - \gamma_{qy}$, which is independent of both $f_q$ and $f$.

\textbf{Part (2): Cross-shell query-noise non-cancellation.}
Substituting $\varepsilon_\angle(q, v) = \eta_q(u_v) + \eta_v(u_q) + \xi_{qv}$ into $\Delta\widehat{\Psi}_{\mathrm{rank}}(q; x, y) = (\widehat{P}_{\mathbb{B}}(q, x) - P_{\mathbb{B}}(q, x)) - (\widehat{P}_{\mathbb{B}}(q, y) - P_{\mathbb{B}}(q, y))$ and writing $W_v = W(r_q, r_v)$:
\begin{equation}
  \begin{aligned}
    &\Delta\widehat{\Psi}_{\mathrm{rank}}(q; x, y) \\
    &= W_y\eta_q(u_y) - W_x\eta_q(u_x) \\
    &\quad+ \bigl(W_y\eta_y(u_q) - W_x\eta_x(u_q)\bigr) \\
    &\quad+ \bigl(W_y\xi_{qy} - W_x\xi_{qx}\bigr) \\
    &= (W_y - W_x)\eta_q(u_x) + W_y\langle e_q, u_y - u_x\rangle \\
    &\quad+ \bigl(W_y\eta_y(u_q) - W_x\eta_x(u_q)\bigr) \\
    &\quad+ \bigl(W_y\xi_{qy} - W_x\xi_{qx}\bigr).
  \end{aligned}
\end{equation}
At $u_x = u_y$, $\langle e_q, u_y - u_x\rangle = 0$.
The first-order query term has variance $(W_y-W_x)^2u_x^\top\Sigma_qu_x$ for $\Sigma_q=\Cov(e_q)$. Under $\Sigma_q=\frac{\kappa_q}{D}(I_D-u_qu_q^\top)$, $\Var_{e_q}(\eta_q(u_x)) = u_x^\top \Cov(e_q) u_x = \frac{\kappa_q}{D}(1 - \gamma_{qx}^2)$; substituting $W_y - W_x = \frac{2}{c}\sinh\rho_q(\sinh\rho_y - \sinh\rho_x)$ proves~\eqref{eq:cross_shell_var}.
When $|\rho_y - \rho_x| \le \delta_\rho$, $|\sinh\rho_y - \sinh\rho_x| = 2\sinh\frac{|\rho_y - \rho_x|}{2}\cosh\frac{\rho_y + \rho_x}{2} = O(\delta_\rho\cosh\rho_x)$, bounding $(W_y - W_x)^2 = O(\delta_\rho^2 W_x^2)$.
When $e_q = 0$, $\eta_q \equiv 0$ and $\xi_{qv} \equiv 0$ for all $(r_x, r_y)$.

\textbf{Part (3): Direct $(D+1)$-D {Lorentz}-MIPS suboptimality.}
On $\Lmod$, $z = (\frac{1}{\sqrt{c}}\cosh\rho_z, \frac{1}{\sqrt{c}}\sinh\rho_z\,u_z)$.
If $\tilde{z}_v = \frac{1}{\sqrt{c}}\sinh\rho_v\,u_v \in \R^D$ is quantized directly by a coordinate quantizer $\hat{\tilde{z}}_v = \tilde{z}_v + \epsilon_v$ without factoring out $\|\tilde{z}_v\|_2 = \frac{1}{\sqrt{c}}\sinh\rho_v$, the stated model $\Cov(\epsilon_v)=\Theta(\|\tilde z_v\|_2^2/D)I_D$ gives $\Var(\langle \tilde{z}_q, \epsilon_v\rangle) = \Theta(\frac{1}{D}\|\tilde{z}_q\|_2^2 \|\tilde{z}_v\|_2^2) = \Theta(\frac{1}{c^2 D}\sinh^2\rho_q\sinh^2\rho_v)$.
In particular, when $u_x = u_y = u_q$ ($\theta_{qx} = \theta_{qy} = 0$) with $\rho_x \ne \rho_y$, polar decoupling has $\langle u_q, \hat{u}_x\rangle/\sigma_x^2 = 1$ (or exact radial comparison $R(r_q, r_v)$), whereas direct Lorentz coordinate quantization suffers $\Theta(e^{\rho_q + \max(\rho_x,\rho_y)}/\sqrt{D})$ noise along the radial direction itself and provides no $O(1)$ radial lower bound $P_{\mathrm{rad}}$ prior to scanning $D$ coordinates.
\end{proof}

\section{Hyperbolic Vamana Pathologies, Stem-Excess Dominance, and Chord-Swap Invariants}
\label{app:vamana_pathologies}

\subsection{Complete Proof of Theorem~\ref{thm:vamana_pathologies}}
\begin{proof}
\textbf{Claim (1): Radial occlusion bound.}
Let $u, p, v \in \Bball$ have geodesic radii $R_u, R_p, R_v$ from the origin $0$, with $\alpha R_p < R_u$.
Applying the metric triangle inequality on $(\Bball, d_{\mathbb{B}})$ through $0$ gives two angle-independent bounds valid for \emph{all} spherical directions $u_u, u_p, u_v \in \Sunit$:
\begin{equation}
  \begin{aligned}
    d_{\mathbb{B}}(p, v) &\le d_{\mathbb{B}}(p, 0) + d_{\mathbb{B}}(0, v) \\
    &= R_p + R_v, \\
    d_{\mathbb{B}}(u, v) &\ge \bigl|d_{\mathbb{B}}(u, 0) - d_{\mathbb{B}}(0, v)\bigr| \\
    &\ge R_u - R_v.
  \end{aligned}
\end{equation}
Whenever $R_v \le \frac{R_u - \alpha R_p}{\alpha + 1}$, rearranging $(\alpha + 1)R_v \le R_u - \alpha R_p$ yields $\alpha(R_p + R_v) \le R_u - R_v$, whence
\begin{equation}
  \alpha\,d_{\mathbb{B}}(p, v) \le \alpha(R_p + R_v) \le R_u - R_v \le d_{\mathbb{B}}(u, v)
\end{equation}
for every $\theta_{pv} \in [0, \pi]$ (attaining worst-case equality on $d_{\mathbb{B}}(p, v) = R_p + R_v$ when $u_v = -u_p$, i.e., $180^\circ$ opposite $p$).

To isolate the curvature-specific angular contraction from this generic metric ball, consider any same-shell pair $u, v \in \Bball$ at geodesic radius $R_u = R_v = R > 0$ ($\rho = \sqrt{c}\,R$) and angle $\theta_{uv} \in [0, \pi]$ relative to the origin neighbor $p = 0$ ($R_p = 0 < R/\alpha$).
Substituting $\rho_u = \rho_v = \rho$ into the warped-product identity $\frac{c}{4}P_{\mathbb{B}}(u, v) = \sinh^2(\frac{\sqrt{c}}{2}d_{\mathbb{B}}(u, v)) = \sinh^2\rho\,\sin^2(\theta_{uv}/2)$ yields $\sinh(\frac{\sqrt{c}}{2}d_{\mathbb{B}}(u, v)) = \sinh\rho\,\sin(\theta_{uv}/2)$.
Because $d_{\mathbb{B}}(0, v) = R$, the pruning test $\alpha\,d_{\mathbb{B}}(0, v) \le d_{\mathbb{B}}(u, v)$ holds if and only if
\begin{equation}
  \sin\!\left(\frac{\theta_{uv}}{2}\right) \ge \frac{\sinh\!\left(\frac{\alpha}{2}\rho\right)}{\sinh\rho}.
\end{equation}
In the flat limit $c \to 0^+$, $\sinh(\frac{\alpha}{2}\rho)/\sinh\rho \to \alpha / 2$, keeping the surviving unoccluded cone constant at $\theta_{\mathrm{surv}} = 2\arcsin(\alpha/2) \approx 73.7^\circ$ for $\alpha = 1.2$.
Under negative curvature $-c < 0$ and $\alpha \in (1, 2)$, $\sinh(\frac{\alpha}{2}\rho)/\sinh\rho \sim e^{-\frac{2-\alpha}{2}\rho}$ decays exponentially with depth $\rho$, shrinking the surviving cone to $23.1^\circ$ at $\rho = 4$ and $10.4^\circ$ at $\rho = 6$.

\textbf{Claim (2): Angular escape condition.}
Since $\arccosh$ is strictly increasing on $[1, \infty)$ and $d_{\mathbb{B}}(q, 0) = R_q = \rho_q / \sqrt{c}$, we have $d_{\mathbb{B}}(q, v) < d_{\mathbb{B}}(q, 0)$ if and only if
\begin{equation}
  \begin{gathered}
    \cosh\rho_q\cosh\rho_v - \sinh\rho_q\sinh\rho_v\,\gamma_{qv} < \cosh\rho_q \\
    \iff \cosh\rho_q(\cosh\rho_v - 1) < \sinh\rho_q\sinh\rho_v\,\gamma_{qv}.
  \end{gathered}
\end{equation}
For $R_q, R_v > 0$, $\sinh\rho_q\sinh\rho_v > 0$; dividing through and using the half-angle identity $\frac{\cosh\rho_v - 1}{\sinh\rho_v} = \frac{2\sinh^2(\rho_v/2)}{2\sinh(\rho_v/2)\cosh(\rho_v/2)} = \tanh(\rho_v/2)$ gives $\gamma_{qv} > \frac{\tanh(\rho_v/2)}{\tanh\rho_q} = \frac{\tanh(\sqrt{c}R_v/2)}{\tanh(\sqrt{c}R_q)}$.

\textbf{Claim (3): Offset-intercept and proxy non-homogeneity distortion.}
If $\hat{d}(x, y) = a\,\tilde{d}(x, y) + b$ with $a > 0, b \ne 0$, then
\begin{equation}
  \hat{d}(u, v) - \alpha\hat{d}(w, v) = a\bigl(\tilde{d}(u, v) - \alpha\tilde{d}(w, v)\bigr) + (1 - \alpha)b.
\end{equation}
Since $\alpha > 1$, $(1-\alpha)b<0$ for $b>0$, making the nonnegative pruning-margin test harder and producing under-pruning. For $b<0$, the test becomes easier and produces over-pruning.
For direct proxy slack $\alpha P_{\mathbb{B}}(w, v) \le P_{\mathbb{B}}(u, v)$, note from~\eqref{eq:psi_prune} that exact geodesic slack requires $P_\alpha(P) = \frac{2}{c}(\cosh(\alpha\arccosh(1 + \frac{c}{2}P)) - 1)$: as $P \to 0^+$, $P_\alpha(P) \sim \alpha^2 P$, whereas as $P \to \infty$, $P_\alpha(P) \sim \frac{1}{c}(c P)^\alpha = c^{\alpha - 1} P^\alpha \gg \alpha P$ since $\alpha > 1$.
The exact pruning test is $P_\alpha(P(w,v))\le P(u,v)$. Since $P(d)=\frac{2}{c}(\cosh(\sqrt c\,d)-1)$ is strictly convex with $P(0)=0$, $P(\alpha d)>\alpha P(d)$ for $d>0$, $\alpha>1$. Replacing $P_\alpha(P)$ by $\alpha P$ expands the set of accepted dominance pairs and produces over-pruning.
\end{proof}

\subsection{Reverse-Triangle Shortcut and Channel-Wise Flat Limits}
\paragraph{Lossless $O(1)$ Reverse-Triangle Shortcut.}
By the reverse triangle inequality on $(\Bball, d_{\mathbb{B}})$, $\tilde{d}(w, v) = \sqrt{c}\,d_{\mathbb{B}}(w, v) \ge |\sqrt{c}\,d_{\mathbb{B}}(0, w) - \sqrt{c}\,d_{\mathbb{B}}(0, v)| = |\rho_w - \rho_v|$.
Therefore, if $\alpha|\rho_w - \rho_v| > \tilde{d}_{\mathrm{eff}}(u, v; w)$, we immediately have $\alpha\tilde{d}(w, v) \ge \alpha|\rho_w - \rho_v| > \tilde{d}_{\mathrm{eff}}(u, v; w)$, proving $\neg\operatorname{Dom}_{\mathrm{geo}}(w \mid u, v)$ with zero false rejections and without computing $\arccosh$.

\paragraph{Constant-Depth and Flat-Curvature Limits.}
For constant depth $\rho_v=\rho_0$, both elevator candidate sets are empty and $\Delta_{\mathrm{stem}}=0$. Ring ordering uses angular loss alone, and every selected neighbor satisfies the radial shield. The geodesic channel retains $\alpha d(w,v)\le d(u,v)$, whereas the ring uses $\alpha\ell_\angle(w,v)\le\ell_\angle(u,v)$. In the limit $c\to0^+$ at fixed Euclidean coordinates, $f_z\to2$ and $d_{\mathbb B}(x,y)\to2\|x-y\|_2$. At constant Euclidean radius, the geodesic-channel condition consequently becomes $\alpha^2\ell_\angle(w,v)\le\ell_\angle(u,v)$, while the ring keeps its angular $\alpha$ test. Both channels select spherical diversity at their respective slack values. \qed

\subsection{Degree-$M_0$ and Reachability Invariant of Chord-Swap Island Repair}
\begin{proposition}[Spanning-Tree Chord-Swap Invariant]
\label{prop:chord_swap_invariant}
Suppose the entry set $\mathcal{E} \subset V$ is non-empty, $(\mathcal{R}, \texttt{parent})$ is initialized by directed {BFS} from $\mathcal{E}$, and Algorithm~\ref{alg:compass_build}'s single repair pass updates edge slots, $\deg^-(\cdot)$, and $\texttt{parent}[\cdot]$ sequentially.
Whenever an unreachable vertex $u \notin \mathcal{R}$ selects a reachable anchor $p \in \mathcal{R}$ that either satisfies $\deg^+(p) < M_0$ or owns a non-tree chord $p \to w^\star$ with $\texttt{parent}[w^\star] \ne p$ and $\deg^-(w^\star) \ge 2$:
(i)~all vertices preserve $\deg^+(v) \le M_0$,
(ii)~every vertex previously having $\deg^-(w) \ge 1$ (including $w^\star$ and $u$) maintains $\deg^-(w) \ge 1$,
(iii)~zero {BFS} spanning-tree edges in $\{(\texttt{parent}[v], v) : v \in \mathcal{R} \setminus \mathcal{E}\}$ are overwritten, so every vertex in $\mathcal{R}$ remains permanently reachable from $\mathcal{E}$, and
(iv)~attaching $p \to u$ plus incremental directed {BFS} from $u$ strictly increases $|\mathcal{R}|$ by at least $1$.
Hence, whenever every remaining island locates a qualifying anchor $p$, the single pass terminates with $100\%$ directed reachability ($|\mathcal{R}| = N$).
\end{proposition}
\begin{proof}
Let $T = \{(\texttt{parent}[v], v) : v \in \mathcal{R} \setminus \mathcal{E}\}$ denote the current directed {BFS} spanning tree rooted at $\mathcal{E}$.
By induction on the {BFS} depth stored in $\texttt{parent}[\cdot]$, every $v \in \mathcal{R}$ retains a directed path from $\mathcal{E}$ inside $T$.
If $\deg^+(p) < M_0$, appending $p \to u$ preserves $\deg^+(p) \le M_0$, deletes no edges of $G$, and sets $\deg^-(u) \ge 1$.
If $\deg^+(p) = M_0$, replacing $p \to w^\star$ with $p \to u$ leaves $\deg^+(p) = M_0$ unchanged, increments $\deg^-(u)$ by one, and decrements $\deg^-(w^\star)$ to $\deg^-(w^\star) - 1 \ge 1$.
Because the explicit guard $\texttt{parent}[w^\star] \ne p$ ensures $(p, w^\star) \notin T$, $T$ remains an intact subgraph of $G$ and no previously reachable vertex in $\mathcal{R}$ can become disconnected through cycle breakage.
Since $p \in \mathcal{R}$ stays reachable and $p \to u$ is present, setting $\texttt{parent}[u] = p$ validly extends $T$ to $u$ and, via incremental directed {BFS} from $u$, to every unvisited vertex reachable from $u$, strictly enlarging $|\mathcal{R}|$.
When every island $u \notin \mathcal{R}$ encountered by the pass obtains a qualifying anchor, no unreachable vertices can remain upon completion.
\end{proof}

\section{Tangent-Cone Recovery, Pool-Dilution Inversion, Off-Axis CJ-Gate, and Sketch Variance}
\label{app:tangent_cone}

\subsection{Complete Proof of Theorem~\ref{thm:tangent_cone}}
\begin{proof}
\textbf{Part (1).}
For equal radial and axial coordinates, subtracting~\eqref{eq:tangent_inner} gives $\gamma_{qy}-\gamma_{qx}=\beta_{q,c}\beta\Delta_{\tilde r}$. Global proxy error is $-Wg_v$, so the comparison-error variance is $W^2\Var(g_y-g_x)=W^2V_g$. Division by its standard deviation proves~\eqref{eq:mnr_collapse}. The covariance term captures the shared query, common bitplanes, and threshold crossings of nearby candidates.

\textbf{Part (2).}
Condition on the retained cone pool and its true-neighbor membership. Under uniform score ranks, every member has probability $B/N_c(L)$ of occupying a rerank slot. Let $I_v$ indicate admission of true neighbor $v$. Linearity of expectation gives
\begin{equation}
 \mathbb E\!\left[\sum_{v\in G_k\cap C_L\cap c}I_v\right]
 =\sum_{v\in G_k\cap C_L\cap c}\frac{B}{N_c(L)}
 =\frac{B k_c(L)}{N_c(L)}.
\end{equation}
Taking the ratio at two widths proves the stated decrease condition. For arbitrary ranks the same identity is $\mathbb E[|G_k\cap F_B|\mid C_L]=\sum_{v\in G_k\cap C_L}\Pr(v\in F_B\mid C_L)$. The implementation orders score ties deterministically, and the four-stage audit measures these actual admission counts.

\textbf{Part (3).}
Subtract~\eqref{eq:tangent_inner} from~\eqref{eq:cone_estimator}. The axial product cancels and yields $\hat\gamma_{\mathrm{cone}}-\gamma=\beta_{q,c}\beta_v e_v$ for every pair. For equal $\beta_x=\beta_y=\beta$, comparison-error variance is $W^2\beta_{q,c}^2\beta^2V_r$. Cancelling $W\beta_{q,c}\beta$ proves~\eqref{eq:mnr_recovery}. Equal marginal variances $\kappa_{\mathrm{SM}}/D$ with correlation $\varrho$ give $V_r=2\kappa_{\mathrm{SM}}(1-\varrho)/D$ by direct substitution.
\end{proof}

\subsection{FP32 Polar and Axial Scalar Perturbations}
\label{app:scalar_rounding}
Polar headers store scalars computed in FP64 and rounded independently to FP32. The implementation stores $F=f/2$, so the following bounds for $f$ apply to twice the stored boundary factor. For a normal finite scalar, round-to-nearest gives $|\delta r|\le u_{32}|r|$ and $|\delta f|\le u_{32}|f|$, where $u_{32}=2^{-24}$. In particular, the stored conformal factor retains relative error at most $u_{32}$ from this final conversion.
For an FP64 query, define $A=r_q^2+r^2-2r_qr\gamma$ and $T=2|r-r_q\gamma|\epsilon_r+\epsilon_r^2$, with $\epsilon_r=u_{32}|r|$ and $\epsilon_f=u_{32}|f|$. Header conversion changes the real-valued proxy by at most
\begin{equation}
 |\widetilde P-P|\le f_q\bigl[fT+\epsilon_f(A+T)\bigr].
 \label{eq:header_rounding}
\end{equation}
Indeed, replacing $r$ by $r+\delta r$ changes $A$ by $2(r-r_q\gamma)\delta r+\delta r^2$, and multiplying by $f+\delta f$ gives the bound. Setting $\gamma=1$ gives the radial-gate perturbation. With the mapped-data evaluation clamp $r\le1-10^{-6}$ at $c=1$, $f\le1{,}000{,}000.501$ and $\epsilon_f\le0.059605$.
For an axial scalar $\tilde\alpha=\alpha+\delta\alpha$ clipped to $[-1,1]$, define $\tilde\beta=\sqrt{1-\tilde\alpha^2}$. Then
\begin{equation}
 |\tilde\beta-\beta|
 =\frac{|2\alpha\delta\alpha+(\delta\alpha)^2|}{\tilde\beta+\beta}
 \label{eq:axial_rounding}
\end{equation}
whenever the denominator is positive. The cone scorer clips the square-root argument at zero and uses the axial product for a degenerate residual. Final reranking evaluates the original FP32 coordinates with FP64 accumulation.

\subsection{Off-Axis CJ-Gate Jacobian and $5.49\times$ Underestimation Counterexample}
From $P_{\mathbb{B}}(q, v) = R(r_q, r_v) + 2 r_q r_v f_q f_v(1 - \gamma_{qv})$ and~\eqref{eq:cone_estimator}, only $\hat{\gamma}_{\mathrm{asym}}(\tilde{r}_{q,c}, \tilde{r}_v)$ carries quantization uncertainty $\sigma_0 = \sqrt{\kappa_{\mathrm{SM}}/D}$, multiplied by $\frac{\partial \hat{\gamma}_{\mathrm{cone}}}{\partial \hat{\gamma}_{\mathrm{asym}}} = \beta_{q,c}\beta_v$ and $\bigl|\frac{\partial P_{\mathbb{B}}}{\partial \gamma_{qv}}\bigr| = 2 r_q r_v f_q f_v$, proving~\eqref{eq:cj_sigma}.
If an implementation approximates $q$ as lying at the cluster center ($\beta_{q,c} \approx \beta_v$), replacing $\beta_{q,c}\beta_v$ by $\beta_v^2$ mis-scales $\sigma_P(q, v)$ by $\frac{\beta_{q,c}}{\beta_v}$: for a boundary-cone query at $\alpha_{q,c} = 0.84$ ($\beta_{q,c} = \sqrt{1 - 0.84^2} \approx 0.5426$) evaluating a near-axis candidate at $\alpha_v = 0.9951$ ($\beta_v = \sqrt{1 - 0.9951^2} \approx 0.0989$), $\frac{\beta_{q,c}}{\beta_v} = \frac{0.5426}{0.0989} \approx 5.49$, underestimating true plateau uncertainty by $5.49\times$.
Using exact $\tilde{\beta}(q, v) = \beta_{q,c}\beta_v$ eliminates this underestimation across all query angles.

\subsection{Binomial Variance Propagation for \texttt{SymSketch256}}
Let $H \in \{0, \dots, B_s\}$ ($B_s = 256$) be the Hamming distance over $B_s$ sampled coordinates with disagreement probability $p \in (0, 1)$, so $\E[H/B_s] = p$ and $\Var(H/B_s) = \frac{p(1 - p)}{B_s}$.
Applying the first-order delta method to $g(p) = \cos(\pi p)$ ($|g'(p)| = \pi\sin(\pi p)$) gives
\begin{equation}
  \sigma_{\mathrm{sk}}(p) = \bigl|g'(p)\bigr|\sqrt{\Var(H/B_s)} = \frac{\pi}{\sqrt{B_s}}\sin(\pi p)\sqrt{p(1 - p)}.
\end{equation}
Define $u(p) = \sin(\pi p)\sqrt{p(1 - p)}$.
At high-similarity reference $p_{\mathrm{ref}} = 48 / 256 = 0.1875$ ($\cos(\pi p_{\mathrm{ref}}) \approx 0.8315$), $u(p_{\mathrm{ref}}) = \sin(0.1875\pi)\sqrt{0.1875 \times 0.8125} \approx 0.5556 \times 0.3903 \approx 0.2168$.
Near orthogonality $p = 115 / 256 \approx 0.4492$ ($\cos(\pi p) \approx 0.159$), $u(p) = \sin(0.4492\pi)\sqrt{0.4492 \times 0.5508} \approx 0.9873 \times 0.4974 \approx 0.4911$, giving $\frac{u(115/256)}{u(48/256)} \approx \frac{0.4911}{0.2168} \approx 2.265$ (peaking at $0.5 / 0.2168 \approx 2.306$ at $p = 0.5$). \qed

\section{Four-Stage Telescoping Recall-Loss Identity and Supplementary Breakdowns}
\label{app:loss_decomposition}

\subsection{Proof of the Four-Stage Telescoping Recall-Loss Identity~\eqref{eq:four_stage_loss}}
\begin{proof}
Because every stage filters the output of the preceding stage ($R_k(q) \subseteq F_B(q) \subseteq C_L(q) \subseteq V(q)$), intersecting with $G_k(q)$ preserves inclusion:
\begin{equation}
  \begin{aligned}
    G_k(q) \cap R_k(q) &\subseteq G_k(q) \cap F_B(q) \\
    &\subseteq G_k(q) \cap C_L(q) \\
    &\subseteq G_k(q) \cap V(q) \subseteq G_k(q).
  \end{aligned}
\end{equation}
For any finite nested chain $A_4 \subseteq A_3 \subseteq A_2 \subseteq A_1 \subseteq A_0$ with $|A_0| = k$, $|A_0| - |A_4| = |A_0 \setminus A_1| + |A_1 \setminus A_2| + |A_2 \setminus A_3| + |A_3 \setminus A_4|$; dividing by $k$ yields~\eqref{eq:four_stage_loss}.
Moreover, suppose exact IEEE-754 FP64 reranking sorts $F_B(q)$ strictly by true geodesic distance $d_{\mathbb{B}}(q, \cdot)$ (with consistent tie-breaking matching $G_k(q)$) and returns the top $k$ elements $R_k(q)$ ($|F_B(q)| \ge k$).
For any $v^\star \in G_k(q) \cap F_B(q)$, at most $k - 1$ points in the entire dataset $\mathcal{X}$---and a fortiori at most $k - 1$ points in $F_B(q) \subseteq \mathcal{X}$---precede $v^\star$ in exact geodesic order, so $v^\star$ must lie in the top $k$ of $F_B(q)$, i.e., $v^\star \in R_k(q)$.
Hence $G_k(q) \cap (F_B(q) \setminus R_k(q)) = \emptyset$ and $\mathcal{L}_{\mathrm{rerank}}(q) \equiv 0$.
\end{proof}

\subsection{Supplementary Stage-Wise Audit and Component Ablation Matrix}

\begin{table}[h]
  \centering
  \caption{Controlled component ablation on {Clustered-Narrow} ($N=1\text{M}, D=512, \rho \in [3.9, 4.1]$, fixed $L=256, B=128$, AMD Zen~4 32T build / 1T search).}
  \label{tab:app_stage_audit}
  \small
  \setlength{\tabcolsep}{3pt}
  \begin{tabularx}{\linewidth}{@{}>{\raggedright\arraybackslash}Xrrrr@{}}
    \toprule
    \textbf{Configuration} & \textbf{Cones} & \textbf{Build (32T)} & \textbf{Recall@10} & \textbf{1T QPS} \\
    \midrule
    Global-SM (No Rescue) & 0 & 21.9\,s & 34.22\% & \textbf{2,071} \\
    + \textsc{CJ-Gate} Only & 0 & 21.9\,s & 50.59\% & 1,884 \\
    + Var.-Stab.\ \texttt{SymSketch256} & 0 & 21.5\,s & 34.22\% & 1,960 \\
    + \texttt{ResidualConeDict} + \textsc{CJ-Gate} & 256 & 25.7\,s & 99.37\% & 1,958 \\
    \textbf{\rquiver{}} & \textbf{256} & 25.3\,s & \textbf{99.39\%} & 1,840 \\
    \bottomrule
  \end{tabularx}
\end{table}

Table~\ref{tab:app_stage_audit} reports the controlled fixed-budget ablation ($L=256, B=128$, independent Zen~4 graph builds per configuration) on {Clustered-Narrow}:
on the diagnostic subset, {Global-SM} traversal without tangent-cone rescue or adaptive gating visits $93.9\%$ of true neighbors ($\mathcal{L}_{\mathrm{reach}} = 6.1\%$) yet loses $44.6\%$ inside the bounded $L$-beam ($\mathcal{L}_{\mathrm{beam}}$) and $16.3\%$ at pool gating ($\mathcal{L}_{\mathrm{gate}}$, $33.0\%$ subset / $34.22\%$ full Recall@10).
The full-query measurements raise recall from $34.22\%$ to $50.59\%$ with \textsc{CJ-Gate} alone ($1{,}884$\,QPS). Residual encoding with \textsc{CJ-Gate} reaches $99.365\%$ ($1{,}958$\,QPS), and full \rquiver{} reaches $99.385\%$ ($1{,}840$\,QPS 1T and $38{,}876$\,QPS 24T). At $L=384,B=128$, full-method recall is $99.80\%$. Residual encoding reuses the existing per-vertex bitplanes and header slots.

\fi

\end{document}